\documentclass[aps,prl,reprint,amsmath,amssymb,superscriptaddress,longbibliography]{revtex4-2}

\usepackage{graphicx}
\usepackage{dcolumn}
\usepackage{bm}
\usepackage{amsmath}
\usepackage{amssymb}
\usepackage{xcolor}
\usepackage{xspace}
\usepackage{xfrac}
\usepackage{color,soul}
\usepackage{graphicx}

\usepackage{ulem} 
\usepackage[colorlinks=true,urlcolor=blue,citecolor=blue,linkcolor=blue,bookmarks=false,pdfstartview={FitH}]{hyperref}
\begin{document}
\preprint{APS/123-QED}
\newcommand{\pv}[1]{\textcolor{black}{#1}}
\newcommand{\AV}[1]{\textcolor{blue}{AV:#1}}
\newcommand{\corr}[1]{\textcolor{blue}{#1}}

\title{Tunable $t-t'-U$ Hubbard models in twisted square homobilayers}

\author{P. Myles Eugenio} 
\email{paul.eugenio@uconn.edu}
\affiliation{Department of Physics, University of Connecticut, Storrs, Connecticut 06269, USA}
\affiliation{Department of Physics, Harvard University, Cambridge, Massachusetts 02138, USA}
\author{Zhu-Xi Luo}
\email{zhuxi\_luo@gatech.edu}
\affiliation{Department of Physics, Harvard University, Cambridge, Massachusetts 02138, USA}
\affiliation{School of Physics, Georgia Institute of Technology, Atlanta, Georgia 30332, USA}
\author{Ashvin Vishwanath}
\email{avishwanath@g.harvard.edu}
\affiliation{Department of Physics, Harvard University, Cambridge, Massachusetts 02138, USA}
\author{Pavel A. Volkov}\email{pavel.volkov@uconn.edu}
\affiliation{Department of Physics, University of Connecticut, Storrs, Connecticut 06269, USA}
\affiliation{Department of Physics, Harvard University, Cambridge, Massachusetts 02138, USA}

\begin{abstract}
Square lattice Hubbard model with tunable hopping ratio $t'/t$ is highly promising for realizing a variety of quantum phases, including higher-temperature superconductivity. We show that twisted square lattice homobilayers offer such tunability when the flat bands originate from the corner of the Brillouin zone. An emergent symmetry, generically present at low twist-angles, enforces $t=0$ for these bands. Breaking it with interlayer displacement field or in-plane magnetic field introduces $t$ and anisotropy, tunable in a wide range for correlated electrons on the moir\'e square lattice.
\end{abstract}
\maketitle

The discovery of two-dimensional Van der Waals materials \cite{liu2016van,doi:10.1126/science.aac9439,Mounet_2018} has recently led to the realization of a number of exotic states of matter in twisted multilayers, which feature moir\'e lattices~\cite{balents2020}. These systems allow to realize electronic correlations \cite{PhysRevLett.121.026402,tang2020simulation} and magnetism \cite{burch2018magnetism}, tunable in-situ by, e.g., gating \cite{PhysRevLett.121.026402,Liu&all&Vafek&Li2021_moire_screening}.


While the majority of the current moir\'e materials are built from hexagonal layers \cite{Kennes}, there is significant interest in achieving highly tunable {\it square lattice} platforms. The square lattice Hubbard model is believed to capture the essential physics of the high-Tc superconductivity \cite{deng2015emergent,huang2017numerical,zheng2017stripe,jiang2019superconductivity,flatiron_2020,tj2021,annurev_num,annurev_an,Chen&Haldane&Sheng2023} in cuprates and possibly nickelates \cite{Anisimov,Pan,Hwang,kitatani2020}. Tunable Hubbard model simulators have been developed in cold-atom systems \cite{BOHRDT}, but accessing low temperature physics there remains an ongoing challenge.  Moir\'e platforms, on the other hand, can combine tunability with elevated electronic energy scales. 



To date, twisted square lattices were explored in certain specific contexts: twisted bilayers of Bravais lattices \cite{Toshi}, flux states \cite{PhysRevB.104.035136,PhysRevB.105.165422}, quadratic band touching systems \cite{PhysRevResearch.4.043151,sarkar2023symmetrybased}, cuprates \cite{Can_2021,zhao2021emergent,PhysRevB.105.L201102,PhysRevB.106.104505,PhysRevB.107.174506,Volkov2022} and FeSe \cite{10.21468/SciPostPhys.15.3.081}. They have also been recently implemented in a number of analogue systems, such as cold atoms \cite{cirac2019,salamon2020,meng2023atomic}, optical \cite{fu2020optical,wang2020localization,
lou2021} and acoustic \cite{phon2022} metamaterials. 

\begin{figure}[t]
    \includegraphics[width=\linewidth]{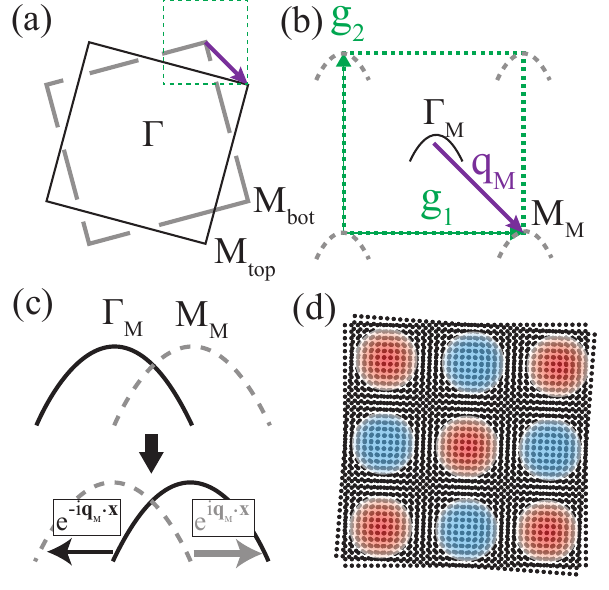}
    \caption{(a) Momentum space of a twisted square lattice bilayer; green dashed line marks the moir\'e Brillouin zone near the microscopic $M$ point. (b) Moir\'e Brillouin zone with marked maxima of the single-layer dispersion. 
    (c,d) Action of the $\mathcal{L}$-symmetry in (c) momentum and (d) real space. (c) Layers are interchanged and then momentum of top (bottom) layer is shifted by -(+) $q_M$. (d) The momentum shifts in (c) lead to an alternation of the wavefunction sign at AA sites, forbidding the nearest-neighbor hopping on the moir\'e lattice.
    }
    \label{Fig:Main:fig1}
\end{figure}

In this work, we reveal that the moir\'e bands of twisted square lattice bilayers formed around the the Brillouin zone corner (Fig. \ref{Fig:Main:fig1} (a,b)) generically possess an approximate symmetry at low twist angles, which we refer to as $\mathcal{L}$- symmetry ($\mathcal{L}$ for ``layer") (Fig. \ref{Fig:Main:fig1} (c)). This symmetry involves interchanging the bands of two layers, but does not reduce to any mirror or rotational symmetries of the bilayer.
In real space, $\mathcal{L}$ alternates the sign of the flat band wave-function between nearest neighbor moir\'e sites (Fig. \ref{Fig:Main:fig1} (d)), forbidding nearest neighbor tunneling $t$. Interlayer displacement field or in-plane magnetic field break the $\mathcal{L}$-symmetry, allowing $t$ to be continuously increased. Tuning the twist angle and displacement field allows for the realization of $t-t'-U$ Hubbard models with widely tunable $t/t'$ ratios, critical for superconductivity \cite{jiang2019superconductivity,white2020,jiang2020,danilov2022degenerate,tj2021,tj2024,qu2024} and frustrated spin states \cite{jiang2012,imada2021,qianj1j22024,ruckriegel2024}.

\begin{figure}[t!]
    \centering
\includegraphics[width=0.45\textwidth]{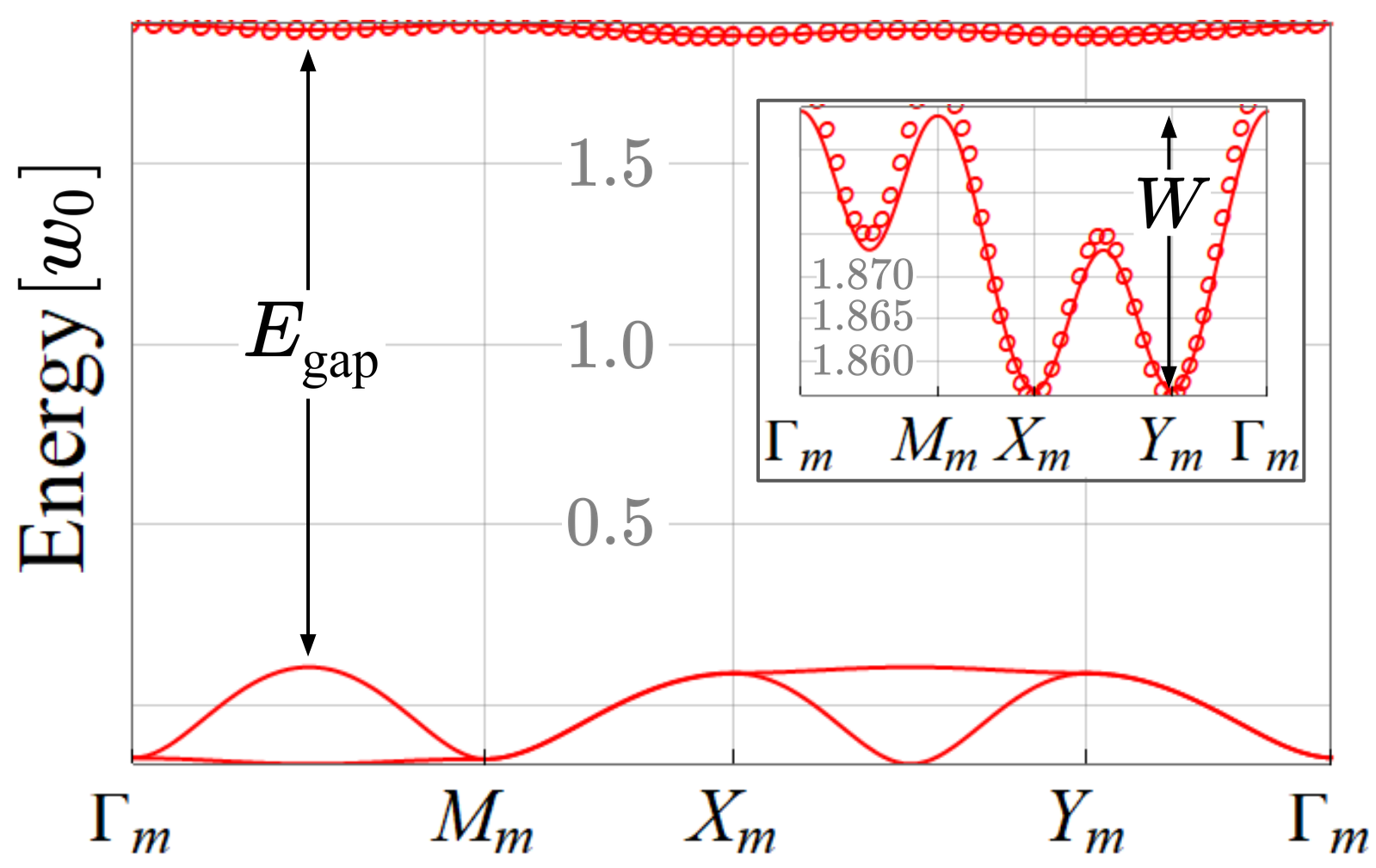}
\includegraphics[width=0.45\textwidth]{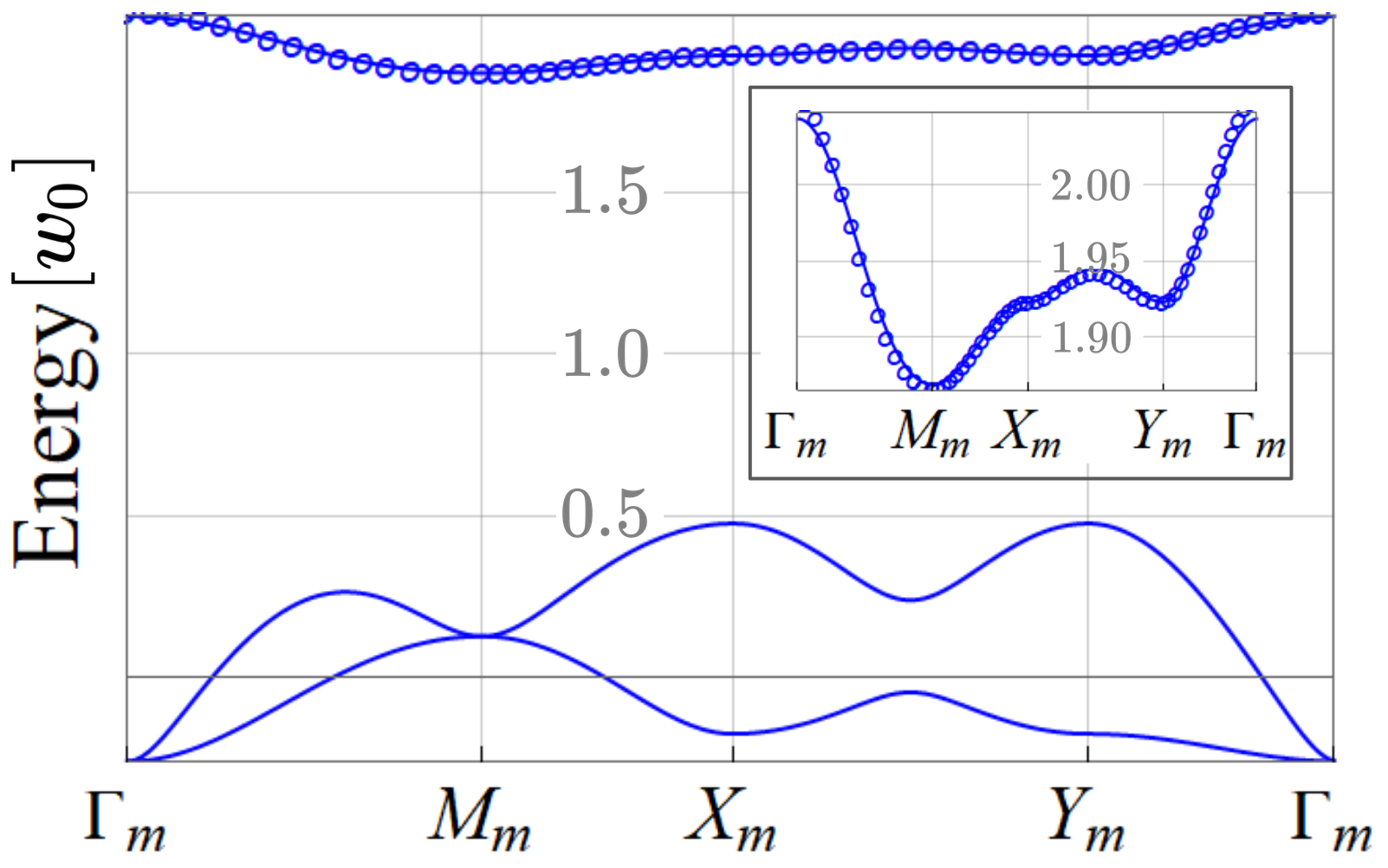}   
    \caption{(a) Band structure (in units of interlayer tunneling amplitude $w_0$) of the model Eq. \eqref{Eqn:Main:H_leading} with $\theta/\theta^*=1/\sqrt{3}$ and (b) with an additional inter-layer displacement field $V=5V_c$ \eqref{vcdef}. Insets show the details of the top flat band, separated by $E_{gap}$ from the rest. 
    Numerical results using plane-wave expansion are shown as continuous lines. Circles represent the analytic tight-binding model for the top band -- see Eqn's \ref{Eqn:Main:epsilonK}, \ref{Eqn:Main:tp}, \& \ref{Eqn:Main:t}; for (a) we find $t=0$, $t'\approx 0.014 w_0$, while in (b) we find $t\approx 0.07$, $t'\approx 0.014 w_0$ giving $t/t'=5$.}
\label{fig:Main:dispersion}
\end{figure}

{\it Model:} We consider twisted homobilayers of square-lattice materials using the continuum description at low twist angles, focusing on the corner ($M$) of the original Brillouin Zone (Fig. \ref{Fig:Main:fig1} (a)). To begin with, a simplified continuum model appropriate for the low-twist angle limit can be obtained by approximating twisting by a momentum shift \footnote{See Supplementary Material that includes Refs. \cite{theory_Vafek&Kang2018_TBGwannier,theory_Vafek&Kang2023_T(u),Ballentine,Soper_notes,glazman2011}}   (Fig. \ref{Fig:Main:fig1} (b)) \cite{bistritzer2011moire,hejazi2019}:
\begin{eqnarray}\label{Eqn:Main:H_leading}
H_0 = 
\begin{pmatrix}
 \mu\boldsymbol{\nabla}_{\bf x}^2  & T({\bf x})\\
T({\bf x})^* & \mu(\boldsymbol{\nabla}_{\bf x}+i{\bf q}_M)^2 \\
\end{pmatrix},
\end{eqnarray}
where $T({\bf x})=w_0(1+e^{i({\bf g}_1+{\bf g}_2)\cdot{\bf x}}+e^{i{\bf g}_1\cdot{\bf x}}+e^{i{\bf g}_2\cdot{\bf x}})$, ${\bf g}_{1,2}$ being the reciprocal vectors of the moir\'e lattice, $|{\bf g}_{1,2}| = \frac{2 \pi}{l_M}$. $l_M = l_a/\theta$ and $l_a$ - are the moir\'e and microscopic lattice constants, and  ${\bf q}_M = ({\bf g}_1+{\bf g}_2)/2$. Comparing the inter-layer tunneling strength $w_0$ to the kinetic energy $\mu q_M^2/4$, we introduce a characteristic twist angle value
\begin{eqnarray}
\theta^* = \sqrt{\frac{w_0}{\frac{1}{4}\mu(\sqrt{2}\pi/l_a)^2}},
\end{eqnarray}
such that emergence of isolated flat bands is expected for $\theta<\theta^*$. An example of the band structure of the model \eqref{Eqn:Main:H_leading} for $\frac{\theta}{\theta^*}=\frac{1}{\sqrt{3}}$ is shown in Fig. \ref{fig:Main:dispersion} (a). The top band has a much smaller bandwidth $W$ compared to others and is separated from them by a gap $E_{gap}>W$, forming an isolated flat band. Its dispersion (Fig. \ref{fig:Main:dispersion} (a), inset) fits well to a tight-binding model:
\begin{eqnarray}\label{Eqn:Main:epsilonK}
&&\epsilon_{\bf k}=2t\big(\cos(k_x)+\cos(k_y)\big) \notag\\
&+&2t'\Big(\cos\big(k_x+k_y\big)+\cos\big(k_x-k_y\big)\Big) ,
\end{eqnarray}
with $t$ equal to zero and $t'\approx 0.014 w_0$. 
In real space, the system thus separates into two decoupled moir\'e sublattices, consistent with the microscopic tight-binding calculations \cite{Toshi}. Unlike that case, Eq. (\ref{Eqn:Main:H_leading}) respects the full moir\'e translation symmetry (see discussion of subleading effects in \cite{Note1}) making the emergence of sublattices surprising, and deserving further investigation.

The origin of $t=0$ for the flat band can be traced to a symmetry of \eqref{Eqn:Main:H_leading} (see Fig. \ref{Fig:Main:fig1} (c) for illustration): $\hat{\mathcal{L}}^\dagger H_0 \hat{\mathcal{L}}=H_0$,  where $\hat{\mathcal{L}} = \sigma_x \exp[-i \sigma_z {\bf q}_M {\bf x}]$, $\sigma_i$ being Pauli matrices in layer space. 
A qualitative argument as to why $\hat{\mathcal{L}}$-symmetry forbids nearest-neighbor hopping is as follows. The states in the flat-band (small $\theta$) limit are localized near AA sites, where $T({\bf x})\approx4 w_0$ is maximal, and are thus approximately in the antibonding $(+1,+1)$ superposition between layers. These sites are located at ${\bf x} = n {\bf R}_1+m {\bf R}_2$, where ${\bf R}_j\cdot{\bf g}_k=2\pi\delta_{jk}$ are moir\'e lattice vectors. There, $\mathcal{L}$-symmetry reduces to $\pm \sigma_x$, where $\sigma_x$ does not affect the $(+1,+1)$ state. The sign of the wavefunction therefore flips between neighboring sites under $\hat{\mathcal{L}}$  (Fig. \ref{Fig:Main:fig1} (c)), forbidding nearest-neighbor hopping.

To appreciate the scope of the $\hat{\mathcal{L}}$ symmetry, we analyze the system beyond the leading-order Eq. \eqref{Eqn:Main:H_leading}. We consider the continuum Hamiltonian in the representation, where the band extrema of the two layers are shifted to the same point  \cite{bistritzer2011moire,balents_scipost,10.21468/SciPostPhys.15.3.081,tarnop2019}:
\begin{eqnarray}\label{Eqn:Main:H_folded}
H_{f} =
h(-i\boldsymbol{\nabla}_{\bf x}^{-\theta \sigma_z /2}) + U({\bf x}) \sigma_x,
\end{eqnarray}
where $h(-i\boldsymbol{\nabla}_{\bf x}^{-\theta \sigma_z /2})$ is the single-layer dispersion and $U({\bf x}) = 2w_0\Big(\cos\Big(\frac{{\bf g}_1+{\bf g}_2}{2}\cdot{\bf x}\Big)+\cos\Big(\frac{{\bf g}_1-{\bf g}_2}{2}\cdot{\bf x}\Big)\Big)$, which can be interpreted as a staggered ``layer Zeeman'' field  [$U({\bf x}+{\bf R}_1) = -U({\bf x})$], restricting electron motion to one sublattice for large $U({\bf x})$. Despite the staggered $U({\bf x})$, the moir\'e translation symmetry is still present in Eq. \eqref{Eqn:Main:H_folded} as a nonsymmorphic one \cite{Note1} $T_{{\bf R}_{1,2}}=e^{i {\bf q}_M {\bf R}_{1,2} \sigma_z/2 } e^{{\bf R}_{1,2} \nabla_x}$.

The $\mathcal{L}$-symmetry emerges in \eqref{Eqn:Main:H_folded} as $[H_f,\sigma_x] = 0$ if the rotation of the intralayer dispersion is neglected: $h(-i\boldsymbol{\nabla}_{\bf x}^{\pm \theta/2})\to h(-i\boldsymbol{\nabla}_{\bf x})$. This symmetry implies that \eqref{Eqn:Main:H_folded}  can be diagonalized with bonding/antibonding states of the form $(1,\pm 1)^T\psi_{\pm}({\bf x})$, where
\begin{eqnarray}\label{Eqn:Main:HarmonicOscillators}
\Big(h(-i\boldsymbol{\nabla}_{\bf x}) \pm U({\bf x})\Big)\psi_{\pm}({\bf x}) = E\psi_{\pm}({\bf x}) .
\end{eqnarray}
Taking into account $U({\bf x}+{\bf R}_1) = -U({\bf x}) $, implies that: (i) the eigen-energies of the +/- sectors are degenerate (ii) the eigenstates of the sectors are related via a translation by a moir\'e site. Therefore, the eigenstates of \eqref{Eqn:Main:H_folded} form two degenerate decoupled bands with Wannier functions localized on two different sublattices of the moir\'e lattice. In the tight-binding language all hoppings between the two sublattices vanish exactly in this case, explaining the structure of the bands in Fig. \ref{fig:Main:dispersion} (a).

Let us now discuss the relation of the $\mathcal{L}$-symmetry to microscopic symmetries of the twisted bilayer. Eq. \eqref{Eqn:Main:HarmonicOscillators} appears to possess all elements of the $D_{4h}$ point group of the untwisted bilayer. However, the $\mathcal{L}$-symmetry $\sigma_x$ does not commute with the translations $T_{{\bf R}_{1,2}}$, notably distinct from the $z\rightarrow-z$ mirror symmetry of the $\theta=0$ bilayer. 
It is the presence of $\mathcal{L}$ and $T_{{\bf R}_{1,2}}$ 
that guarantees the double degeneracy of the eigenvalues of \eqref{Eqn:Main:HarmonicOscillators}, translating into the formation of decoupled sublattices. All other symmetries of Eqn. \ref{Eqn:Main:H_folded}, not being local in $k$-space, can not result in such degeneracies.

The rotations of momenta $i\boldsymbol{\nabla}_{\bf x}^{\pm\theta/2}$ in Eq. \eqref{Eqn:Main:H_folded} break the $\mathcal{L}$-symmetry, reducing the point-group to $D_4$ --- the actual point-group of the twisted bilayer. However, the $\mathcal{L}$-breaking corrections only arise in the 4th order in the expansion of $h({\bf k})$ in ${\bf k}$: $\delta h({\bf k}) \propto  \theta k_xk_y(k_x^2-k_y^2)\sigma_z$. Their magnitude is of the order $\mathcal{O}(\theta^5)$, demonstrating that the $\mathcal{L}$-symmetry is remarkably robust. In addition to the terms considered in \eqref{Eqn:Main:H_unfolded}, the twisted layers also produce a potential that amounts to a term $U_0({\bf x}) \sigma_0$\cite{PhysRevLett.121.026402,10.21468/SciPostPhys.15.3.081}, that is periodic under a moir\'e translation. This term does not change any of the arguments above and does not break the $\mathcal{L}$-symmetry. If the magnitude of this term is smaller than $E_{gap}$, it will not alter the band structure qualitatively; we thus leave the detailed study of the effects of $U_0({\bf x})$ to future work. 

The above implies that for small twist angles, the M-point flat bands are characterized by an approximate symmetry group that is distinct from (and not a subgroup of) the space group of the twisted bilayer, and enforces the unexpected pattern of band degeneracies.

{\it Tunable $t-t'-U$ Hubbard models:} Identifying the $\mathcal{L}$-symmetry as the origin of the moir\'e sublattice decoupling ($t=0$) opens the possibility to create a tunable $t$ by breaking it with, e.g., displacement field $V \sigma_z$. In the context of Eqn. \eqref{Eqn:Main:H_folded}, it produces a ``layer Zeeman'' field along $z$, `canting' the `layer spin' away from $\pm x$, which permits hopping between sublattices. In Fig. \ref{fig:Main:dispersion} (b), we show the bands of Eqn. \eqref{Eqn:Main:H_leading} in presence of a finite $V$, all other parameters taken the same as in Fig. \ref{fig:Main:dispersion} (a). The top band is well fit by \eqref{Eqn:Main:epsilonK} with $t/t'=5$ and remains flat and isolated demonstrating excellent tunability. 

To realize the physics of tunable t-t'-U Hubbard model in strongly correlated regime, the bandwidth $W=\text{max}\{4t+8t',8t\}$ should be smaller then both $E_{gap}$ (Fig. \ref{fig:Main:dispersion} (a), equal to $4w_0\theta/\theta^*$ for $\theta\ll \theta^*$ \cite{Note1}) and the interaction strength. Expressions for $t$ \& $t'$ in Eq. \eqref{Eqn:Main:epsilonK} can be obtained using exact solutions of Eqn. \eqref{Eqn:Main:HarmonicOscillators} for $h(-i\boldsymbol{\nabla}_{\bf x}) \approx \mu\nabla_{\bf x}^2$. After variable separation $\psi_{\pm}({\bf x})=\psi_{X,\pm}(\frac{x+y}{\sqrt{2}})\psi_{Y,\pm}(\frac{x-y}{\sqrt{2}})$, Eqn. \eqref{Eqn:Main:HarmonicOscillators} reduces to 1D Mathieu equations 
so that $t'$ and $t$ are given by \cite{Note1}:  
\begin{eqnarray}\label{Eqn:Main:tp}
t' = \frac{w_0 \theta^2}{4 \theta^{*2}} f_{t'}\left(\frac{\theta^*}{\theta}\right) \approx_{_{\theta\ll \theta^*}} w_0 \frac{4^{7/4}}{\sqrt{\pi}}\Big(\frac{\theta}{\theta^*}\Big)^{1/2}e^{-4\frac{\theta^*}{\theta}} 
\end{eqnarray}
where $f_{t'}\left(z\right) =  a_1\left(1,z^2\right)-a_1\left(0,z^2\right)$, $a_1$ being the characteristic value of the 1D Mathieu equation,and
\begin{eqnarray}\label{Eqn:Main:t}
t = \frac{V g_{t}^2\left(\frac{\theta^*}{\theta}\right)}{\pi^2} \approx_{_{\theta\ll \theta^*}}V\frac{\tan^2\left(\frac{\pi}{8}\right)}{2^{-7/2}} e^{-4\left(2-\sqrt{2}\right) \frac{\theta^*}{\theta}} . 
\end{eqnarray}
where $g_t(z)=\int_0^\pi dx \, ce_0(x,z)\, ce_0(x+\pi/2,z)$, $ce_0$ being the periodic Mathieu functions \cite{Note1}.

\begin{figure}
    \centering
    \includegraphics[width=\linewidth]{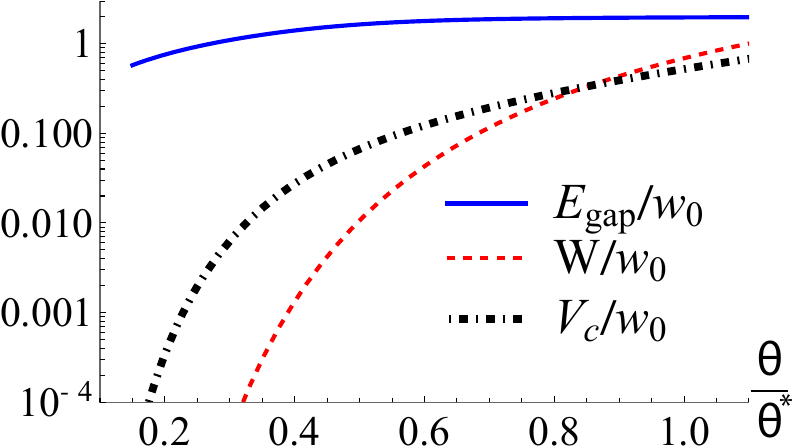}
    \caption{Dependence of the moir\'e flat band parameters [Fig. \ref{fig:Main:dispersion}] on $\theta$: gap to the remote bands ($E_{\text{gap}}$), bandwidth ($W$ with $V=0$)  and displacement field necessary to achieve $t/t'=1$ ($V_c$, Eq. \eqref{vcdef} ), all normalized to $w_0$. At $\theta<\theta^*$ one has $E_{gap}\gg W, V_c$, such that an isolated flat band with widely tunable $t'/t$ is realized.}
    \label{Fig:Main:moneyplot}
\end{figure}

We can estimate now the displacement field strength needed to tune the $t/t'$ ratio in a wide range. Consider the value $V_c$, where nearest-neighbor tunneling becomes comparable to the next-nearest neighbor $t(V_c)/t'=1$:
\begin{eqnarray}\label{vcdef}
V_c = \frac{t'}{(t/V)}\approx_{_{\theta\ll \theta^*}}  w_0\frac{\sqrt{\theta/\theta^*}}{\tan^2(\frac{\pi}{8})\sqrt{\pi}}e^{-4(\sqrt{2}-1)\frac{\theta^*}{\theta}} .
\end{eqnarray}
Crucially, $V_c(\theta)$ is an exponential function of $\theta$; therefore, displacement fields much smaller than the band gap $V\ll E_{gap}$ are sufficient to induce strong nearest neighbor hopping, retaining the isolated flat band.

The effects of Coulomb interactions can be estimated in the limit $\theta\ll\theta^*$, where states can be approximated as Gaussians localized near AA moir\'e sites \cite{Note1}. The interaction for electrons at different sites is of the order $U(r)\sim e_0^2/(2\epsilon r) \propto \theta$, where $\epsilon$ is the permittivity, which can be controlled through screening via an electrostatic gate \cite{Liu&all&Vafek&Li2021_moire_screening}. The on-site interaction energy ($r=0$) \cite{Note1}
\begin{eqnarray}
\label{eq:coul}
U_0 = \frac{e_0^2}{4\pi\epsilon l_a}\frac{\pi^{3/2}}{2^{3/2}}\sqrt{\theta\theta^*} , 
\end{eqnarray}
is parametrically larger, suggesting that a Hubbard-like description is appropriate at low $\theta$. Formally, $U_0(\theta=\theta^*)/w_0$ can be expressed via the dimensionless Coulomb electron correlation parameter $r_s^*$, for a single layer at density $(\theta^*/l_a)^2$:
$\frac{U(\theta=\theta^*)}{w_0}\approx \frac{r_s^*}{\sqrt{8}\pi}$. 

We summarize the energy scales discussed above as a function of twist angle in Fig. \ref{Fig:Main:moneyplot}. The results demonstrate that for $\theta\ll\theta^*$, an isolated ($W\ll E_{\text{gap}}$) flat ($W\ll w_0$) band emerges. At low $\theta$, $W$ is exponentially suppressed (\ref{Eqn:Main:tp},\ref{Eqn:Main:t}) in contrast to the interaction \eqref{eq:coul}, leading to the strongly interacting regime $U_0\gg W$. Finally, the relevant value of displacement field $V_c$ \eqref{vcdef} is suppressed exponentially at low $\theta$, guaranteeing $W(V_c)\ll E_{\text{gap}}$ as $\theta\rightarrow 0$. Thus the band will remain flat and isolated even when $t$ is dominant (e.g., $t/t'=5$ in Fig \ref{fig:Main:dispersion} (b)), allowing to traverse the entire $t'/t$ phase diagram of the Hubbard model in the strong coupling regime by tuning $\theta$ \& $V$.

In addition to the above, a significant anisotropy of $t$ can be introduced via in-plane magnetic field, $-i\boldsymbol\nabla_{\bf x} \to -i\boldsymbol\nabla_{\bf x} - \frac{e_0}{2c}{\bf B} d_{\|} \sigma_z$ \footnote{The Zeeman effect produces a spin splitting only.}, where $d_{\|}$ is the distance between two layers and $e_0$ is the electron charge. Perturbatively, it introduces an anisotropic nearest-neighbor hopping, that can be computed analogously to Eqn. \eqref{Eqn:Main:t}:
\begin{eqnarray}\label{Eqn:Main:tpx/y}
t_{x/y} &\approx& i4\sqrt{\mu w_0}\frac{e_0B_{x/y}d_{\|}}{2c}\frac{\tan^2\left(\frac{\pi}{8}\right)}{2^{-7/2}} e^{-4\left(2-\sqrt{2}\right) \frac{\theta^*}{\theta}}.
\end{eqnarray}
The relative effect of this term compared to that of $V$ can be estimated as $t_{x/y}^B/t^V \sim \frac{w_0 \theta}{V \theta^*} \frac{B d_\parallel l_a}{\theta\Phi_0}$, where $\frac{B d_\parallel l_a}{\theta\Phi_0}$ is the magnetic flux per lateral moir\'e unit cell which can reach values $\sim 0.1$ for fields $\sim$ 10 T and $d\sim 1$ nm, $l_a/\theta \sim 10$ nm. The stronger exponential dependence of $t'(\theta)$ \eqref{Eqn:Main:tp} suggests that $t_{x/y}$ can still become comparable to bandwidth at low enough twist angles.

{\it Discussion and Outlook:} The emergence of moir\'e sublattices and tunability of $t'/t$ for flat bands around the $M$ point can be contrasted with those near $\Gamma$ \cite{10.21468/SciPostPhys.15.3.081}. There, the dominant term in the tunneling is a constant ``layer Zeeman'' field along $x$, resulting in a usual tight-binding description, where $t\gg t'$ is expected.

The ratio $t'/t$ is believed to be crucial for realizing superconductivity in square lattice Hubbard\cite{jiang2019superconductivity,white2020,jiang2020,danilov2022degenerate} and t-t'-J models \cite{tj2021,tj2024,qu2024}, to the extent that for $t'=0$ no superconductivity appears \cite{jiang2019superconductivity,flatiron_2020}. The exploration of a wide range of $t'/t$ ratios may therefore test the ultimate bounds on $T_c$ in the Hubbard model. For $t'/t\sim 1$, this model has been argued to exhibit a number of exotic phenomena, including a topological $d_{x^2-y^2}\pm id_{xy}$ pairing state \cite{sachdev2002,nica2021multiorbital}, and, in the Mott insulating regime at half filling, possible spin liquid phase (\cite{jiang2012,imada2021,qianj1j22024,ruckriegel2024} and references therein). Beyond the Hubbard physics, twisted square lattices can realize spin-spin interactions beyond the superexchange ones. In particular, nearest-neighbor direct exchange interactions can become larger than $t'$-induced superexchange at low $\theta$, due to larger overlap between nearest-neighbor wavefunctions \cite{Note1}. Since the sign of direct exchange interactions is opposite to the superexchange ones, inducing nearest-neighbor superexchange with $V$ allows to tune the sign of the nearest-neighbor exchange interactions. Finally, anisotropy of $t_{x,y}$ induced by an in-plane magnetic field can further clarify the role of nematic order on the square lattice.

Realization of the $t-t'-U$ model on the moir\'e scale is also promising for the study of correlated topological states, as large values of flux per unit cell can be achieved with realistic magnetic fields \cite{efetovflux}. For flux $\Phi_0/2$ per moir\'e unit cell the system becomes a $|C|=1$ Chern insulator \cite{affleck1988,wen1989,hatsugai1990}, which can give rise to fractional Chern insulators \cite{neupert2011}. To realize the latter, tunable $t/t'$ ratio is crucial (with optimal value $\frac{t}{t'}=\sqrt{2}$ \cite{neupert2011}). More generally, interplay between flux and interactions can lead to unconventional superconducting \cite{scalettarpairing} and correlated states \cite{scalettar2012,feng2022,ding2022thermodynamics}.

The dependence of the band structure on the  interlayer voltage implies a nontrivial contribution to quantum capacitance from the flat band $C_q^{FB} = -e^2 \frac{d^2 \Omega_{FB}}{d V^2} =- e^2\alpha^2(\theta) \frac{d^2 \Omega_{FB}}{d t^2}$ \cite{Note1}, where $\Omega_{FB}$ - is the grand potential and $\alpha(\theta) = t(\theta)/V$, Eq. \eqref{Eqn:Main:t}. A usual problem is that the quantum capacitance acts in series with the geometric one $\frac{1}{C_{tot}} =  \frac{1}{C_q}+\frac{1}{C_{geom}}$. Per moir'e unit cell, we can estimate $C_{geom}$ as $\frac{l_a^2/\theta^2}{4 \pi d}$. For $C_q^{FB}$, the estimate can be obtained for non-interacting case  and is of the order $e^2\alpha^2(\theta)/W$ \cite{Note1}. In contrast to usual case, the capacitance is strongly reduced by the factor $\alpha^2(\theta)$, such that $C_q^{FB}/C_{geom} \sim \frac{e^2 d}{W l_M^2}4\pi \alpha^2(\theta)$. Taking typical values of $l_M\sim 10$ nm, $d\sim 1$ nm, $W\sim 0.001 E_h$ the ratio can be estimated to be about $1$ at $\theta/\theta^*\approx$ 0.86 and becomes 0.1 at $\theta/\theta^*\approx 0.6$. Thus, $C_q^{FB}$ dominates at low twist angles and should be experimentally accessible. Discontinuities of $C_q^{FB}(V)$ provide a thermodynamic probe of phase transitions as a function of $t/t'$, which is typically difficult to achieve in 2D materials.

Let us now discuss the potential material platforms. We stress that our proposal requires neither special band touching nor non-trivial topology for individual layers, utilizing instead the conventional band extremum at the Brillouin zone corner expected generically. Recent computational searches of 2D materials \cite{Mounet_2018,campi2023expansion,jiang2024_2dtheoreticallytwistablematerial} reveal a number of square-lattice insulators with valence band maxima at $M$, including but not limited to: SnI$_3$ \cite{campi2023expansion}, PbF$_4$ \cite{Mounet_2018,campi2023expansion}, ZnF$_2$ \cite{campi2023expansion,jiang2024_2dtheoreticallytwistablematerial}, Cl$_2$Ge, SnF$_2$, SnBr$_2$, TlO$_4$P \cite{jiang2024_2dtheoreticallytwistablematerial}.
Of these, ZnF$_2$ \& TlO$_4$P have been predicted to be exfoliable \cite{jiang2024_2dtheoreticallytwistablematerial}. Additionally, bilayers of SnO, which has been synthesized in mono- or few-layer form \cite{daeneke2017wafer,jiang2023mechanical}, have a conduction band minimum at the $M$ point \cite{daeneke2017wafer,jiang2023mechanical}.

As we show in the End Matter, the main results of our analysis (formation of independent moir\'e sublattices) carry over to $S$-point materials with rectangular lattices, adding several candidates \cite{jiang2024_2dtheoreticallytwistablematerial}. One example is pentagonal PdS$_2$ \cite{Mounet_2018,campi2023expansion} where conduction band minimum is at the $S$ point \cite{wang2015not,Mounet_2018,campi2023expansion}. It has been synthesized in few-layer form \cite{zhang2021centimeter} with monolayers of the chemically similar PdSe$_2$ already available \cite{oyedele2017pdse2}.

Apart from materials exhibiting two-dimensionality in bulk, monolayers of cubic perovskites
\cite{ricciardulli2021emerging}, such as SrTiO$_3$ \cite{ji2019freestanding}, possess square lattices and valence band top is typically at the $M$ point \cite{xiao2021freestanding,sto_dft}. Additionally, monolayers of square lattice cuprates \cite{Yu2019}, FeSe \cite{exp_FeSe/SrTiO3_DiracSemimetal,exp_FeSe/SrTiO3_ZX} and FeS \cite{exp_Shigekawa&Sato2019_monolayerFeSe&FeTe&FeS} have all been experimentally realized. Unlike the prior materials, they are metals with Fermi level away from $M$ points; a substantial doping would be required for them to realize the physics discussed here.

For several materials mentioned above (SnO, PbF$_4$, SrTiO$_3$, FeSe, FeS), the bands at $M$ point have an additional orbital degeneracy. In the End Matter, we consider such case in detail. In the presence of spin-orbit coupling $\lambda$, the resulting model realizes two phases depending on the relative strength of intra- and inter- orbital tunneling $w_0$ and $w_1$, respectively. When $w_0$ is dominant, a tunable Hubbard model arises, while increasing $w_1$ drives a transition into a spin Chern number $|C_s|=2$ topological phase. In the latter, each sublattice contributes $|C_s|=1$ \cite{sarkar2023symmetrybased,10.21468/SciPostPhys.15.3.081}, an example of higher-Chern number band \cite{PhysRevResearch.4.043151,sarkar2023symmetrybased}, in a system with time-reversal symmetry. While $\mathbb{Z}_2$ invariant is zero in this case, the $\mathcal{L}$-symmetry forbids scattering between sublattices, leading to 4 helical edge modes.


Finally, our main prediction of the emergent sublattices can be tested in analogue simulators, such as cold atoms \cite{cirac2019,salamon2020,meng2023atomic}, optical \cite{fu2020optical,wang2020localization,
lou2021} and acoustic \cite{phon2022} metamaterials by studying the wave propagation from a state localized on one sublattice only.

To conclude, we have revealed an approximate $\mathcal{L}$-symmetry in twisted square lattices that leads to the emergence of decoupled sublattices for the flat bands. Introducing layer asymmetry with external fields couples them in a controllable way, opening the path towards the realization of electronic $t-t'-U$ Hubbard model physics with a widely tunable $t'/t$ ratio. Our results demonstrate the important role that emergent symmetries beyond the space group play in moir\'e systems.

\begin{acknowledgments}
We thank  Leon Balents, Donna N. Sheng and Eslam Khalaf for comments and discussions. A. V. and Z.-X. L. would like to thank Toshikaze Kariyado for an earlier related collaboration and for discussions. This research was supported in part by the Simons Collaboration on Ultra-Quantum Matter which is a grant from the Simons Foundation (651440, Z.-X. L.) and by the Center for Advancement of Topological Semimetals, an Energy Frontier Research Center funded by the
US Department of Energy Office of Science, Office of Basic Energy Sciences, through the Ames Laboratory under contract No. DEAC02-07CH11358.
\end{acknowledgments}

\bibliography{biblio}

\begin{thebibliography}{94}%
\makeatletter
\providecommand \@ifxundefined [1]{%
 \@ifx{#1\undefined}
}%
\providecommand \@ifnum [1]{%
 \ifnum #1\expandafter \@firstoftwo
 \else \expandafter \@secondoftwo
 \fi
}%
\providecommand \@ifx [1]{%
 \ifx #1\expandafter \@firstoftwo
 \else \expandafter \@secondoftwo
 \fi
}%
\providecommand \natexlab [1]{#1}%
\providecommand \enquote  [1]{``#1''}%
\providecommand \bibnamefont  [1]{#1}%
\providecommand \bibfnamefont [1]{#1}%
\providecommand \citenamefont [1]{#1}%
\providecommand \href@noop [0]{\@secondoftwo}%
\providecommand \href [0]{\begingroup \@sanitize@url \@href}%
\providecommand \@href[1]{\@@startlink{#1}\@@href}%
\providecommand \@@href[1]{\endgroup#1\@@endlink}%
\providecommand \@sanitize@url [0]{\catcode `\\12\catcode `\$12\catcode
  `\&12\catcode `\#12\catcode `\^12\catcode `\_12\catcode `\%12\relax}%
\providecommand \@@startlink[1]{}%
\providecommand \@@endlink[0]{}%
\providecommand \url  [0]{\begingroup\@sanitize@url \@url }%
\providecommand \@url [1]{\endgroup\@href {#1}{\urlprefix }}%
\providecommand \urlprefix  [0]{URL }%
\providecommand \Eprint [0]{\href }%
\providecommand \doibase [0]{https://doi.org/}%
\providecommand \selectlanguage [0]{\@gobble}%
\providecommand \bibinfo  [0]{\@secondoftwo}%
\providecommand \bibfield  [0]{\@secondoftwo}%
\providecommand \translation [1]{[#1]}%
\providecommand \BibitemOpen [0]{}%
\providecommand \bibitemStop [0]{}%
\providecommand \bibitemNoStop [0]{.\EOS\space}%
\providecommand \EOS [0]{\spacefactor3000\relax}%
\providecommand \BibitemShut  [1]{\csname bibitem#1\endcsname}%
\let\auto@bib@innerbib\@empty
\bibitem [{\citenamefont {Liu}\ \emph {et~al.}(2016)\citenamefont {Liu},
  \citenamefont {Weiss}, \citenamefont {Duan}, \citenamefont {Cheng},
  \citenamefont {Huang},\ and\ \citenamefont {Duan}}]{liu2016van}%
  \BibitemOpen
  \bibfield  {author} {\bibinfo {author} {\bibfnamefont {Y.}~\bibnamefont
  {Liu}}, \bibinfo {author} {\bibfnamefont {N.~O.}\ \bibnamefont {Weiss}},
  \bibinfo {author} {\bibfnamefont {X.}~\bibnamefont {Duan}}, \bibinfo {author}
  {\bibfnamefont {H.-C.}\ \bibnamefont {Cheng}}, \bibinfo {author}
  {\bibfnamefont {Y.}~\bibnamefont {Huang}},\ and\ \bibinfo {author}
  {\bibfnamefont {X.}~\bibnamefont {Duan}},\ }\bibfield  {title} {\bibinfo
  {title} {Van der waals heterostructures and devices},\ }\href@noop {}
  {\bibfield  {journal} {\bibinfo  {journal} {Nature Reviews Materials}\
  }\textbf {\bibinfo {volume} {1}},\ \bibinfo {pages} {1} (\bibinfo {year}
  {2016})}\BibitemShut {NoStop}%
\bibitem [{\citenamefont {Novoselov}\ \emph {et~al.}(2016)\citenamefont
  {Novoselov}, \citenamefont {Mishchenko}, \citenamefont {Carvalho},\ and\
  \citenamefont {Neto}}]{doi:10.1126/science.aac9439}%
  \BibitemOpen
  \bibfield  {author} {\bibinfo {author} {\bibfnamefont {K.~S.}\ \bibnamefont
  {Novoselov}}, \bibinfo {author} {\bibfnamefont {A.}~\bibnamefont
  {Mishchenko}}, \bibinfo {author} {\bibfnamefont {A.}~\bibnamefont
  {Carvalho}},\ and\ \bibinfo {author} {\bibfnamefont {A.~H.~C.}\ \bibnamefont
  {Neto}},\ }\bibfield  {title} {\bibinfo {title} {2d materials and van der
  waals heterostructures},\ }\href {https://doi.org/10.1126/science.aac9439}
  {\bibfield  {journal} {\bibinfo  {journal} {Science}\ }\textbf {\bibinfo
  {volume} {353}},\ \bibinfo {pages} {aac9439} (\bibinfo {year} {2016})},\
  \Eprint
  {https://arxiv.org/abs/https://www.science.org/doi/pdf/10.1126/science.aac9439}
  {https://www.science.org/doi/pdf/10.1126/science.aac9439} \BibitemShut
  {NoStop}%
\bibitem [{\citenamefont {Mounet}\ \emph {et~al.}(2018)\citenamefont {Mounet},
  \citenamefont {Gibertini}, \citenamefont {Schwaller}, \citenamefont {Campi},
  \citenamefont {Merkys}, \citenamefont {Marrazzo}, \citenamefont {Sohier},
  \citenamefont {Castelli}, \citenamefont {Cepellotti}, \citenamefont {Pizzi},\
  and\ \citenamefont {Marzari}}]{Mounet_2018}%
  \BibitemOpen
  \bibfield  {author} {\bibinfo {author} {\bibfnamefont {N.}~\bibnamefont
  {Mounet}}, \bibinfo {author} {\bibfnamefont {M.}~\bibnamefont {Gibertini}},
  \bibinfo {author} {\bibfnamefont {P.}~\bibnamefont {Schwaller}}, \bibinfo
  {author} {\bibfnamefont {D.}~\bibnamefont {Campi}}, \bibinfo {author}
  {\bibfnamefont {A.}~\bibnamefont {Merkys}}, \bibinfo {author} {\bibfnamefont
  {A.}~\bibnamefont {Marrazzo}}, \bibinfo {author} {\bibfnamefont
  {T.}~\bibnamefont {Sohier}}, \bibinfo {author} {\bibfnamefont {I.~E.}\
  \bibnamefont {Castelli}}, \bibinfo {author} {\bibfnamefont {A.}~\bibnamefont
  {Cepellotti}}, \bibinfo {author} {\bibfnamefont {G.}~\bibnamefont {Pizzi}},\
  and\ \bibinfo {author} {\bibfnamefont {N.}~\bibnamefont {Marzari}},\
  }\bibfield  {title} {\bibinfo {title} {Two-dimensional materials from
  high-throughput computational exfoliation of experimentally known
  compounds},\ }\href {https://doi.org/10.1038/s41565-017-0035-5} {\bibfield
  {journal} {\bibinfo  {journal} {Nature Nanotechnology}\ }\textbf {\bibinfo
  {volume} {13}},\ \bibinfo {pages} {246–252} (\bibinfo {year}
  {2018})}\BibitemShut {NoStop}%
\bibitem [{\citenamefont {Balents}\ \emph {et~al.}(2020)\citenamefont
  {Balents}, \citenamefont {Dean}, \citenamefont {Efetov},\ and\ \citenamefont
  {Young}}]{balents2020}%
  \BibitemOpen
  \bibfield  {author} {\bibinfo {author} {\bibfnamefont {L.}~\bibnamefont
  {Balents}}, \bibinfo {author} {\bibfnamefont {C.~R.}\ \bibnamefont {Dean}},
  \bibinfo {author} {\bibfnamefont {D.~K.}\ \bibnamefont {Efetov}},\ and\
  \bibinfo {author} {\bibfnamefont {A.~F.}\ \bibnamefont {Young}},\ }\bibfield
  {title} {\bibinfo {title} {Superconductivity and strong correlations in
  moir{\'e} flat bands},\ }\href {https://doi.org/10.1038/s41567-020-0906-9}
  {\bibfield  {journal} {\bibinfo  {journal} {Nat. Phys.}\ }\textbf {\bibinfo
  {volume} {16}},\ \bibinfo {pages} {725} (\bibinfo {year} {2020})}\BibitemShut
  {NoStop}%
\bibitem [{\citenamefont {Wu}\ \emph {et~al.}(2018)\citenamefont {Wu},
  \citenamefont {Lovorn}, \citenamefont {Tutuc},\ and\ \citenamefont
  {MacDonald}}]{PhysRevLett.121.026402}%
  \BibitemOpen
  \bibfield  {author} {\bibinfo {author} {\bibfnamefont {F.}~\bibnamefont
  {Wu}}, \bibinfo {author} {\bibfnamefont {T.}~\bibnamefont {Lovorn}}, \bibinfo
  {author} {\bibfnamefont {E.}~\bibnamefont {Tutuc}},\ and\ \bibinfo {author}
  {\bibfnamefont {A.~H.}\ \bibnamefont {MacDonald}},\ }\bibfield  {title}
  {\bibinfo {title} {Hubbard model physics in transition metal dichalcogenide
  moir\'e bands},\ }\href {https://doi.org/10.1103/PhysRevLett.121.026402}
  {\bibfield  {journal} {\bibinfo  {journal} {Phys. Rev. Lett.}\ }\textbf
  {\bibinfo {volume} {121}},\ \bibinfo {pages} {026402} (\bibinfo {year}
  {2018})}\BibitemShut {NoStop}%
\bibitem [{\citenamefont {Tang}\ \emph {et~al.}(2020)\citenamefont {Tang},
  \citenamefont {Li}, \citenamefont {Li}, \citenamefont {Xu}, \citenamefont
  {Liu}, \citenamefont {Barmak}, \citenamefont {Watanabe}, \citenamefont
  {Taniguchi}, \citenamefont {MacDonald}, \citenamefont {Shan} \emph
  {et~al.}}]{tang2020simulation}%
  \BibitemOpen
  \bibfield  {author} {\bibinfo {author} {\bibfnamefont {Y.}~\bibnamefont
  {Tang}}, \bibinfo {author} {\bibfnamefont {L.}~\bibnamefont {Li}}, \bibinfo
  {author} {\bibfnamefont {T.}~\bibnamefont {Li}}, \bibinfo {author}
  {\bibfnamefont {Y.}~\bibnamefont {Xu}}, \bibinfo {author} {\bibfnamefont
  {S.}~\bibnamefont {Liu}}, \bibinfo {author} {\bibfnamefont {K.}~\bibnamefont
  {Barmak}}, \bibinfo {author} {\bibfnamefont {K.}~\bibnamefont {Watanabe}},
  \bibinfo {author} {\bibfnamefont {T.}~\bibnamefont {Taniguchi}}, \bibinfo
  {author} {\bibfnamefont {A.~H.}\ \bibnamefont {MacDonald}}, \bibinfo {author}
  {\bibfnamefont {J.}~\bibnamefont {Shan}}, \emph {et~al.},\ }\bibfield
  {title} {\bibinfo {title} {Simulation of hubbard model physics in wse2/ws2
  moir{\'e} superlattices},\ }\href@noop {} {\bibfield  {journal} {\bibinfo
  {journal} {Nature}\ }\textbf {\bibinfo {volume} {579}},\ \bibinfo {pages}
  {353} (\bibinfo {year} {2020})}\BibitemShut {NoStop}%
\bibitem [{\citenamefont {Burch}\ \emph {et~al.}(2018)\citenamefont {Burch},
  \citenamefont {Mandrus},\ and\ \citenamefont {Park}}]{burch2018magnetism}%
  \BibitemOpen
  \bibfield  {author} {\bibinfo {author} {\bibfnamefont {K.~S.}\ \bibnamefont
  {Burch}}, \bibinfo {author} {\bibfnamefont {D.}~\bibnamefont {Mandrus}},\
  and\ \bibinfo {author} {\bibfnamefont {J.-G.}\ \bibnamefont {Park}},\
  }\bibfield  {title} {\bibinfo {title} {Magnetism in two-dimensional van der
  waals materials},\ }\href@noop {} {\bibfield  {journal} {\bibinfo  {journal}
  {Nature}\ }\textbf {\bibinfo {volume} {563}},\ \bibinfo {pages} {47}
  (\bibinfo {year} {2018})}\BibitemShut {NoStop}%
\bibitem [{\citenamefont {Liu}\ \emph {et~al.}(2021)\citenamefont {Liu},
  \citenamefont {Wang}, \citenamefont {Watanabe}, \citenamefont {Taniguchi},
  \citenamefont {Vafek},\ and\ \citenamefont
  {Li}}]{Liu&all&Vafek&Li2021_moire_screening}%
  \BibitemOpen
  \bibfield  {author} {\bibinfo {author} {\bibfnamefont {X.}~\bibnamefont
  {Liu}}, \bibinfo {author} {\bibfnamefont {Z.}~\bibnamefont {Wang}}, \bibinfo
  {author} {\bibfnamefont {K.}~\bibnamefont {Watanabe}}, \bibinfo {author}
  {\bibfnamefont {T.}~\bibnamefont {Taniguchi}}, \bibinfo {author}
  {\bibfnamefont {O.}~\bibnamefont {Vafek}},\ and\ \bibinfo {author}
  {\bibfnamefont {J.~I.~A.}\ \bibnamefont {Li}},\ }\bibfield  {title} {\bibinfo
  {title} {Tuning electron correlation in magic-angle twisted bilayer graphene
  using coulomb screening},\ }\href {https://doi.org/10.1126/science.abb8754}
  {\bibfield  {journal} {\bibinfo  {journal} {Science}\ }\textbf {\bibinfo
  {volume} {371}},\ \bibinfo {pages} {1261} (\bibinfo {year} {2021})},\ \Eprint
  {https://arxiv.org/abs/https://www.science.org/doi/pdf/10.1126/science.abb8754}
  {https://www.science.org/doi/pdf/10.1126/science.abb8754} \BibitemShut
  {NoStop}%
\bibitem [{\citenamefont {Kennes}\ \emph {et~al.}(2021)\citenamefont {Kennes},
  \citenamefont {Claassen}, \citenamefont {Xian}, \citenamefont {Georges},
  \citenamefont {Millis}, \citenamefont {Hone}, \citenamefont {Dean},
  \citenamefont {Basov}, \citenamefont {Pasupathy},\ and\ \citenamefont
  {Rubio}}]{Kennes}%
  \BibitemOpen
  \bibfield  {author} {\bibinfo {author} {\bibfnamefont {D.~M.}\ \bibnamefont
  {Kennes}}, \bibinfo {author} {\bibfnamefont {M.}~\bibnamefont {Claassen}},
  \bibinfo {author} {\bibfnamefont {L.}~\bibnamefont {Xian}}, \bibinfo {author}
  {\bibfnamefont {A.}~\bibnamefont {Georges}}, \bibinfo {author} {\bibfnamefont
  {A.~J.}\ \bibnamefont {Millis}}, \bibinfo {author} {\bibfnamefont
  {J.}~\bibnamefont {Hone}}, \bibinfo {author} {\bibfnamefont {C.~R.}\
  \bibnamefont {Dean}}, \bibinfo {author} {\bibfnamefont {D.~N.}\ \bibnamefont
  {Basov}}, \bibinfo {author} {\bibfnamefont {A.~N.}\ \bibnamefont
  {Pasupathy}},\ and\ \bibinfo {author} {\bibfnamefont {A.}~\bibnamefont
  {Rubio}},\ }\bibfield  {title} {\bibinfo {title} {Moir{\'e}heterostructures
  as a condensed-matter quantum simulator},\ }\href
  {https://doi.org/10.1038/s41567-020-01154-3} {\bibfield  {journal} {\bibinfo
  {journal} {Nature Physics}\ }\textbf {\bibinfo {volume} {17}},\ \bibinfo
  {pages} {155} (\bibinfo {year} {2021})}\BibitemShut {NoStop}%
\bibitem [{\citenamefont {Deng}\ \emph {et~al.}(2015)\citenamefont {Deng},
  \citenamefont {Kozik}, \citenamefont {Prokof'Ev},\ and\ \citenamefont
  {Svistunov}}]{deng2015emergent}%
  \BibitemOpen
  \bibfield  {author} {\bibinfo {author} {\bibfnamefont {Y.}~\bibnamefont
  {Deng}}, \bibinfo {author} {\bibfnamefont {E.}~\bibnamefont {Kozik}},
  \bibinfo {author} {\bibfnamefont {N.~V.}\ \bibnamefont {Prokof'Ev}},\ and\
  \bibinfo {author} {\bibfnamefont {B.~V.}\ \bibnamefont {Svistunov}},\
  }\bibfield  {title} {\bibinfo {title} {Emergent bcs regime of the
  two-dimensional fermionic hubbard model: Ground-state phase diagram},\
  }\href@noop {} {\bibfield  {journal} {\bibinfo  {journal} {Europhysics
  Letters}\ }\textbf {\bibinfo {volume} {110}},\ \bibinfo {pages} {57001}
  (\bibinfo {year} {2015})}\BibitemShut {NoStop}%
\bibitem [{\citenamefont {Huang}\ \emph {et~al.}(2017)\citenamefont {Huang},
  \citenamefont {Mendl}, \citenamefont {Liu}, \citenamefont {Johnston},
  \citenamefont {Jiang}, \citenamefont {Moritz},\ and\ \citenamefont
  {Devereaux}}]{huang2017numerical}%
  \BibitemOpen
  \bibfield  {author} {\bibinfo {author} {\bibfnamefont {E.~W.}\ \bibnamefont
  {Huang}}, \bibinfo {author} {\bibfnamefont {C.~B.}\ \bibnamefont {Mendl}},
  \bibinfo {author} {\bibfnamefont {S.}~\bibnamefont {Liu}}, \bibinfo {author}
  {\bibfnamefont {S.}~\bibnamefont {Johnston}}, \bibinfo {author}
  {\bibfnamefont {H.-C.}\ \bibnamefont {Jiang}}, \bibinfo {author}
  {\bibfnamefont {B.}~\bibnamefont {Moritz}},\ and\ \bibinfo {author}
  {\bibfnamefont {T.~P.}\ \bibnamefont {Devereaux}},\ }\bibfield  {title}
  {\bibinfo {title} {Numerical evidence of fluctuating stripes in the normal
  state of high-t c cuprate superconductors},\ }\href@noop {} {\bibfield
  {journal} {\bibinfo  {journal} {Science}\ }\textbf {\bibinfo {volume}
  {358}},\ \bibinfo {pages} {1161} (\bibinfo {year} {2017})}\BibitemShut
  {NoStop}%
\bibitem [{\citenamefont {Zheng}\ \emph {et~al.}(2017)\citenamefont {Zheng},
  \citenamefont {Chung}, \citenamefont {Corboz}, \citenamefont {Ehlers},
  \citenamefont {Qin}, \citenamefont {Noack}, \citenamefont {Shi},
  \citenamefont {White}, \citenamefont {Zhang},\ and\ \citenamefont
  {Chan}}]{zheng2017stripe}%
  \BibitemOpen
  \bibfield  {author} {\bibinfo {author} {\bibfnamefont {B.-X.}\ \bibnamefont
  {Zheng}}, \bibinfo {author} {\bibfnamefont {C.-M.}\ \bibnamefont {Chung}},
  \bibinfo {author} {\bibfnamefont {P.}~\bibnamefont {Corboz}}, \bibinfo
  {author} {\bibfnamefont {G.}~\bibnamefont {Ehlers}}, \bibinfo {author}
  {\bibfnamefont {M.-P.}\ \bibnamefont {Qin}}, \bibinfo {author} {\bibfnamefont
  {R.~M.}\ \bibnamefont {Noack}}, \bibinfo {author} {\bibfnamefont
  {H.}~\bibnamefont {Shi}}, \bibinfo {author} {\bibfnamefont {S.~R.}\
  \bibnamefont {White}}, \bibinfo {author} {\bibfnamefont {S.}~\bibnamefont
  {Zhang}},\ and\ \bibinfo {author} {\bibfnamefont {G.~K.-L.}\ \bibnamefont
  {Chan}},\ }\bibfield  {title} {\bibinfo {title} {Stripe order in the
  underdoped region of the two-dimensional hubbard model},\ }\href@noop {}
  {\bibfield  {journal} {\bibinfo  {journal} {Science}\ }\textbf {\bibinfo
  {volume} {358}},\ \bibinfo {pages} {1155} (\bibinfo {year}
  {2017})}\BibitemShut {NoStop}%
\bibitem [{\citenamefont {Jiang}\ and\ \citenamefont
  {Devereaux}(2019)}]{jiang2019superconductivity}%
  \BibitemOpen
  \bibfield  {author} {\bibinfo {author} {\bibfnamefont {H.-C.}\ \bibnamefont
  {Jiang}}\ and\ \bibinfo {author} {\bibfnamefont {T.~P.}\ \bibnamefont
  {Devereaux}},\ }\bibfield  {title} {\bibinfo {title} {{Superconductivity in
  the doped Hubbard model and its interplay with next-nearest hopping t$'$}},\
  }\href@noop {} {\bibfield  {journal} {\bibinfo  {journal} {Science}\ }\textbf
  {\bibinfo {volume} {365}},\ \bibinfo {pages} {1424} (\bibinfo {year}
  {2019})}\BibitemShut {NoStop}%
\bibitem [{\citenamefont {Qin}\ \emph {et~al.}(2020)\citenamefont {Qin},
  \citenamefont {Chung}, \citenamefont {Shi}, \citenamefont {Vitali},
  \citenamefont {Hubig}, \citenamefont {Schollw\"ock}, \citenamefont {White},\
  and\ \citenamefont {Zhang}}]{flatiron_2020}%
  \BibitemOpen
  \bibfield  {author} {\bibinfo {author} {\bibfnamefont {M.}~\bibnamefont
  {Qin}}, \bibinfo {author} {\bibfnamefont {C.-M.}\ \bibnamefont {Chung}},
  \bibinfo {author} {\bibfnamefont {H.}~\bibnamefont {Shi}}, \bibinfo {author}
  {\bibfnamefont {E.}~\bibnamefont {Vitali}}, \bibinfo {author} {\bibfnamefont
  {C.}~\bibnamefont {Hubig}}, \bibinfo {author} {\bibfnamefont
  {U.}~\bibnamefont {Schollw\"ock}}, \bibinfo {author} {\bibfnamefont {S.~R.}\
  \bibnamefont {White}},\ and\ \bibinfo {author} {\bibfnamefont
  {S.}~\bibnamefont {Zhang}} (\bibinfo {collaboration} {Simons Collaboration on
  the Many-Electron Problem}),\ }\bibfield  {title} {\bibinfo {title} {Absence
  of superconductivity in the pure two-dimensional hubbard model},\ }\href
  {https://doi.org/10.1103/PhysRevX.10.031016} {\bibfield  {journal} {\bibinfo
  {journal} {Phys. Rev. X}\ }\textbf {\bibinfo {volume} {10}},\ \bibinfo
  {pages} {031016} (\bibinfo {year} {2020})}\BibitemShut {NoStop}%
\bibitem [{\citenamefont {Jiang}\ \emph {et~al.}(2021)\citenamefont {Jiang},
  \citenamefont {Scalapino},\ and\ \citenamefont {White}}]{tj2021}%
  \BibitemOpen
  \bibfield  {author} {\bibinfo {author} {\bibfnamefont {S.}~\bibnamefont
  {Jiang}}, \bibinfo {author} {\bibfnamefont {D.~J.}\ \bibnamefont
  {Scalapino}},\ and\ \bibinfo {author} {\bibfnamefont {S.~R.}\ \bibnamefont
  {White}},\ }\bibfield  {title} {\bibinfo {title} {{Ground-state phase diagram
  of the $t-t'-J$ model}},\ }\href {https://doi.org/10.1073/pnas.2109978118}
  {\bibfield  {journal} {\bibinfo  {journal} {Proceedings of the National
  Academy of Sciences}\ }\textbf {\bibinfo {volume} {118}},\ \bibinfo {pages}
  {e2109978118} (\bibinfo {year} {2021})}\BibitemShut {NoStop}%
\bibitem [{\citenamefont {Qin}\ \emph {et~al.}(2022)\citenamefont {Qin},
  \citenamefont {Schäfer}, \citenamefont {Andergassen}, \citenamefont
  {Corboz},\ and\ \citenamefont {Gull}}]{annurev_num}%
  \BibitemOpen
  \bibfield  {author} {\bibinfo {author} {\bibfnamefont {M.}~\bibnamefont
  {Qin}}, \bibinfo {author} {\bibfnamefont {T.}~\bibnamefont {Schäfer}},
  \bibinfo {author} {\bibfnamefont {S.}~\bibnamefont {Andergassen}}, \bibinfo
  {author} {\bibfnamefont {P.}~\bibnamefont {Corboz}},\ and\ \bibinfo {author}
  {\bibfnamefont {E.}~\bibnamefont {Gull}},\ }\bibfield  {title} {\bibinfo
  {title} {The hubbard model: A computational perspective},\ }\href
  {https://doi.org/https://doi.org/10.1146/annurev-conmatphys-090921-033948}
  {\bibfield  {journal} {\bibinfo  {journal} {Annual Review of Condensed Matter
  Physics}\ }\textbf {\bibinfo {volume} {13}},\ \bibinfo {pages} {275}
  (\bibinfo {year} {2022})}\BibitemShut {NoStop}%
\bibitem [{\citenamefont {Arovas}\ \emph {et~al.}(2022)\citenamefont {Arovas},
  \citenamefont {Berg}, \citenamefont {Kivelson},\ and\ \citenamefont
  {Raghu}}]{annurev_an}%
  \BibitemOpen
  \bibfield  {author} {\bibinfo {author} {\bibfnamefont {D.~P.}\ \bibnamefont
  {Arovas}}, \bibinfo {author} {\bibfnamefont {E.}~\bibnamefont {Berg}},
  \bibinfo {author} {\bibfnamefont {S.~A.}\ \bibnamefont {Kivelson}},\ and\
  \bibinfo {author} {\bibfnamefont {S.}~\bibnamefont {Raghu}},\ }\bibfield
  {title} {\bibinfo {title} {The hubbard model},\ }\href
  {https://doi.org/https://doi.org/10.1146/annurev-conmatphys-031620-102024}
  {\bibfield  {journal} {\bibinfo  {journal} {Annual Review of Condensed Matter
  Physics}\ }\textbf {\bibinfo {volume} {13}},\ \bibinfo {pages} {239}
  (\bibinfo {year} {2022})}\BibitemShut {NoStop}%
\bibitem [{\citenamefont {Chen}\ \emph {et~al.}(2023)\citenamefont {Chen},
  \citenamefont {Haldane},\ and\ \citenamefont
  {Sheng}}]{Chen&Haldane&Sheng2023}%
  \BibitemOpen
  \bibfield  {author} {\bibinfo {author} {\bibfnamefont {F.}~\bibnamefont
  {Chen}}, \bibinfo {author} {\bibfnamefont {F.~D.~M.}\ \bibnamefont
  {Haldane}},\ and\ \bibinfo {author} {\bibfnamefont {D.~N.}\ \bibnamefont
  {Sheng}},\ }\href@noop {} {\bibinfo {title} {D-wave and pair-density-wave
  superconductivity in the square-lattice t-j model}} (\bibinfo {year}
  {2023}),\ \Eprint {https://arxiv.org/abs/2311.15092} {arXiv:2311.15092}
  \BibitemShut {NoStop}%
\bibitem [{\citenamefont {Anisimov}\ \emph {et~al.}(1999)\citenamefont
  {Anisimov}, \citenamefont {Bukhvalov},\ and\ \citenamefont
  {Rice}}]{Anisimov}%
  \BibitemOpen
  \bibfield  {author} {\bibinfo {author} {\bibfnamefont {V.~I.}\ \bibnamefont
  {Anisimov}}, \bibinfo {author} {\bibfnamefont {D.}~\bibnamefont
  {Bukhvalov}},\ and\ \bibinfo {author} {\bibfnamefont {T.~M.}\ \bibnamefont
  {Rice}},\ }\bibfield  {title} {\bibinfo {title} {Electronic structure of
  possible nickelate analogs to the cuprates},\ }\href
  {https://doi.org/10.1103/PhysRevB.59.7901} {\bibfield  {journal} {\bibinfo
  {journal} {Phys. Rev. B}\ }\textbf {\bibinfo {volume} {59}},\ \bibinfo
  {pages} {7901} (\bibinfo {year} {1999})}\BibitemShut {NoStop}%
\bibitem [{\citenamefont {Pan}\ \emph {et~al.}(2022)\citenamefont {Pan},
  \citenamefont {Ferenc~Segedin}, \citenamefont {LaBollita}, \citenamefont
  {Song}, \citenamefont {Nica}, \citenamefont {Goodge}, \citenamefont {Pierce},
  \citenamefont {Doyle}, \citenamefont {Novakov}, \citenamefont
  {C{\'o}rdova~Carrizales}, \citenamefont {N'Diaye}, \citenamefont {Shafer},
  \citenamefont {Paik}, \citenamefont {Heron}, \citenamefont {Mason},
  \citenamefont {Yacoby}, \citenamefont {Kourkoutis}, \citenamefont {Erten},
  \citenamefont {Brooks}, \citenamefont {Botana},\ and\ \citenamefont
  {Mundy}}]{Pan}%
  \BibitemOpen
  \bibfield  {author} {\bibinfo {author} {\bibfnamefont {G.~A.}\ \bibnamefont
  {Pan}}, \bibinfo {author} {\bibfnamefont {D.}~\bibnamefont {Ferenc~Segedin}},
  \bibinfo {author} {\bibfnamefont {H.}~\bibnamefont {LaBollita}}, \bibinfo
  {author} {\bibfnamefont {Q.}~\bibnamefont {Song}}, \bibinfo {author}
  {\bibfnamefont {E.~M.}\ \bibnamefont {Nica}}, \bibinfo {author}
  {\bibfnamefont {B.~H.}\ \bibnamefont {Goodge}}, \bibinfo {author}
  {\bibfnamefont {A.~T.}\ \bibnamefont {Pierce}}, \bibinfo {author}
  {\bibfnamefont {S.}~\bibnamefont {Doyle}}, \bibinfo {author} {\bibfnamefont
  {S.}~\bibnamefont {Novakov}}, \bibinfo {author} {\bibfnamefont
  {D.}~\bibnamefont {C{\'o}rdova~Carrizales}}, \bibinfo {author} {\bibfnamefont
  {A.~T.}\ \bibnamefont {N'Diaye}}, \bibinfo {author} {\bibfnamefont
  {P.}~\bibnamefont {Shafer}}, \bibinfo {author} {\bibfnamefont
  {H.}~\bibnamefont {Paik}}, \bibinfo {author} {\bibfnamefont {J.~T.}\
  \bibnamefont {Heron}}, \bibinfo {author} {\bibfnamefont {J.~A.}\ \bibnamefont
  {Mason}}, \bibinfo {author} {\bibfnamefont {A.}~\bibnamefont {Yacoby}},
  \bibinfo {author} {\bibfnamefont {L.~F.}\ \bibnamefont {Kourkoutis}},
  \bibinfo {author} {\bibfnamefont {O.}~\bibnamefont {Erten}}, \bibinfo
  {author} {\bibfnamefont {C.~M.}\ \bibnamefont {Brooks}}, \bibinfo {author}
  {\bibfnamefont {A.~S.}\ \bibnamefont {Botana}},\ and\ \bibinfo {author}
  {\bibfnamefont {J.~A.}\ \bibnamefont {Mundy}},\ }\bibfield  {title} {\bibinfo
  {title} {Superconductivity in a quintuple-layer square-planar nickelate},\
  }\href {https://doi.org/10.1038/s41563-021-01142-9} {\bibfield  {journal}
  {\bibinfo  {journal} {Nature Materials}\ }\textbf {\bibinfo {volume} {21}},\
  \bibinfo {pages} {160} (\bibinfo {year} {2022})}\BibitemShut {NoStop}%
\bibitem [{\citenamefont {Li}\ \emph {et~al.}(2019)\citenamefont {Li},
  \citenamefont {Lee}, \citenamefont {Wang}, \citenamefont {Osada},
  \citenamefont {Crossley}, \citenamefont {Lee}, \citenamefont {Cui},
  \citenamefont {Hikita},\ and\ \citenamefont {Hwang}}]{Hwang}%
  \BibitemOpen
  \bibfield  {author} {\bibinfo {author} {\bibfnamefont {D.}~\bibnamefont
  {Li}}, \bibinfo {author} {\bibfnamefont {K.}~\bibnamefont {Lee}}, \bibinfo
  {author} {\bibfnamefont {B.~Y.}\ \bibnamefont {Wang}}, \bibinfo {author}
  {\bibfnamefont {M.}~\bibnamefont {Osada}}, \bibinfo {author} {\bibfnamefont
  {S.}~\bibnamefont {Crossley}}, \bibinfo {author} {\bibfnamefont {H.~R.}\
  \bibnamefont {Lee}}, \bibinfo {author} {\bibfnamefont {Y.}~\bibnamefont
  {Cui}}, \bibinfo {author} {\bibfnamefont {Y.}~\bibnamefont {Hikita}},\ and\
  \bibinfo {author} {\bibfnamefont {H.~Y.}\ \bibnamefont {Hwang}},\ }\bibfield
  {title} {\bibinfo {title} {Superconductivity in an infinite-layer
  nickelate},\ }\href {https://doi.org/10.1038/s41586-019-1496-5} {\bibfield
  {journal} {\bibinfo  {journal} {Nature}\ }\textbf {\bibinfo {volume} {572}},\
  \bibinfo {pages} {624} (\bibinfo {year} {2019})}\BibitemShut {NoStop}%
\bibitem [{\citenamefont {Kitatani}\ \emph {et~al.}(2020)\citenamefont
  {Kitatani}, \citenamefont {Si}, \citenamefont {Janson}, \citenamefont
  {Arita}, \citenamefont {Zhong},\ and\ \citenamefont {Held}}]{kitatani2020}%
  \BibitemOpen
  \bibfield  {author} {\bibinfo {author} {\bibfnamefont {M.}~\bibnamefont
  {Kitatani}}, \bibinfo {author} {\bibfnamefont {L.}~\bibnamefont {Si}},
  \bibinfo {author} {\bibfnamefont {O.}~\bibnamefont {Janson}}, \bibinfo
  {author} {\bibfnamefont {R.}~\bibnamefont {Arita}}, \bibinfo {author}
  {\bibfnamefont {Z.}~\bibnamefont {Zhong}},\ and\ \bibinfo {author}
  {\bibfnamefont {K.}~\bibnamefont {Held}},\ }\bibfield  {title} {\bibinfo
  {title} {Nickelate superconductors—a renaissance of the one-band hubbard
  model},\ }\href@noop {} {\bibfield  {journal} {\bibinfo  {journal} {npj
  Quantum Materials}\ }\textbf {\bibinfo {volume} {5}},\ \bibinfo {pages} {59}
  (\bibinfo {year} {2020})}\BibitemShut {NoStop}%
\bibitem [{\citenamefont {Bohrdt}\ \emph {et~al.}(2021)\citenamefont {Bohrdt},
  \citenamefont {Homeier}, \citenamefont {Reinmoser}, \citenamefont {Demler},\
  and\ \citenamefont {Grusdt}}]{BOHRDT}%
  \BibitemOpen
  \bibfield  {author} {\bibinfo {author} {\bibfnamefont {A.}~\bibnamefont
  {Bohrdt}}, \bibinfo {author} {\bibfnamefont {L.}~\bibnamefont {Homeier}},
  \bibinfo {author} {\bibfnamefont {C.}~\bibnamefont {Reinmoser}}, \bibinfo
  {author} {\bibfnamefont {E.}~\bibnamefont {Demler}},\ and\ \bibinfo {author}
  {\bibfnamefont {F.}~\bibnamefont {Grusdt}},\ }\bibfield  {title} {\bibinfo
  {title} {Exploration of doped quantum magnets with ultracold atoms},\ }\href
  {https://doi.org/https://doi.org/10.1016/j.aop.2021.168651} {\bibfield
  {journal} {\bibinfo  {journal} {Annals of Physics}\ }\textbf {\bibinfo
  {volume} {435}},\ \bibinfo {pages} {168651} (\bibinfo {year} {2021})},\
  \bibinfo {note} {special issue on Philip W. Anderson}\BibitemShut {NoStop}%
\bibitem [{\citenamefont {Kariyado}\ and\ \citenamefont
  {Vishwanath}(2019)}]{Toshi}%
  \BibitemOpen
  \bibfield  {author} {\bibinfo {author} {\bibfnamefont {T.}~\bibnamefont
  {Kariyado}}\ and\ \bibinfo {author} {\bibfnamefont {A.}~\bibnamefont
  {Vishwanath}},\ }\bibfield  {title} {\bibinfo {title} {Flat band in twisted
  bilayer bravais lattices},\ }\href
  {https://doi.org/10.1103/PhysRevResearch.1.033076} {\bibfield  {journal}
  {\bibinfo  {journal} {Phys. Rev. Res.}\ }\textbf {\bibinfo {volume} {1}},\
  \bibinfo {pages} {033076} (\bibinfo {year} {2019})}\BibitemShut {NoStop}%
\bibitem [{\citenamefont {Luo}\ \emph {et~al.}(2021)\citenamefont {Luo},
  \citenamefont {Xu},\ and\ \citenamefont {Jian}}]{PhysRevB.104.035136}%
  \BibitemOpen
  \bibfield  {author} {\bibinfo {author} {\bibfnamefont {Z.-X.}\ \bibnamefont
  {Luo}}, \bibinfo {author} {\bibfnamefont {C.}~\bibnamefont {Xu}},\ and\
  \bibinfo {author} {\bibfnamefont {C.-M.}\ \bibnamefont {Jian}},\ }\bibfield
  {title} {\bibinfo {title} {Magic continuum in a twisted bilayer square
  lattice with staggered flux},\ }\href
  {https://doi.org/10.1103/PhysRevB.104.035136} {\bibfield  {journal} {\bibinfo
   {journal} {Phys. Rev. B}\ }\textbf {\bibinfo {volume} {104}},\ \bibinfo
  {pages} {035136} (\bibinfo {year} {2021})}\BibitemShut {NoStop}%
\bibitem [{\citenamefont {Soeda}\ \emph {et~al.}(2022)\citenamefont {Soeda},
  \citenamefont {Asaga},\ and\ \citenamefont {Fukui}}]{PhysRevB.105.165422}%
  \BibitemOpen
  \bibfield  {author} {\bibinfo {author} {\bibfnamefont {Y.}~\bibnamefont
  {Soeda}}, \bibinfo {author} {\bibfnamefont {K.}~\bibnamefont {Asaga}},\ and\
  \bibinfo {author} {\bibfnamefont {T.}~\bibnamefont {Fukui}},\ }\bibfield
  {title} {\bibinfo {title} {Moir\'e landau levels of a ${C}_{4}$-symmetric
  twisted bilayer system in the absence of a magnetic field},\ }\href
  {https://doi.org/10.1103/PhysRevB.105.165422} {\bibfield  {journal} {\bibinfo
   {journal} {Phys. Rev. B}\ }\textbf {\bibinfo {volume} {105}},\ \bibinfo
  {pages} {165422} (\bibinfo {year} {2022})}\BibitemShut {NoStop}%
\bibitem [{\citenamefont {Li}\ \emph {et~al.}(2022)\citenamefont {Li},
  \citenamefont {He},\ and\ \citenamefont {Yao}}]{PhysRevResearch.4.043151}%
  \BibitemOpen
  \bibfield  {author} {\bibinfo {author} {\bibfnamefont {M.-R.}\ \bibnamefont
  {Li}}, \bibinfo {author} {\bibfnamefont {A.-L.}\ \bibnamefont {He}},\ and\
  \bibinfo {author} {\bibfnamefont {H.}~\bibnamefont {Yao}},\ }\bibfield
  {title} {\bibinfo {title} {Magic-angle twisted bilayer systems with quadratic
  band touching: Exactly flat bands with high chern number},\ }\href
  {https://doi.org/10.1103/PhysRevResearch.4.043151} {\bibfield  {journal}
  {\bibinfo  {journal} {Phys. Rev. Res.}\ }\textbf {\bibinfo {volume} {4}},\
  \bibinfo {pages} {043151} (\bibinfo {year} {2022})}\BibitemShut {NoStop}%
\bibitem [{\citenamefont {Sarkar}\ \emph {et~al.}(2023)\citenamefont {Sarkar},
  \citenamefont {Wan}, \citenamefont {Lin},\ and\ \citenamefont
  {Sun}}]{sarkar2023symmetrybased}%
  \BibitemOpen
  \bibfield  {author} {\bibinfo {author} {\bibfnamefont {S.}~\bibnamefont
  {Sarkar}}, \bibinfo {author} {\bibfnamefont {X.}~\bibnamefont {Wan}},
  \bibinfo {author} {\bibfnamefont {S.-Z.}\ \bibnamefont {Lin}},\ and\ \bibinfo
  {author} {\bibfnamefont {K.}~\bibnamefont {Sun}},\ }\href@noop {} {\bibinfo
  {title} {Symmetry-based classification of exact flat bands in single and
  bilayer moir\'e systems}} (\bibinfo {year} {2023}),\ \Eprint
  {https://arxiv.org/abs/2310.02218} {arXiv:2310.02218 [cond-mat.mes-hall]}
  \BibitemShut {NoStop}%
\bibitem [{\citenamefont {Can}\ \emph {et~al.}(2021)\citenamefont {Can},
  \citenamefont {Tummuru}, \citenamefont {Day}, \citenamefont {Elfimov},
  \citenamefont {Damascelli},\ and\ \citenamefont {Franz}}]{Can_2021}%
  \BibitemOpen
  \bibfield  {author} {\bibinfo {author} {\bibfnamefont {O.}~\bibnamefont
  {Can}}, \bibinfo {author} {\bibfnamefont {T.}~\bibnamefont {Tummuru}},
  \bibinfo {author} {\bibfnamefont {R.~P.}\ \bibnamefont {Day}}, \bibinfo
  {author} {\bibfnamefont {I.}~\bibnamefont {Elfimov}}, \bibinfo {author}
  {\bibfnamefont {A.}~\bibnamefont {Damascelli}},\ and\ \bibinfo {author}
  {\bibfnamefont {M.}~\bibnamefont {Franz}},\ }\bibfield  {title} {\bibinfo
  {title} {High-temperature topological superconductivity in twisted
  double-layer copper oxides},\ }\href
  {https://doi.org/10.1038/s41567-020-01142-7} {\bibfield  {journal} {\bibinfo
  {journal} {Nature Physics}\ }\textbf {\bibinfo {volume} {17}},\ \bibinfo
  {pages} {519–524} (\bibinfo {year} {2021})}\BibitemShut {NoStop}%
\bibitem [{\citenamefont {Zhao}\ \emph {et~al.}(2023)\citenamefont {Zhao},
  \citenamefont {Cui}, \citenamefont {Volkov}, \citenamefont {Yoo},
  \citenamefont {Lee}, \citenamefont {Gardener}, \citenamefont {Akey},
  \citenamefont {Engelke}, \citenamefont {Ronen}, \citenamefont {Zhong} \emph
  {et~al.}}]{zhao2021emergent}%
  \BibitemOpen
  \bibfield  {author} {\bibinfo {author} {\bibfnamefont {S.~F.}\ \bibnamefont
  {Zhao}}, \bibinfo {author} {\bibfnamefont {X.}~\bibnamefont {Cui}}, \bibinfo
  {author} {\bibfnamefont {P.~A.}\ \bibnamefont {Volkov}}, \bibinfo {author}
  {\bibfnamefont {H.}~\bibnamefont {Yoo}}, \bibinfo {author} {\bibfnamefont
  {S.}~\bibnamefont {Lee}}, \bibinfo {author} {\bibfnamefont {J.~A.}\
  \bibnamefont {Gardener}}, \bibinfo {author} {\bibfnamefont {A.~J.}\
  \bibnamefont {Akey}}, \bibinfo {author} {\bibfnamefont {R.}~\bibnamefont
  {Engelke}}, \bibinfo {author} {\bibfnamefont {Y.}~\bibnamefont {Ronen}},
  \bibinfo {author} {\bibfnamefont {R.}~\bibnamefont {Zhong}}, \emph {et~al.},\
  }\bibfield  {title} {\bibinfo {title} {Time-reversal symmetry breaking
  superconductivity between twisted cuprate superconductors},\ }\href@noop {}
  {\bibfield  {journal} {\bibinfo  {journal} {Science}\ }\textbf {\bibinfo
  {volume} {382}},\ \bibinfo {pages} {1422} (\bibinfo {year}
  {2023})}\BibitemShut {NoStop}%
\bibitem [{\citenamefont {Song}\ \emph {et~al.}(2022)\citenamefont {Song},
  \citenamefont {Zhang},\ and\ \citenamefont
  {Vishwanath}}]{PhysRevB.105.L201102}%
  \BibitemOpen
  \bibfield  {author} {\bibinfo {author} {\bibfnamefont {X.-Y.}\ \bibnamefont
  {Song}}, \bibinfo {author} {\bibfnamefont {Y.-H.}\ \bibnamefont {Zhang}},\
  and\ \bibinfo {author} {\bibfnamefont {A.}~\bibnamefont {Vishwanath}},\
  }\bibfield  {title} {\bibinfo {title} {Doping a moir\'e mott insulator: A
  $t\ensuremath{-}j$ model study of twisted cuprates},\ }\href
  {https://doi.org/10.1103/PhysRevB.105.L201102} {\bibfield  {journal}
  {\bibinfo  {journal} {Phys. Rev. B}\ }\textbf {\bibinfo {volume} {105}},\
  \bibinfo {pages} {L201102} (\bibinfo {year} {2022})}\BibitemShut {NoStop}%
\bibitem [{\citenamefont {Haenel}\ \emph {et~al.}(2022)\citenamefont {Haenel},
  \citenamefont {Tummuru},\ and\ \citenamefont {Franz}}]{PhysRevB.106.104505}%
  \BibitemOpen
  \bibfield  {author} {\bibinfo {author} {\bibfnamefont {R.}~\bibnamefont
  {Haenel}}, \bibinfo {author} {\bibfnamefont {T.}~\bibnamefont {Tummuru}},\
  and\ \bibinfo {author} {\bibfnamefont {M.}~\bibnamefont {Franz}},\ }\bibfield
   {title} {\bibinfo {title} {Incoherent tunneling and topological
  superconductivity in twisted cuprate bilayers},\ }\href
  {https://doi.org/10.1103/PhysRevB.106.104505} {\bibfield  {journal} {\bibinfo
   {journal} {Phys. Rev. B}\ }\textbf {\bibinfo {volume} {106}},\ \bibinfo
  {pages} {104505} (\bibinfo {year} {2022})}\BibitemShut {NoStop}%
\bibitem [{\citenamefont {Volkov}\ \emph
  {et~al.}(2023{\natexlab{a}})\citenamefont {Volkov}, \citenamefont {Wilson},
  \citenamefont {Lucht},\ and\ \citenamefont {Pixley}}]{PhysRevB.107.174506}%
  \BibitemOpen
  \bibfield  {author} {\bibinfo {author} {\bibfnamefont {P.~A.}\ \bibnamefont
  {Volkov}}, \bibinfo {author} {\bibfnamefont {J.~H.}\ \bibnamefont {Wilson}},
  \bibinfo {author} {\bibfnamefont {K.~P.}\ \bibnamefont {Lucht}},\ and\
  \bibinfo {author} {\bibfnamefont {J.~H.}\ \bibnamefont {Pixley}},\ }\bibfield
   {title} {\bibinfo {title} {Magic angles and correlations in twisted nodal
  superconductors},\ }\href {https://doi.org/10.1103/PhysRevB.107.174506}
  {\bibfield  {journal} {\bibinfo  {journal} {Phys. Rev. B}\ }\textbf {\bibinfo
  {volume} {107}},\ \bibinfo {pages} {174506} (\bibinfo {year}
  {2023}{\natexlab{a}})}\BibitemShut {NoStop}%
\bibitem [{\citenamefont {Volkov}\ \emph
  {et~al.}(2023{\natexlab{b}})\citenamefont {Volkov}, \citenamefont {Wilson},
  \citenamefont {Lucht},\ and\ \citenamefont {Pixley}}]{Volkov2022}%
  \BibitemOpen
  \bibfield  {author} {\bibinfo {author} {\bibfnamefont {P.~A.}\ \bibnamefont
  {Volkov}}, \bibinfo {author} {\bibfnamefont {J.~H.}\ \bibnamefont {Wilson}},
  \bibinfo {author} {\bibfnamefont {K.~P.}\ \bibnamefont {Lucht}},\ and\
  \bibinfo {author} {\bibfnamefont {J.~H.}\ \bibnamefont {Pixley}},\ }\bibfield
   {title} {\bibinfo {title} {Current- and field-induced topology in twisted
  nodal superconductors},\ }\href
  {https://doi.org/10.1103/PhysRevLett.130.186001} {\bibfield  {journal}
  {\bibinfo  {journal} {Phys. Rev. Lett.}\ }\textbf {\bibinfo {volume} {130}},\
  \bibinfo {pages} {186001} (\bibinfo {year} {2023}{\natexlab{b}})}\BibitemShut
  {NoStop}%
\bibitem [{\citenamefont {Eugenio}\ and\ \citenamefont
  {Vafek}(2023{\natexlab{a}})}]{10.21468/SciPostPhys.15.3.081}%
  \BibitemOpen
  \bibfield  {author} {\bibinfo {author} {\bibfnamefont {P.~M.}\ \bibnamefont
  {Eugenio}}\ and\ \bibinfo {author} {\bibfnamefont {O.}~\bibnamefont
  {Vafek}},\ }\bibfield  {title} {\bibinfo {title} {{Twisted-bilayer FeSe and
  the Fe-based superlattices}},\ }\href
  {https://doi.org/10.21468/SciPostPhys.15.3.081} {\bibfield  {journal}
  {\bibinfo  {journal} {SciPost Phys.}\ }\textbf {\bibinfo {volume} {15}},\
  \bibinfo {pages} {081} (\bibinfo {year} {2023}{\natexlab{a}})}\BibitemShut
  {NoStop}%
\bibitem [{\citenamefont {Gonz\'alez-Tudela}\ and\ \citenamefont
  {Cirac}(2019)}]{cirac2019}%
  \BibitemOpen
  \bibfield  {author} {\bibinfo {author} {\bibfnamefont {A.}~\bibnamefont
  {Gonz\'alez-Tudela}}\ and\ \bibinfo {author} {\bibfnamefont {J.~I.}\
  \bibnamefont {Cirac}},\ }\bibfield  {title} {\bibinfo {title} {Cold atoms in
  twisted-bilayer optical potentials},\ }\href
  {https://doi.org/10.1103/PhysRevA.100.053604} {\bibfield  {journal} {\bibinfo
   {journal} {Phys. Rev. A}\ }\textbf {\bibinfo {volume} {100}},\ \bibinfo
  {pages} {053604} (\bibinfo {year} {2019})}\BibitemShut {NoStop}%
\bibitem [{\citenamefont {Salamon}\ \emph {et~al.}(2020)\citenamefont
  {Salamon}, \citenamefont {Celi}, \citenamefont {Chhajlany}, \citenamefont
  {Fr\'erot}, \citenamefont {Lewenstein}, \citenamefont {Tarruell},\ and\
  \citenamefont {Rakshit}}]{salamon2020}%
  \BibitemOpen
  \bibfield  {author} {\bibinfo {author} {\bibfnamefont {T.}~\bibnamefont
  {Salamon}}, \bibinfo {author} {\bibfnamefont {A.}~\bibnamefont {Celi}},
  \bibinfo {author} {\bibfnamefont {R.~W.}\ \bibnamefont {Chhajlany}}, \bibinfo
  {author} {\bibfnamefont {I.}~\bibnamefont {Fr\'erot}}, \bibinfo {author}
  {\bibfnamefont {M.}~\bibnamefont {Lewenstein}}, \bibinfo {author}
  {\bibfnamefont {L.}~\bibnamefont {Tarruell}},\ and\ \bibinfo {author}
  {\bibfnamefont {D.}~\bibnamefont {Rakshit}},\ }\bibfield  {title} {\bibinfo
  {title} {Simulating twistronics without a twist},\ }\href
  {https://doi.org/10.1103/PhysRevLett.125.030504} {\bibfield  {journal}
  {\bibinfo  {journal} {Phys. Rev. Lett.}\ }\textbf {\bibinfo {volume} {125}},\
  \bibinfo {pages} {030504} (\bibinfo {year} {2020})}\BibitemShut {NoStop}%
\bibitem [{\citenamefont {Meng}\ \emph {et~al.}(2023)\citenamefont {Meng},
  \citenamefont {Wang}, \citenamefont {Han}, \citenamefont {Liu}, \citenamefont
  {Wen}, \citenamefont {Gao}, \citenamefont {Wang}, \citenamefont {Chin},\ and\
  \citenamefont {Zhang}}]{meng2023atomic}%
  \BibitemOpen
  \bibfield  {author} {\bibinfo {author} {\bibfnamefont {Z.}~\bibnamefont
  {Meng}}, \bibinfo {author} {\bibfnamefont {L.}~\bibnamefont {Wang}}, \bibinfo
  {author} {\bibfnamefont {W.}~\bibnamefont {Han}}, \bibinfo {author}
  {\bibfnamefont {F.}~\bibnamefont {Liu}}, \bibinfo {author} {\bibfnamefont
  {K.}~\bibnamefont {Wen}}, \bibinfo {author} {\bibfnamefont {C.}~\bibnamefont
  {Gao}}, \bibinfo {author} {\bibfnamefont {P.}~\bibnamefont {Wang}}, \bibinfo
  {author} {\bibfnamefont {C.}~\bibnamefont {Chin}},\ and\ \bibinfo {author}
  {\bibfnamefont {J.}~\bibnamefont {Zhang}},\ }\bibfield  {title} {\bibinfo
  {title} {Atomic bose--einstein condensate in twisted-bilayer optical
  lattices},\ }\href@noop {} {\bibfield  {journal} {\bibinfo  {journal}
  {Nature}\ }\textbf {\bibinfo {volume} {615}},\ \bibinfo {pages} {231}
  (\bibinfo {year} {2023})}\BibitemShut {NoStop}%
\bibitem [{\citenamefont {Fu}\ \emph {et~al.}(2020)\citenamefont {Fu},
  \citenamefont {Wang}, \citenamefont {Huang}, \citenamefont {Kartashov},
  \citenamefont {Torner}, \citenamefont {Konotop},\ and\ \citenamefont
  {Ye}}]{fu2020optical}%
  \BibitemOpen
  \bibfield  {author} {\bibinfo {author} {\bibfnamefont {Q.}~\bibnamefont
  {Fu}}, \bibinfo {author} {\bibfnamefont {P.}~\bibnamefont {Wang}}, \bibinfo
  {author} {\bibfnamefont {C.}~\bibnamefont {Huang}}, \bibinfo {author}
  {\bibfnamefont {Y.~V.}\ \bibnamefont {Kartashov}}, \bibinfo {author}
  {\bibfnamefont {L.}~\bibnamefont {Torner}}, \bibinfo {author} {\bibfnamefont
  {V.~V.}\ \bibnamefont {Konotop}},\ and\ \bibinfo {author} {\bibfnamefont
  {F.}~\bibnamefont {Ye}},\ }\bibfield  {title} {\bibinfo {title} {Optical
  soliton formation controlled by angle twisting in photonic moir{\'e}
  lattices},\ }\href@noop {} {\bibfield  {journal} {\bibinfo  {journal} {Nature
  Photonics}\ }\textbf {\bibinfo {volume} {14}},\ \bibinfo {pages} {663}
  (\bibinfo {year} {2020})}\BibitemShut {NoStop}%
\bibitem [{\citenamefont {Wang}\ \emph {et~al.}(2020)\citenamefont {Wang},
  \citenamefont {Zheng}, \citenamefont {Chen}, \citenamefont {Huang},
  \citenamefont {Kartashov}, \citenamefont {Torner}, \citenamefont {Konotop},\
  and\ \citenamefont {Ye}}]{wang2020localization}%
  \BibitemOpen
  \bibfield  {author} {\bibinfo {author} {\bibfnamefont {P.}~\bibnamefont
  {Wang}}, \bibinfo {author} {\bibfnamefont {Y.}~\bibnamefont {Zheng}},
  \bibinfo {author} {\bibfnamefont {X.}~\bibnamefont {Chen}}, \bibinfo {author}
  {\bibfnamefont {C.}~\bibnamefont {Huang}}, \bibinfo {author} {\bibfnamefont
  {Y.~V.}\ \bibnamefont {Kartashov}}, \bibinfo {author} {\bibfnamefont
  {L.}~\bibnamefont {Torner}}, \bibinfo {author} {\bibfnamefont {V.~V.}\
  \bibnamefont {Konotop}},\ and\ \bibinfo {author} {\bibfnamefont
  {F.}~\bibnamefont {Ye}},\ }\bibfield  {title} {\bibinfo {title} {Localization
  and delocalization of light in photonic moir{\'e} lattices},\ }\href@noop {}
  {\bibfield  {journal} {\bibinfo  {journal} {Nature}\ }\textbf {\bibinfo
  {volume} {577}},\ \bibinfo {pages} {42} (\bibinfo {year} {2020})}\BibitemShut
  {NoStop}%
\bibitem [{\citenamefont {Lou}\ \emph {et~al.}(2021)\citenamefont {Lou},
  \citenamefont {Zhao}, \citenamefont {Minkov}, \citenamefont {Guo},
  \citenamefont {Orenstein},\ and\ \citenamefont {Fan}}]{lou2021}%
  \BibitemOpen
  \bibfield  {author} {\bibinfo {author} {\bibfnamefont {B.}~\bibnamefont
  {Lou}}, \bibinfo {author} {\bibfnamefont {N.}~\bibnamefont {Zhao}}, \bibinfo
  {author} {\bibfnamefont {M.}~\bibnamefont {Minkov}}, \bibinfo {author}
  {\bibfnamefont {C.}~\bibnamefont {Guo}}, \bibinfo {author} {\bibfnamefont
  {M.}~\bibnamefont {Orenstein}},\ and\ \bibinfo {author} {\bibfnamefont
  {S.}~\bibnamefont {Fan}},\ }\bibfield  {title} {\bibinfo {title} {Theory for
  twisted bilayer photonic crystal slabs},\ }\href
  {https://doi.org/10.1103/PhysRevLett.126.136101} {\bibfield  {journal}
  {\bibinfo  {journal} {Phys. Rev. Lett.}\ }\textbf {\bibinfo {volume} {126}},\
  \bibinfo {pages} {136101} (\bibinfo {year} {2021})}\BibitemShut {NoStop}%
\bibitem [{\citenamefont {Jiang}\ \emph {et~al.}(2022)\citenamefont {Jiang},
  \citenamefont {Liu}, \citenamefont {Zheng}, \citenamefont {Duan},\ and\
  \citenamefont {Xia}}]{phon2022}%
  \BibitemOpen
  \bibfield  {author} {\bibinfo {author} {\bibfnamefont {Z.}~\bibnamefont
  {Jiang}}, \bibinfo {author} {\bibfnamefont {J.}~\bibnamefont {Liu}}, \bibinfo
  {author} {\bibfnamefont {S.}~\bibnamefont {Zheng}}, \bibinfo {author}
  {\bibfnamefont {G.}~\bibnamefont {Duan}},\ and\ \bibinfo {author}
  {\bibfnamefont {B.}~\bibnamefont {Xia}},\ }\bibfield  {title} {\bibinfo
  {title} {{Phononic twisted moiré lattice with quasicrystalline patterns}},\
  }\href {https://doi.org/10.1063/5.0109404} {\bibfield  {journal} {\bibinfo
  {journal} {Applied Physics Letters}\ }\textbf {\bibinfo {volume} {121}},\
  \bibinfo {pages} {142202} (\bibinfo {year} {2022})},\ \Eprint
  {https://arxiv.org/abs/https://pubs.aip.org/aip/apl/article-pdf/doi/10.1063/5.0109404/16483255/142202\_1\_online.pdf}
  {https://pubs.aip.org/aip/apl/article-pdf/doi/10.1063/5.0109404/16483255/142202\_1\_online.pdf}
  \BibitemShut {NoStop}%
\bibitem [{\citenamefont {Chung}\ \emph {et~al.}(2020)\citenamefont {Chung},
  \citenamefont {Qin}, \citenamefont {Zhang}, \citenamefont {Schollw\"ock},\
  and\ \citenamefont {White}}]{white2020}%
  \BibitemOpen
  \bibfield  {author} {\bibinfo {author} {\bibfnamefont {C.-M.}\ \bibnamefont
  {Chung}}, \bibinfo {author} {\bibfnamefont {M.}~\bibnamefont {Qin}}, \bibinfo
  {author} {\bibfnamefont {S.}~\bibnamefont {Zhang}}, \bibinfo {author}
  {\bibfnamefont {U.}~\bibnamefont {Schollw\"ock}},\ and\ \bibinfo {author}
  {\bibfnamefont {S.~R.}\ \bibnamefont {White}} (\bibinfo {collaboration} {The
  Simons Collaboration on the Many-Electron Problem}),\ }\bibfield  {title}
  {\bibinfo {title} {Plaquette versus ordinary $d$-wave pairing in the
  ${t}^{\ensuremath{'}}$-hubbard model on a width-4 cylinder},\ }\href
  {https://doi.org/10.1103/PhysRevB.102.041106} {\bibfield  {journal} {\bibinfo
   {journal} {Phys. Rev. B}\ }\textbf {\bibinfo {volume} {102}},\ \bibinfo
  {pages} {041106} (\bibinfo {year} {2020})}\BibitemShut {NoStop}%
\bibitem [{\citenamefont {Jiang}\ \emph {et~al.}(2020)\citenamefont {Jiang},
  \citenamefont {Zaanen}, \citenamefont {Devereaux},\ and\ \citenamefont
  {Jiang}}]{jiang2020}%
  \BibitemOpen
  \bibfield  {author} {\bibinfo {author} {\bibfnamefont {Y.-F.}\ \bibnamefont
  {Jiang}}, \bibinfo {author} {\bibfnamefont {J.}~\bibnamefont {Zaanen}},
  \bibinfo {author} {\bibfnamefont {T.~P.}\ \bibnamefont {Devereaux}},\ and\
  \bibinfo {author} {\bibfnamefont {H.-C.}\ \bibnamefont {Jiang}},\ }\bibfield
  {title} {\bibinfo {title} {Ground state phase diagram of the doped hubbard
  model on the four-leg cylinder},\ }\href
  {https://doi.org/10.1103/PhysRevResearch.2.033073} {\bibfield  {journal}
  {\bibinfo  {journal} {Phys. Rev. Res.}\ }\textbf {\bibinfo {volume} {2}},\
  \bibinfo {pages} {033073} (\bibinfo {year} {2020})}\BibitemShut {NoStop}%
\bibitem [{\citenamefont {Danilov}\ \emph {et~al.}(2022)\citenamefont
  {Danilov}, \citenamefont {van Loon}, \citenamefont {Brener}, \citenamefont
  {Iskakov}, \citenamefont {Katsnelson},\ and\ \citenamefont
  {Lichtenstein}}]{danilov2022degenerate}%
  \BibitemOpen
  \bibfield  {author} {\bibinfo {author} {\bibfnamefont {M.}~\bibnamefont
  {Danilov}}, \bibinfo {author} {\bibfnamefont {E.~G.}\ \bibnamefont {van
  Loon}}, \bibinfo {author} {\bibfnamefont {S.}~\bibnamefont {Brener}},
  \bibinfo {author} {\bibfnamefont {S.}~\bibnamefont {Iskakov}}, \bibinfo
  {author} {\bibfnamefont {M.~I.}\ \bibnamefont {Katsnelson}},\ and\ \bibinfo
  {author} {\bibfnamefont {A.~I.}\ \bibnamefont {Lichtenstein}},\ }\bibfield
  {title} {\bibinfo {title} {Degenerate plaquette physics as key ingredient of
  high-temperature superconductivity in cuprates},\ }\href@noop {} {\bibfield
  {journal} {\bibinfo  {journal} {npj Quantum Materials}\ }\textbf {\bibinfo
  {volume} {7}},\ \bibinfo {pages} {50} (\bibinfo {year} {2022})}\BibitemShut
  {NoStop}%
\bibitem [{\citenamefont {Lu}\ \emph {et~al.}(2024)\citenamefont {Lu},
  \citenamefont {Chen}, \citenamefont {Zhu}, \citenamefont {Sheng},\ and\
  \citenamefont {Gong}}]{tj2024}%
  \BibitemOpen
  \bibfield  {author} {\bibinfo {author} {\bibfnamefont {X.}~\bibnamefont
  {Lu}}, \bibinfo {author} {\bibfnamefont {F.}~\bibnamefont {Chen}}, \bibinfo
  {author} {\bibfnamefont {W.}~\bibnamefont {Zhu}}, \bibinfo {author}
  {\bibfnamefont {D.~N.}\ \bibnamefont {Sheng}},\ and\ \bibinfo {author}
  {\bibfnamefont {S.-S.}\ \bibnamefont {Gong}},\ }\bibfield  {title} {\bibinfo
  {title} {{Emergent Superconductivity and Competing Charge Orders in
  Hole-Doped Square-Lattice $t\text{\ensuremath{-}}J$ Model}},\ }\href
  {https://doi.org/10.1103/PhysRevLett.132.066002} {\bibfield  {journal}
  {\bibinfo  {journal} {Phys. Rev. Lett.}\ }\textbf {\bibinfo {volume} {132}},\
  \bibinfo {pages} {066002} (\bibinfo {year} {2024})}\BibitemShut {NoStop}%
\bibitem [{\citenamefont {Qu}\ \emph {et~al.}(2024)\citenamefont {Qu},
  \citenamefont {Li}, \citenamefont {Gong}, \citenamefont {Qi}, \citenamefont
  {Li},\ and\ \citenamefont {Su}}]{qu2024}%
  \BibitemOpen
  \bibfield  {author} {\bibinfo {author} {\bibfnamefont {D.-W.}\ \bibnamefont
  {Qu}}, \bibinfo {author} {\bibfnamefont {Q.}~\bibnamefont {Li}}, \bibinfo
  {author} {\bibfnamefont {S.-S.}\ \bibnamefont {Gong}}, \bibinfo {author}
  {\bibfnamefont {Y.}~\bibnamefont {Qi}}, \bibinfo {author} {\bibfnamefont
  {W.}~\bibnamefont {Li}},\ and\ \bibinfo {author} {\bibfnamefont
  {G.}~\bibnamefont {Su}},\ }\bibfield  {title} {\bibinfo {title} {Phase
  diagram, $d$-wave superconductivity, and pseudogap of the
  $t\ensuremath{-}{t}^{\ensuremath{'}}\ensuremath{-}j$ model at finite
  temperature},\ }\href {https://doi.org/10.1103/PhysRevLett.133.256003}
  {\bibfield  {journal} {\bibinfo  {journal} {Phys. Rev. Lett.}\ }\textbf
  {\bibinfo {volume} {133}},\ \bibinfo {pages} {256003} (\bibinfo {year}
  {2024})}\BibitemShut {NoStop}%
\bibitem [{\citenamefont {Jiang}\ \emph {et~al.}(2012)\citenamefont {Jiang},
  \citenamefont {Yao},\ and\ \citenamefont {Balents}}]{jiang2012}%
  \BibitemOpen
  \bibfield  {author} {\bibinfo {author} {\bibfnamefont {H.-C.}\ \bibnamefont
  {Jiang}}, \bibinfo {author} {\bibfnamefont {H.}~\bibnamefont {Yao}},\ and\
  \bibinfo {author} {\bibfnamefont {L.}~\bibnamefont {Balents}},\ }\bibfield
  {title} {\bibinfo {title} {Spin liquid ground state of the spin-$\frac{1}{2}$
  square ${J}_{1}$-${J}_{2}$ heisenberg model},\ }\href
  {https://doi.org/10.1103/PhysRevB.86.024424} {\bibfield  {journal} {\bibinfo
  {journal} {Phys. Rev. B}\ }\textbf {\bibinfo {volume} {86}},\ \bibinfo
  {pages} {024424} (\bibinfo {year} {2012})}\BibitemShut {NoStop}%
\bibitem [{\citenamefont {Nomura}\ and\ \citenamefont
  {Imada}(2021)}]{imada2021}%
  \BibitemOpen
  \bibfield  {author} {\bibinfo {author} {\bibfnamefont {Y.}~\bibnamefont
  {Nomura}}\ and\ \bibinfo {author} {\bibfnamefont {M.}~\bibnamefont {Imada}},\
  }\bibfield  {title} {\bibinfo {title} {Dirac-type nodal spin liquid revealed
  by refined quantum many-body solver using neural-network wave function,
  correlation ratio, and level spectroscopy},\ }\href
  {https://doi.org/10.1103/PhysRevX.11.031034} {\bibfield  {journal} {\bibinfo
  {journal} {Phys. Rev. X}\ }\textbf {\bibinfo {volume} {11}},\ \bibinfo
  {pages} {031034} (\bibinfo {year} {2021})}\BibitemShut {NoStop}%
\bibitem [{\citenamefont {Qian}\ and\ \citenamefont
  {Qin}(2024)}]{qianj1j22024}%
  \BibitemOpen
  \bibfield  {author} {\bibinfo {author} {\bibfnamefont {X.}~\bibnamefont
  {Qian}}\ and\ \bibinfo {author} {\bibfnamefont {M.}~\bibnamefont {Qin}},\
  }\bibfield  {title} {\bibinfo {title} {Absence of spin liquid phase in the
  ${J}_{1}\ensuremath{-}{J}_{2}$ heisenberg model on the square lattice},\
  }\href {https://doi.org/10.1103/PhysRevB.109.L161103} {\bibfield  {journal}
  {\bibinfo  {journal} {Phys. Rev. B}\ }\textbf {\bibinfo {volume} {109}},\
  \bibinfo {pages} {L161103} (\bibinfo {year} {2024})}\BibitemShut {NoStop}%
\bibitem [{\citenamefont {Rückriegel}\ \emph {et~al.}(2024)\citenamefont
  {Rückriegel}, \citenamefont {Tarasevych},\ and\ \citenamefont
  {Kopietz}}]{ruckriegel2024}%
  \BibitemOpen
  \bibfield  {author} {\bibinfo {author} {\bibfnamefont {A.}~\bibnamefont
  {Rückriegel}}, \bibinfo {author} {\bibfnamefont {D.}~\bibnamefont
  {Tarasevych}},\ and\ \bibinfo {author} {\bibfnamefont {P.}~\bibnamefont
  {Kopietz}},\ }\href@noop {} {\bibinfo {title} {Phase diagram of the
  $j_1$-$j_2$ quantum heisenberg model for arbitrary spin}} (\bibinfo {year}
  {2024}),\ \Eprint {https://arxiv.org/abs/2403.08713} {arXiv:2403.08713
  [cond-mat.str-el]} \BibitemShut {NoStop}%
\bibitem [{Note1()}]{Note1}%
  \BibitemOpen
  \bibinfo {note} {See Supplementary Material that includes Refs. \cite
  {theory_Vafek&Kang2018_TBGwannier,theory_Vafek&Kang2023_T(u),Ballentine,Soper_notes,glazman2011}}\BibitemShut
  {NoStop}%
\bibitem [{\citenamefont {Bistritzer}\ and\ \citenamefont
  {MacDonald}(2011)}]{bistritzer2011moire}%
  \BibitemOpen
  \bibfield  {author} {\bibinfo {author} {\bibfnamefont {R.}~\bibnamefont
  {Bistritzer}}\ and\ \bibinfo {author} {\bibfnamefont {A.~H.}\ \bibnamefont
  {MacDonald}},\ }\bibfield  {title} {\bibinfo {title} {Moir{\'e} bands in
  twisted double-layer graphene},\ }\href@noop {} {\bibfield  {journal}
  {\bibinfo  {journal} {Proceedings of the National Academy of Sciences}\
  }\textbf {\bibinfo {volume} {108}},\ \bibinfo {pages} {12233} (\bibinfo
  {year} {2011})}\BibitemShut {NoStop}%
\bibitem [{\citenamefont {Hejazi}\ \emph {et~al.}(2019)\citenamefont {Hejazi},
  \citenamefont {Liu}, \citenamefont {Shapourian}, \citenamefont {Chen},\ and\
  \citenamefont {Balents}}]{hejazi2019}%
  \BibitemOpen
  \bibfield  {author} {\bibinfo {author} {\bibfnamefont {K.}~\bibnamefont
  {Hejazi}}, \bibinfo {author} {\bibfnamefont {C.}~\bibnamefont {Liu}},
  \bibinfo {author} {\bibfnamefont {H.}~\bibnamefont {Shapourian}}, \bibinfo
  {author} {\bibfnamefont {X.}~\bibnamefont {Chen}},\ and\ \bibinfo {author}
  {\bibfnamefont {L.}~\bibnamefont {Balents}},\ }\bibfield  {title} {\bibinfo
  {title} {Multiple topological transitions in twisted bilayer graphene near
  the first magic angle},\ }\href {https://doi.org/10.1103/PhysRevB.99.035111}
  {\bibfield  {journal} {\bibinfo  {journal} {Phys. Rev. B}\ }\textbf {\bibinfo
  {volume} {99}},\ \bibinfo {pages} {035111} (\bibinfo {year}
  {2019})}\BibitemShut {NoStop}%
\bibitem [{\citenamefont {Balents}(2019)}]{balents_scipost}%
  \BibitemOpen
  \bibfield  {author} {\bibinfo {author} {\bibfnamefont {L.}~\bibnamefont
  {Balents}},\ }\bibfield  {title} {\bibinfo {title} {{General continuum model
  for twisted bilayer graphene and arbitrary smooth deformations}},\ }\href
  {https://doi.org/10.21468/SciPostPhys.7.4.048} {\bibfield  {journal}
  {\bibinfo  {journal} {SciPost Phys.}\ }\textbf {\bibinfo {volume} {7}},\
  \bibinfo {pages} {048} (\bibinfo {year} {2019})}\BibitemShut {NoStop}%
\bibitem [{\citenamefont {Tarnopolsky}\ \emph {et~al.}(2019)\citenamefont
  {Tarnopolsky}, \citenamefont {Kruchkov},\ and\ \citenamefont
  {Vishwanath}}]{tarnop2019}%
  \BibitemOpen
  \bibfield  {author} {\bibinfo {author} {\bibfnamefont {G.}~\bibnamefont
  {Tarnopolsky}}, \bibinfo {author} {\bibfnamefont {A.~J.}\ \bibnamefont
  {Kruchkov}},\ and\ \bibinfo {author} {\bibfnamefont {A.}~\bibnamefont
  {Vishwanath}},\ }\bibfield  {title} {\bibinfo {title} {Origin of magic angles
  in twisted bilayer graphene},\ }\href
  {https://doi.org/10.1103/PhysRevLett.122.106405} {\bibfield  {journal}
  {\bibinfo  {journal} {Phys. Rev. Lett.}\ }\textbf {\bibinfo {volume} {122}},\
  \bibinfo {pages} {106405} (\bibinfo {year} {2019})}\BibitemShut {NoStop}%
\bibitem [{Note2()}]{Note2}%
  \BibitemOpen
  \bibinfo {note} {The Zeeman effect produces a spin splitting
  only.}\BibitemShut {Stop}%
\bibitem [{\citenamefont {Sachdev}(2002)}]{sachdev2002}%
  \BibitemOpen
  \bibfield  {author} {\bibinfo {author} {\bibfnamefont {S.}~\bibnamefont
  {Sachdev}},\ }\bibfield  {title} {\bibinfo {title} {Quantum phase transitions
  of correlated electrons in two dimensions},\ }\href
  {https://doi.org/https://doi.org/10.1016/S0378-4371(02)01040-3} {\bibfield
  {journal} {\bibinfo  {journal} {Physica A: Statistical Mechanics and its
  Applications}\ }\textbf {\bibinfo {volume} {313}},\ \bibinfo {pages} {252}
  (\bibinfo {year} {2002})},\ \bibinfo {note} {fundamental Problems in
  Statistical Physics}\BibitemShut {NoStop}%
\bibitem [{\citenamefont {Nica}\ and\ \citenamefont
  {Si}(2021)}]{nica2021multiorbital}%
  \BibitemOpen
  \bibfield  {author} {\bibinfo {author} {\bibfnamefont {E.~M.}\ \bibnamefont
  {Nica}}\ and\ \bibinfo {author} {\bibfnamefont {Q.}~\bibnamefont {Si}},\
  }\bibfield  {title} {\bibinfo {title} {Multiorbital singlet pairing and d+ d
  superconductivity},\ }\href@noop {} {\bibfield  {journal} {\bibinfo
  {journal} {npj Quantum Materials}\ }\textbf {\bibinfo {volume} {6}},\
  \bibinfo {pages} {3} (\bibinfo {year} {2021})}\BibitemShut {NoStop}%
\bibitem [{\citenamefont {Das}\ \emph {et~al.}(2022)\citenamefont {Das},
  \citenamefont {Shen}, \citenamefont {Jaoui}, \citenamefont
  {Herzog-Arbeitman}, \citenamefont {Chew}, \citenamefont {Cho}, \citenamefont
  {Watanabe}, \citenamefont {Taniguchi}, \citenamefont {Piot}, \citenamefont
  {Bernevig},\ and\ \citenamefont {Efetov}}]{efetovflux}%
  \BibitemOpen
  \bibfield  {author} {\bibinfo {author} {\bibfnamefont {I.}~\bibnamefont
  {Das}}, \bibinfo {author} {\bibfnamefont {C.}~\bibnamefont {Shen}}, \bibinfo
  {author} {\bibfnamefont {A.}~\bibnamefont {Jaoui}}, \bibinfo {author}
  {\bibfnamefont {J.}~\bibnamefont {Herzog-Arbeitman}}, \bibinfo {author}
  {\bibfnamefont {A.}~\bibnamefont {Chew}}, \bibinfo {author} {\bibfnamefont
  {C.-W.}\ \bibnamefont {Cho}}, \bibinfo {author} {\bibfnamefont
  {K.}~\bibnamefont {Watanabe}}, \bibinfo {author} {\bibfnamefont
  {T.}~\bibnamefont {Taniguchi}}, \bibinfo {author} {\bibfnamefont {B.~A.}\
  \bibnamefont {Piot}}, \bibinfo {author} {\bibfnamefont {B.~A.}\ \bibnamefont
  {Bernevig}},\ and\ \bibinfo {author} {\bibfnamefont {D.~K.}\ \bibnamefont
  {Efetov}},\ }\bibfield  {title} {\bibinfo {title} {Observation of reentrant
  correlated insulators and interaction-driven fermi-surface reconstructions at
  one magnetic flux quantum per moir\'e unit cell in magic-angle twisted
  bilayer graphene},\ }\href {https://doi.org/10.1103/PhysRevLett.128.217701}
  {\bibfield  {journal} {\bibinfo  {journal} {Phys. Rev. Lett.}\ }\textbf
  {\bibinfo {volume} {128}},\ \bibinfo {pages} {217701} (\bibinfo {year}
  {2022})}\BibitemShut {NoStop}%
\bibitem [{\citenamefont {Affleck}\ and\ \citenamefont
  {Marston}(1988)}]{affleck1988}%
  \BibitemOpen
  \bibfield  {author} {\bibinfo {author} {\bibfnamefont {I.}~\bibnamefont
  {Affleck}}\ and\ \bibinfo {author} {\bibfnamefont {J.~B.}\ \bibnamefont
  {Marston}},\ }\bibfield  {title} {\bibinfo {title} {Large-n limit of the
  heisenberg-hubbard model: Implications for high-${T}_{c}$ superconductors},\
  }\href {https://doi.org/10.1103/PhysRevB.37.3774} {\bibfield  {journal}
  {\bibinfo  {journal} {Phys. Rev. B}\ }\textbf {\bibinfo {volume} {37}},\
  \bibinfo {pages} {3774} (\bibinfo {year} {1988})}\BibitemShut {NoStop}%
\bibitem [{\citenamefont {Wen}\ \emph {et~al.}(1989)\citenamefont {Wen},
  \citenamefont {Wilczek},\ and\ \citenamefont {Zee}}]{wen1989}%
  \BibitemOpen
  \bibfield  {author} {\bibinfo {author} {\bibfnamefont {X.~G.}\ \bibnamefont
  {Wen}}, \bibinfo {author} {\bibfnamefont {F.}~\bibnamefont {Wilczek}},\ and\
  \bibinfo {author} {\bibfnamefont {A.}~\bibnamefont {Zee}},\ }\bibfield
  {title} {\bibinfo {title} {Chiral spin states and superconductivity},\ }\href
  {https://doi.org/10.1103/PhysRevB.39.11413} {\bibfield  {journal} {\bibinfo
  {journal} {Phys. Rev. B}\ }\textbf {\bibinfo {volume} {39}},\ \bibinfo
  {pages} {11413} (\bibinfo {year} {1989})}\BibitemShut {NoStop}%
\bibitem [{\citenamefont {Hatsugai}\ and\ \citenamefont
  {Kohmoto}(1990)}]{hatsugai1990}%
  \BibitemOpen
  \bibfield  {author} {\bibinfo {author} {\bibfnamefont {Y.}~\bibnamefont
  {Hatsugai}}\ and\ \bibinfo {author} {\bibfnamefont {M.}~\bibnamefont
  {Kohmoto}},\ }\bibfield  {title} {\bibinfo {title} {Energy spectrum and the
  quantum hall effect on the square lattice with next-nearest-neighbor
  hopping},\ }\href {https://doi.org/10.1103/PhysRevB.42.8282} {\bibfield
  {journal} {\bibinfo  {journal} {Phys. Rev. B}\ }\textbf {\bibinfo {volume}
  {42}},\ \bibinfo {pages} {8282} (\bibinfo {year} {1990})}\BibitemShut
  {NoStop}%
\bibitem [{\citenamefont {Neupert}\ \emph {et~al.}(2011)\citenamefont
  {Neupert}, \citenamefont {Santos}, \citenamefont {Chamon},\ and\
  \citenamefont {Mudry}}]{neupert2011}%
  \BibitemOpen
  \bibfield  {author} {\bibinfo {author} {\bibfnamefont {T.}~\bibnamefont
  {Neupert}}, \bibinfo {author} {\bibfnamefont {L.}~\bibnamefont {Santos}},
  \bibinfo {author} {\bibfnamefont {C.}~\bibnamefont {Chamon}},\ and\ \bibinfo
  {author} {\bibfnamefont {C.}~\bibnamefont {Mudry}},\ }\bibfield  {title}
  {\bibinfo {title} {Fractional quantum hall states at zero magnetic field},\
  }\href {https://doi.org/10.1103/PhysRevLett.106.236804} {\bibfield  {journal}
  {\bibinfo  {journal} {Phys. Rev. Lett.}\ }\textbf {\bibinfo {volume} {106}},\
  \bibinfo {pages} {236804} (\bibinfo {year} {2011})}\BibitemShut {NoStop}%
\bibitem [{\citenamefont {Guo}\ \emph {et~al.}(2018)\citenamefont {Guo},
  \citenamefont {Khatami}, \citenamefont {Wang}, \citenamefont {Devereaux},
  \citenamefont {Singh},\ and\ \citenamefont {Scalettar}}]{scalettarpairing}%
  \BibitemOpen
  \bibfield  {author} {\bibinfo {author} {\bibfnamefont {H.}~\bibnamefont
  {Guo}}, \bibinfo {author} {\bibfnamefont {E.}~\bibnamefont {Khatami}},
  \bibinfo {author} {\bibfnamefont {Y.}~\bibnamefont {Wang}}, \bibinfo {author}
  {\bibfnamefont {T.~P.}\ \bibnamefont {Devereaux}}, \bibinfo {author}
  {\bibfnamefont {R.~R.~P.}\ \bibnamefont {Singh}},\ and\ \bibinfo {author}
  {\bibfnamefont {R.~T.}\ \bibnamefont {Scalettar}},\ }\bibfield  {title}
  {\bibinfo {title} {Unconventional pairing symmetry of interacting dirac
  fermions on a $\ensuremath{\pi}$-flux lattice},\ }\href
  {https://doi.org/10.1103/PhysRevB.97.155146} {\bibfield  {journal} {\bibinfo
  {journal} {Phys. Rev. B}\ }\textbf {\bibinfo {volume} {97}},\ \bibinfo
  {pages} {155146} (\bibinfo {year} {2018})}\BibitemShut {NoStop}%
\bibitem [{\citenamefont {Chang}\ and\ \citenamefont
  {Scalettar}(2012)}]{scalettar2012}%
  \BibitemOpen
  \bibfield  {author} {\bibinfo {author} {\bibfnamefont {C.-C.}\ \bibnamefont
  {Chang}}\ and\ \bibinfo {author} {\bibfnamefont {R.~T.}\ \bibnamefont
  {Scalettar}},\ }\bibfield  {title} {\bibinfo {title} {Quantum disordered
  phase near the mott transition in the staggered-flux hubbard model on a
  square lattice},\ }\href {https://doi.org/10.1103/PhysRevLett.109.026404}
  {\bibfield  {journal} {\bibinfo  {journal} {Phys. Rev. Lett.}\ }\textbf
  {\bibinfo {volume} {109}},\ \bibinfo {pages} {026404} (\bibinfo {year}
  {2012})}\BibitemShut {NoStop}%
\bibitem [{\citenamefont {Zhu}\ \emph {et~al.}(2022)\citenamefont {Zhu},
  \citenamefont {Huang}, \citenamefont {Guo},\ and\ \citenamefont
  {Feng}}]{feng2022}%
  \BibitemOpen
  \bibfield  {author} {\bibinfo {author} {\bibfnamefont {X.}~\bibnamefont
  {Zhu}}, \bibinfo {author} {\bibfnamefont {Y.}~\bibnamefont {Huang}}, \bibinfo
  {author} {\bibfnamefont {H.}~\bibnamefont {Guo}},\ and\ \bibinfo {author}
  {\bibfnamefont {S.}~\bibnamefont {Feng}},\ }\bibfield  {title} {\bibinfo
  {title} {Quantum phase transitions from competing short- and long-range
  interactions on a $\ensuremath{\pi}$-flux lattice},\ }\href
  {https://doi.org/10.1103/PhysRevB.106.075109} {\bibfield  {journal} {\bibinfo
   {journal} {Phys. Rev. B}\ }\textbf {\bibinfo {volume} {106}},\ \bibinfo
  {pages} {075109} (\bibinfo {year} {2022})}\BibitemShut {NoStop}%
\bibitem [{\citenamefont {Ding}\ \emph {et~al.}(2022)\citenamefont {Ding},
  \citenamefont {Wang}, \citenamefont {Moritz}, \citenamefont {Schattner},
  \citenamefont {Huang},\ and\ \citenamefont
  {Devereaux}}]{ding2022thermodynamics}%
  \BibitemOpen
  \bibfield  {author} {\bibinfo {author} {\bibfnamefont {J.~K.}\ \bibnamefont
  {Ding}}, \bibinfo {author} {\bibfnamefont {W.~O.}\ \bibnamefont {Wang}},
  \bibinfo {author} {\bibfnamefont {B.}~\bibnamefont {Moritz}}, \bibinfo
  {author} {\bibfnamefont {Y.}~\bibnamefont {Schattner}}, \bibinfo {author}
  {\bibfnamefont {E.~W.}\ \bibnamefont {Huang}},\ and\ \bibinfo {author}
  {\bibfnamefont {T.~P.}\ \bibnamefont {Devereaux}},\ }\bibfield  {title}
  {\bibinfo {title} {Thermodynamics of correlated electrons in a magnetic
  field},\ }\href@noop {} {\bibfield  {journal} {\bibinfo  {journal}
  {Communications Physics}\ }\textbf {\bibinfo {volume} {5}},\ \bibinfo {pages}
  {204} (\bibinfo {year} {2022})}\BibitemShut {NoStop}%
\bibitem [{\citenamefont {Campi}\ \emph {et~al.}(2023)\citenamefont {Campi},
  \citenamefont {Mounet}, \citenamefont {Gibertini}, \citenamefont {Pizzi},\
  and\ \citenamefont {Marzari}}]{campi2023expansion}%
  \BibitemOpen
  \bibfield  {author} {\bibinfo {author} {\bibfnamefont {D.}~\bibnamefont
  {Campi}}, \bibinfo {author} {\bibfnamefont {N.}~\bibnamefont {Mounet}},
  \bibinfo {author} {\bibfnamefont {M.}~\bibnamefont {Gibertini}}, \bibinfo
  {author} {\bibfnamefont {G.}~\bibnamefont {Pizzi}},\ and\ \bibinfo {author}
  {\bibfnamefont {N.}~\bibnamefont {Marzari}},\ }\bibfield  {title} {\bibinfo
  {title} {Expansion of the materials cloud 2d database},\ }\href@noop {}
  {\bibfield  {journal} {\bibinfo  {journal} {ACS nano}\ }\textbf {\bibinfo
  {volume} {17}},\ \bibinfo {pages} {11268} (\bibinfo {year}
  {2023})}\BibitemShut {NoStop}%
\bibitem [{\citenamefont {Jiang}\ \emph {et~al.}(2024)\citenamefont {Jiang},
  \citenamefont {Petralanda}, \citenamefont {Skorupskii}, \citenamefont {Xu},
  \citenamefont {Pi}, \citenamefont {Călugăru}, \citenamefont {Hu},
  \citenamefont {Xie}, \citenamefont {Mustaf}, \citenamefont {Höhn},
  \citenamefont {Haase}, \citenamefont {Vergniory}, \citenamefont {Claassen},
  \citenamefont {Elcoro}, \citenamefont {Regnault}, \citenamefont {Shan},
  \citenamefont {Mak}, \citenamefont {Efetov}, \citenamefont {Morosan},
  \citenamefont {Kennes}, \citenamefont {Rubio}, \citenamefont {Xian},
  \citenamefont {Felser}, \citenamefont {Schoop},\ and\ \citenamefont
  {Bernevig}}]{jiang2024_2dtheoreticallytwistablematerial}%
  \BibitemOpen
  \bibfield  {author} {\bibinfo {author} {\bibfnamefont {Y.}~\bibnamefont
  {Jiang}}, \bibinfo {author} {\bibfnamefont {U.}~\bibnamefont {Petralanda}},
  \bibinfo {author} {\bibfnamefont {G.}~\bibnamefont {Skorupskii}}, \bibinfo
  {author} {\bibfnamefont {Q.}~\bibnamefont {Xu}}, \bibinfo {author}
  {\bibfnamefont {H.}~\bibnamefont {Pi}}, \bibinfo {author} {\bibfnamefont
  {D.}~\bibnamefont {Călugăru}}, \bibinfo {author} {\bibfnamefont
  {H.}~\bibnamefont {Hu}}, \bibinfo {author} {\bibfnamefont {J.}~\bibnamefont
  {Xie}}, \bibinfo {author} {\bibfnamefont {R.~A.}\ \bibnamefont {Mustaf}},
  \bibinfo {author} {\bibfnamefont {P.}~\bibnamefont {Höhn}}, \bibinfo
  {author} {\bibfnamefont {V.}~\bibnamefont {Haase}}, \bibinfo {author}
  {\bibfnamefont {M.~G.}\ \bibnamefont {Vergniory}}, \bibinfo {author}
  {\bibfnamefont {M.}~\bibnamefont {Claassen}}, \bibinfo {author}
  {\bibfnamefont {L.}~\bibnamefont {Elcoro}}, \bibinfo {author} {\bibfnamefont
  {N.}~\bibnamefont {Regnault}}, \bibinfo {author} {\bibfnamefont
  {J.}~\bibnamefont {Shan}}, \bibinfo {author} {\bibfnamefont {K.~F.}\
  \bibnamefont {Mak}}, \bibinfo {author} {\bibfnamefont {D.~K.}\ \bibnamefont
  {Efetov}}, \bibinfo {author} {\bibfnamefont {E.}~\bibnamefont {Morosan}},
  \bibinfo {author} {\bibfnamefont {D.~M.}\ \bibnamefont {Kennes}}, \bibinfo
  {author} {\bibfnamefont {A.}~\bibnamefont {Rubio}}, \bibinfo {author}
  {\bibfnamefont {L.}~\bibnamefont {Xian}}, \bibinfo {author} {\bibfnamefont
  {C.}~\bibnamefont {Felser}}, \bibinfo {author} {\bibfnamefont {L.~M.}\
  \bibnamefont {Schoop}},\ and\ \bibinfo {author} {\bibfnamefont {B.~A.}\
  \bibnamefont {Bernevig}},\ }\href {https://arxiv.org/abs/2411.09741}
  {\bibinfo {title} {2d theoretically twistable material database}} (\bibinfo
  {year} {2024}),\ \Eprint {https://arxiv.org/abs/2411.09741} {arXiv:2411.09741
  [cond-mat.mtrl-sci]} \BibitemShut {NoStop}%
\bibitem [{\citenamefont {Daeneke}\ \emph {et~al.}(2017)\citenamefont
  {Daeneke}, \citenamefont {Atkin}, \citenamefont {Orrell-Trigg}, \citenamefont
  {Zavabeti}, \citenamefont {Ahmed}, \citenamefont {Walia}, \citenamefont
  {Liu}, \citenamefont {Tachibana}, \citenamefont {Javaid}, \citenamefont
  {Greentree} \emph {et~al.}}]{daeneke2017wafer}%
  \BibitemOpen
  \bibfield  {author} {\bibinfo {author} {\bibfnamefont {T.}~\bibnamefont
  {Daeneke}}, \bibinfo {author} {\bibfnamefont {P.}~\bibnamefont {Atkin}},
  \bibinfo {author} {\bibfnamefont {R.}~\bibnamefont {Orrell-Trigg}}, \bibinfo
  {author} {\bibfnamefont {A.}~\bibnamefont {Zavabeti}}, \bibinfo {author}
  {\bibfnamefont {T.}~\bibnamefont {Ahmed}}, \bibinfo {author} {\bibfnamefont
  {S.}~\bibnamefont {Walia}}, \bibinfo {author} {\bibfnamefont
  {M.}~\bibnamefont {Liu}}, \bibinfo {author} {\bibfnamefont {Y.}~\bibnamefont
  {Tachibana}}, \bibinfo {author} {\bibfnamefont {M.}~\bibnamefont {Javaid}},
  \bibinfo {author} {\bibfnamefont {A.~D.}\ \bibnamefont {Greentree}}, \emph
  {et~al.},\ }\bibfield  {title} {\bibinfo {title} {Wafer-scale synthesis of
  semiconducting sno monolayers from interfacial oxide layers of metallic
  liquid tin},\ }\href@noop {} {\bibfield  {journal} {\bibinfo  {journal} {ACS
  nano}\ }\textbf {\bibinfo {volume} {11}},\ \bibinfo {pages} {10974} (\bibinfo
  {year} {2017})}\BibitemShut {NoStop}%
\bibitem [{\citenamefont {Jiang}\ \emph {et~al.}(2023)\citenamefont {Jiang},
  \citenamefont {Ji}, \citenamefont {Gong}, \citenamefont {Ding}, \citenamefont
  {Li}, \citenamefont {Li}, \citenamefont {Li},\ and\ \citenamefont
  {Geng}}]{jiang2023mechanical}%
  \BibitemOpen
  \bibfield  {author} {\bibinfo {author} {\bibfnamefont {K.}~\bibnamefont
  {Jiang}}, \bibinfo {author} {\bibfnamefont {J.}~\bibnamefont {Ji}}, \bibinfo
  {author} {\bibfnamefont {W.}~\bibnamefont {Gong}}, \bibinfo {author}
  {\bibfnamefont {L.}~\bibnamefont {Ding}}, \bibinfo {author} {\bibfnamefont
  {J.}~\bibnamefont {Li}}, \bibinfo {author} {\bibfnamefont {P.}~\bibnamefont
  {Li}}, \bibinfo {author} {\bibfnamefont {B.}~\bibnamefont {Li}},\ and\
  \bibinfo {author} {\bibfnamefont {F.}~\bibnamefont {Geng}},\ }\bibfield
  {title} {\bibinfo {title} {Mechanical cleavage of non-van der waals
  structures towards two-dimensional crystals},\ }\href@noop {} {\bibfield
  {journal} {\bibinfo  {journal} {Nature Synthesis}\ }\textbf {\bibinfo
  {volume} {2}},\ \bibinfo {pages} {58} (\bibinfo {year} {2023})}\BibitemShut
  {NoStop}%
\bibitem [{\citenamefont {Wang}\ \emph {et~al.}(2015)\citenamefont {Wang},
  \citenamefont {Li},\ and\ \citenamefont {Chen}}]{wang2015not}%
  \BibitemOpen
  \bibfield  {author} {\bibinfo {author} {\bibfnamefont {Y.}~\bibnamefont
  {Wang}}, \bibinfo {author} {\bibfnamefont {Y.}~\bibnamefont {Li}},\ and\
  \bibinfo {author} {\bibfnamefont {Z.}~\bibnamefont {Chen}},\ }\bibfield
  {title} {\bibinfo {title} {Not your familiar two dimensional transition metal
  disulfide: structural and electronic properties of the pds 2 monolayer},\
  }\href@noop {} {\bibfield  {journal} {\bibinfo  {journal} {Journal of
  Materials Chemistry C}\ }\textbf {\bibinfo {volume} {3}},\ \bibinfo {pages}
  {9603} (\bibinfo {year} {2015})}\BibitemShut {NoStop}%
\bibitem [{\citenamefont {Zhang}\ \emph {et~al.}(2021)\citenamefont {Zhang},
  \citenamefont {Su}, \citenamefont {Lu}, \citenamefont {Yang}, \citenamefont
  {Zhuang}, \citenamefont {Han}, \citenamefont {Wang}, \citenamefont {Wan},
  \citenamefont {Yu},\ and\ \citenamefont {Yang}}]{zhang2021centimeter}%
  \BibitemOpen
  \bibfield  {author} {\bibinfo {author} {\bibfnamefont {X.}~\bibnamefont
  {Zhang}}, \bibinfo {author} {\bibfnamefont {G.}~\bibnamefont {Su}}, \bibinfo
  {author} {\bibfnamefont {J.}~\bibnamefont {Lu}}, \bibinfo {author}
  {\bibfnamefont {W.}~\bibnamefont {Yang}}, \bibinfo {author} {\bibfnamefont
  {W.}~\bibnamefont {Zhuang}}, \bibinfo {author} {\bibfnamefont
  {K.}~\bibnamefont {Han}}, \bibinfo {author} {\bibfnamefont {X.}~\bibnamefont
  {Wang}}, \bibinfo {author} {\bibfnamefont {Y.}~\bibnamefont {Wan}}, \bibinfo
  {author} {\bibfnamefont {X.}~\bibnamefont {Yu}},\ and\ \bibinfo {author}
  {\bibfnamefont {P.}~\bibnamefont {Yang}},\ }\bibfield  {title} {\bibinfo
  {title} {Centimeter-scale few-layer pds2: fabrication and physical
  properties},\ }\href@noop {} {\bibfield  {journal} {\bibinfo  {journal} {ACS
  applied materials \& interfaces}\ }\textbf {\bibinfo {volume} {13}},\
  \bibinfo {pages} {43063} (\bibinfo {year} {2021})}\BibitemShut {NoStop}%
\bibitem [{\citenamefont {Oyedele}\ \emph {et~al.}(2017)\citenamefont
  {Oyedele}, \citenamefont {Yang}, \citenamefont {Liang}, \citenamefont
  {Puretzky}, \citenamefont {Wang}, \citenamefont {Zhang}, \citenamefont {Yu},
  \citenamefont {Pudasaini}, \citenamefont {Ghosh}, \citenamefont {Liu} \emph
  {et~al.}}]{oyedele2017pdse2}%
  \BibitemOpen
  \bibfield  {author} {\bibinfo {author} {\bibfnamefont {A.~D.}\ \bibnamefont
  {Oyedele}}, \bibinfo {author} {\bibfnamefont {S.}~\bibnamefont {Yang}},
  \bibinfo {author} {\bibfnamefont {L.}~\bibnamefont {Liang}}, \bibinfo
  {author} {\bibfnamefont {A.~A.}\ \bibnamefont {Puretzky}}, \bibinfo {author}
  {\bibfnamefont {K.}~\bibnamefont {Wang}}, \bibinfo {author} {\bibfnamefont
  {J.}~\bibnamefont {Zhang}}, \bibinfo {author} {\bibfnamefont
  {P.}~\bibnamefont {Yu}}, \bibinfo {author} {\bibfnamefont {P.~R.}\
  \bibnamefont {Pudasaini}}, \bibinfo {author} {\bibfnamefont {A.~W.}\
  \bibnamefont {Ghosh}}, \bibinfo {author} {\bibfnamefont {Z.}~\bibnamefont
  {Liu}}, \emph {et~al.},\ }\bibfield  {title} {\bibinfo {title} {Pdse2:
  pentagonal two-dimensional layers with high air stability for electronics},\
  }\href@noop {} {\bibfield  {journal} {\bibinfo  {journal} {Journal of the
  American Chemical Society}\ }\textbf {\bibinfo {volume} {139}},\ \bibinfo
  {pages} {14090} (\bibinfo {year} {2017})}\BibitemShut {NoStop}%
\bibitem [{\citenamefont {Ricciardulli}\ \emph {et~al.}(2021)\citenamefont
  {Ricciardulli}, \citenamefont {Yang}, \citenamefont {Smet},\ and\
  \citenamefont {Saliba}}]{ricciardulli2021emerging}%
  \BibitemOpen
  \bibfield  {author} {\bibinfo {author} {\bibfnamefont {A.~G.}\ \bibnamefont
  {Ricciardulli}}, \bibinfo {author} {\bibfnamefont {S.}~\bibnamefont {Yang}},
  \bibinfo {author} {\bibfnamefont {J.~H.}\ \bibnamefont {Smet}},\ and\
  \bibinfo {author} {\bibfnamefont {M.}~\bibnamefont {Saliba}},\ }\bibfield
  {title} {\bibinfo {title} {Emerging perovskite monolayers},\ }\href@noop {}
  {\bibfield  {journal} {\bibinfo  {journal} {Nature Materials}\ }\textbf
  {\bibinfo {volume} {20}},\ \bibinfo {pages} {1325} (\bibinfo {year}
  {2021})}\BibitemShut {NoStop}%
\bibitem [{\citenamefont {Ji}\ \emph {et~al.}(2019)\citenamefont {Ji},
  \citenamefont {Cai}, \citenamefont {Paudel}, \citenamefont {Sun},
  \citenamefont {Zhang}, \citenamefont {Han}, \citenamefont {Wei},
  \citenamefont {Zang}, \citenamefont {Gu}, \citenamefont {Zhang} \emph
  {et~al.}}]{ji2019freestanding}%
  \BibitemOpen
  \bibfield  {author} {\bibinfo {author} {\bibfnamefont {D.}~\bibnamefont
  {Ji}}, \bibinfo {author} {\bibfnamefont {S.}~\bibnamefont {Cai}}, \bibinfo
  {author} {\bibfnamefont {T.~R.}\ \bibnamefont {Paudel}}, \bibinfo {author}
  {\bibfnamefont {H.}~\bibnamefont {Sun}}, \bibinfo {author} {\bibfnamefont
  {C.}~\bibnamefont {Zhang}}, \bibinfo {author} {\bibfnamefont
  {L.}~\bibnamefont {Han}}, \bibinfo {author} {\bibfnamefont {Y.}~\bibnamefont
  {Wei}}, \bibinfo {author} {\bibfnamefont {Y.}~\bibnamefont {Zang}}, \bibinfo
  {author} {\bibfnamefont {M.}~\bibnamefont {Gu}}, \bibinfo {author}
  {\bibfnamefont {Y.}~\bibnamefont {Zhang}}, \emph {et~al.},\ }\bibfield
  {title} {\bibinfo {title} {Freestanding crystalline oxide perovskites down to
  the monolayer limit},\ }\href@noop {} {\bibfield  {journal} {\bibinfo
  {journal} {Nature}\ }\textbf {\bibinfo {volume} {570}},\ \bibinfo {pages}
  {87} (\bibinfo {year} {2019})}\BibitemShut {NoStop}%
\bibitem [{\citenamefont {Xiao}\ and\ \citenamefont
  {Liu}(2021)}]{xiao2021freestanding}%
  \BibitemOpen
  \bibfield  {author} {\bibinfo {author} {\bibfnamefont {X.-B.}\ \bibnamefont
  {Xiao}}\ and\ \bibinfo {author} {\bibfnamefont {B.-G.}\ \bibnamefont {Liu}},\
  }\bibfield  {title} {\bibinfo {title} {Freestanding perovskite oxide
  monolayers as two-dimensional semiconductors},\ }\href@noop {} {\bibfield
  {journal} {\bibinfo  {journal} {Nanotechnology}\ }\textbf {\bibinfo {volume}
  {32}},\ \bibinfo {pages} {145705} (\bibinfo {year} {2021})}\BibitemShut
  {NoStop}%
\bibitem [{\citenamefont {Varrassi}\ \emph {et~al.}(2024)\citenamefont
  {Varrassi}, \citenamefont {Liu},\ and\ \citenamefont {Franchini}}]{sto_dft}%
  \BibitemOpen
  \bibfield  {author} {\bibinfo {author} {\bibfnamefont {L.}~\bibnamefont
  {Varrassi}}, \bibinfo {author} {\bibfnamefont {P.}~\bibnamefont {Liu}},\ and\
  \bibinfo {author} {\bibfnamefont {C.}~\bibnamefont {Franchini}},\ }\bibfield
  {title} {\bibinfo {title} {Quasiparticle and excitonic properties of
  monolayer ${\mathrm{srtio}}_{3}$},\ }\href
  {https://doi.org/10.1103/PhysRevMaterials.8.024001} {\bibfield  {journal}
  {\bibinfo  {journal} {Phys. Rev. Mater.}\ }\textbf {\bibinfo {volume} {8}},\
  \bibinfo {pages} {024001} (\bibinfo {year} {2024})}\BibitemShut {NoStop}%
\bibitem [{\citenamefont {Yu}\ \emph {et~al.}(2019)\citenamefont {Yu},
  \citenamefont {Ma}, \citenamefont {Cai}, \citenamefont {Zhong}, \citenamefont
  {Ye}, \citenamefont {Shen}, \citenamefont {Gu}, \citenamefont {Chen},\ and\
  \citenamefont {Zhang}}]{Yu2019}%
  \BibitemOpen
  \bibfield  {author} {\bibinfo {author} {\bibfnamefont {Y.}~\bibnamefont
  {Yu}}, \bibinfo {author} {\bibfnamefont {L.}~\bibnamefont {Ma}}, \bibinfo
  {author} {\bibfnamefont {P.}~\bibnamefont {Cai}}, \bibinfo {author}
  {\bibfnamefont {R.}~\bibnamefont {Zhong}}, \bibinfo {author} {\bibfnamefont
  {C.}~\bibnamefont {Ye}}, \bibinfo {author} {\bibfnamefont {J.}~\bibnamefont
  {Shen}}, \bibinfo {author} {\bibfnamefont {G.~D.}\ \bibnamefont {Gu}},
  \bibinfo {author} {\bibfnamefont {X.~H.}\ \bibnamefont {Chen}},\ and\
  \bibinfo {author} {\bibfnamefont {Y.}~\bibnamefont {Zhang}},\ }\bibfield
  {title} {\bibinfo {title} {High-temperature superconductivity in monolayer
  bi2sr2cacu2o8+$\delta$},\ }\href {https://doi.org/10.1038/s41586-019-1718-x}
  {\bibfield  {journal} {\bibinfo  {journal} {Nature}\ }\textbf {\bibinfo
  {volume} {575}},\ \bibinfo {pages} {156} (\bibinfo {year}
  {2019})}\BibitemShut {NoStop}%
\bibitem [{\citenamefont {Kanayama}\ \emph {et~al.}(2017)\citenamefont
  {Kanayama}, \citenamefont {Nakayama}, \citenamefont {Phan}, \citenamefont
  {Kuno}, \citenamefont {Sugawara}, \citenamefont {Takahashi},\ and\
  \citenamefont {Sato}}]{exp_FeSe/SrTiO3_DiracSemimetal}%
  \BibitemOpen
  \bibfield  {author} {\bibinfo {author} {\bibfnamefont {S.}~\bibnamefont
  {Kanayama}}, \bibinfo {author} {\bibfnamefont {K.}~\bibnamefont {Nakayama}},
  \bibinfo {author} {\bibfnamefont {G.~N.}\ \bibnamefont {Phan}}, \bibinfo
  {author} {\bibfnamefont {M.}~\bibnamefont {Kuno}}, \bibinfo {author}
  {\bibfnamefont {K.}~\bibnamefont {Sugawara}}, \bibinfo {author}
  {\bibfnamefont {T.}~\bibnamefont {Takahashi}},\ and\ \bibinfo {author}
  {\bibfnamefont {T.}~\bibnamefont {Sato}},\ }\bibfield  {title} {\bibinfo
  {title} {Two-dimensional dirac semimetal phase in undoped one-monolayer
  {FeSe} film},\ }\href {https://doi.org/10.1103/PhysRevB.96.220509} {\bibfield
   {journal} {\bibinfo  {journal} {Phys. Rev. B}\ }\textbf {\bibinfo {volume}
  {96}},\ \bibinfo {pages} {220509} (\bibinfo {year} {2017})}\BibitemShut
  {NoStop}%
\bibitem [{\citenamefont {Zhang}\ \emph {et~al.}(2016)\citenamefont {Zhang},
  \citenamefont {Lee}, \citenamefont {Moore}, \citenamefont {Li}, \citenamefont
  {Yi}, \citenamefont {Hashimoto}, \citenamefont {Lu}, \citenamefont
  {Devereaux}, \citenamefont {Lee},\ and\ \citenamefont
  {Shen}}]{exp_FeSe/SrTiO3_ZX}%
  \BibitemOpen
  \bibfield  {author} {\bibinfo {author} {\bibfnamefont {Y.}~\bibnamefont
  {Zhang}}, \bibinfo {author} {\bibfnamefont {J.~J.}\ \bibnamefont {Lee}},
  \bibinfo {author} {\bibfnamefont {R.~G.}\ \bibnamefont {Moore}}, \bibinfo
  {author} {\bibfnamefont {W.}~\bibnamefont {Li}}, \bibinfo {author}
  {\bibfnamefont {M.}~\bibnamefont {Yi}}, \bibinfo {author} {\bibfnamefont
  {M.}~\bibnamefont {Hashimoto}}, \bibinfo {author} {\bibfnamefont {D.~H.}\
  \bibnamefont {Lu}}, \bibinfo {author} {\bibfnamefont {T.~P.}\ \bibnamefont
  {Devereaux}}, \bibinfo {author} {\bibfnamefont {D.-H.}\ \bibnamefont {Lee}},\
  and\ \bibinfo {author} {\bibfnamefont {Z.-X.}\ \bibnamefont {Shen}},\
  }\bibfield  {title} {\bibinfo {title} {Superconducting gap anisotropy in
  monolayer {FeSe} thin film},\ }\href
  {https://doi.org/10.1103/PhysRevLett.117.117001} {\bibfield  {journal}
  {\bibinfo  {journal} {Phys. Rev. Lett.}\ }\textbf {\bibinfo {volume} {117}},\
  \bibinfo {pages} {117001} (\bibinfo {year} {2016})}\BibitemShut {NoStop}%
\bibitem [{\citenamefont {Shigekawa}\ \emph {et~al.}(2019)\citenamefont
  {Shigekawa}, \citenamefont {Nakayama}, \citenamefont {Kuno}, \citenamefont
  {Phan}, \citenamefont {Owada}, \citenamefont {Sugawara}, \citenamefont
  {Takahashi},\ and\ \citenamefont
  {Sato}}]{exp_Shigekawa&Sato2019_monolayerFeSe&FeTe&FeS}%
  \BibitemOpen
  \bibfield  {author} {\bibinfo {author} {\bibfnamefont {K.}~\bibnamefont
  {Shigekawa}}, \bibinfo {author} {\bibfnamefont {K.}~\bibnamefont {Nakayama}},
  \bibinfo {author} {\bibfnamefont {M.}~\bibnamefont {Kuno}}, \bibinfo {author}
  {\bibfnamefont {G.~N.}\ \bibnamefont {Phan}}, \bibinfo {author}
  {\bibfnamefont {K.}~\bibnamefont {Owada}}, \bibinfo {author} {\bibfnamefont
  {K.}~\bibnamefont {Sugawara}}, \bibinfo {author} {\bibfnamefont
  {T.}~\bibnamefont {Takahashi}},\ and\ \bibinfo {author} {\bibfnamefont
  {T.}~\bibnamefont {Sato}},\ }\bibfield  {title} {\bibinfo {title} {Dichotomy
  of superconductivity between monolayer fes and fese},\ }\href@noop {}
  {\bibfield  {journal} {\bibinfo  {journal} {Proceedings of the National
  Academy of Sciences}\ }\textbf {\bibinfo {volume} {116}},\ \bibinfo {pages}
  {24470} (\bibinfo {year} {2019})}\BibitemShut {NoStop}%
\bibitem [{\citenamefont {Kang}\ and\ \citenamefont
  {Vafek}(2018)}]{theory_Vafek&Kang2018_TBGwannier}%
  \BibitemOpen
  \bibfield  {author} {\bibinfo {author} {\bibfnamefont {J.}~\bibnamefont
  {Kang}}\ and\ \bibinfo {author} {\bibfnamefont {O.}~\bibnamefont {Vafek}},\
  }\bibfield  {title} {\bibinfo {title} {Symmetry, maximally localized wannier
  states, and a low-energy model for twisted bilayer graphene narrow bands},\
  }\href {https://doi.org/10.1103/PhysRevX.8.031088} {\bibfield  {journal}
  {\bibinfo  {journal} {Phys. Rev. X}\ }\textbf {\bibinfo {volume} {8}},\
  \bibinfo {pages} {031088} (\bibinfo {year} {2018})}\BibitemShut {NoStop}%
\bibitem [{\citenamefont {Vafek}\ and\ \citenamefont
  {Kang}(2023)}]{theory_Vafek&Kang2023_T(u)}%
  \BibitemOpen
  \bibfield  {author} {\bibinfo {author} {\bibfnamefont {O.}~\bibnamefont
  {Vafek}}\ and\ \bibinfo {author} {\bibfnamefont {J.}~\bibnamefont {Kang}},\
  }\bibfield  {title} {\bibinfo {title} {Continuum effective hamiltonian for
  graphene bilayers for an arbitrary smooth lattice deformation from
  microscopic theories},\ }\href {https://doi.org/10.1103/PhysRevB.107.075123}
  {\bibfield  {journal} {\bibinfo  {journal} {Phys. Rev. B}\ }\textbf {\bibinfo
  {volume} {107}},\ \bibinfo {pages} {075123} (\bibinfo {year}
  {2023})}\BibitemShut {NoStop}%
\bibitem [{\citenamefont {Ballentine}(2003)}]{Ballentine}%
  \BibitemOpen
  \bibfield  {author} {\bibinfo {author} {\bibfnamefont {L.~E.}\ \bibnamefont
  {Ballentine}},\ }\href@noop {} {\emph {\bibinfo {title} {Quantum Mechanics: A
  Modern Development}}}\ (\bibinfo  {publisher} {World Scientific Publishing
  Co. Pte Ltd.},\ \bibinfo {year} {2003})\BibitemShut {NoStop}%
\bibitem [{\citenamefont {Soper}(2011)}]{Soper_notes}%
  \BibitemOpen
  \bibfield  {author} {\bibinfo {author} {\bibfnamefont {D.~E.}\ \bibnamefont
  {Soper}},\ }\href
  {https://pages.uoregon.edu/soper/QuantumMechanics/boosts.pdf} {\bibinfo
  {title} {Galilean boost symmetry (lecture notes)}} (\bibinfo {year} {2011}),\
  \bibinfo {note} {publicly available notes on the instantaneous Galilean
  boost.}\BibitemShut {Stop}%
\bibitem [{\citenamefont {Catelani}\ \emph {et~al.}(2011)\citenamefont
  {Catelani}, \citenamefont {Schoelkopf}, \citenamefont {Devoret},\ and\
  \citenamefont {Glazman}}]{glazman2011}%
  \BibitemOpen
  \bibfield  {author} {\bibinfo {author} {\bibfnamefont {G.}~\bibnamefont
  {Catelani}}, \bibinfo {author} {\bibfnamefont {R.~J.}\ \bibnamefont
  {Schoelkopf}}, \bibinfo {author} {\bibfnamefont {M.~H.}\ \bibnamefont
  {Devoret}},\ and\ \bibinfo {author} {\bibfnamefont {L.~I.}\ \bibnamefont
  {Glazman}},\ }\bibfield  {title} {\bibinfo {title} {Relaxation and frequency
  shifts induced by quasiparticles in superconducting qubits},\ }\href
  {https://doi.org/10.1103/PhysRevB.84.064517} {\bibfield  {journal} {\bibinfo
  {journal} {Phys. Rev. B}\ }\textbf {\bibinfo {volume} {84}},\ \bibinfo
  {pages} {064517} (\bibinfo {year} {2011})}\BibitemShut {NoStop}%
\bibitem [{\citenamefont {Cvetkovic}\ and\ \citenamefont
  {Vafek}(2013)}]{cvetkovic2013}%
  \BibitemOpen
  \bibfield  {author} {\bibinfo {author} {\bibfnamefont {V.}~\bibnamefont
  {Cvetkovic}}\ and\ \bibinfo {author} {\bibfnamefont {O.}~\bibnamefont
  {Vafek}},\ }\bibfield  {title} {\bibinfo {title} {Space group symmetry,
  spin-orbit coupling, and the low-energy effective hamiltonian for iron-based
  superconductors},\ }\href {https://doi.org/10.1103/PhysRevB.88.134510}
  {\bibfield  {journal} {\bibinfo  {journal} {Phys. Rev. B}\ }\textbf {\bibinfo
  {volume} {88}},\ \bibinfo {pages} {134510} (\bibinfo {year}
  {2013})}\BibitemShut {NoStop}%
\bibitem [{\citenamefont {Eugenio}\ and\ \citenamefont
  {Vafek}(2023{\natexlab{b}})}]{paul_fese}%
  \BibitemOpen
  \bibfield  {author} {\bibinfo {author} {\bibfnamefont {P.~M.}\ \bibnamefont
  {Eugenio}}\ and\ \bibinfo {author} {\bibfnamefont {O.}~\bibnamefont
  {Vafek}},\ }\bibfield  {title} {\bibinfo {title} {{Twisted-bilayer FeSe and
  the Fe-based superlattices}},\ }\href
  {https://doi.org/10.21468/SciPostPhys.15.3.081} {\bibfield  {journal}
  {\bibinfo  {journal} {SciPost Phys.}\ }\textbf {\bibinfo {volume} {15}},\
  \bibinfo {pages} {081} (\bibinfo {year} {2023}{\natexlab{b}})}\BibitemShut
  {NoStop}%
\bibitem [{Note3()}]{Note3}%
  \BibitemOpen
  \bibinfo {note} {{We stress that the ``boost" operation described above,
  which we wrote $e^{i{\protect \bf q}\cdot \protect \hat {\protect \bf x}}$,
  is formally the instantaneous ($t=0$) Galileian boost when referring to a
  system of like-mass particles. (See {\protect \it Ballentine} \cite
  {Ballentine} for a complete treatise on the Galilei group.) It is not a
  symmetry of the Hamiltonian \cite {Soper_notes,Ballentine}, but an operation
  which connects physically identical representations of motion. It should not
  be confused with the notion of Galilean symmetry, which is an invariance of
  the Schr\"{o}dinger equation valid for free particles \cite
  {Ballentine}.}}\BibitemShut {Stop}%
\bibitem [{\citenamefont {Arzamasovs}\ and\ \citenamefont
  {Liu}(2017)}]{arzamasovs2017tight}%
  \BibitemOpen
  \bibfield  {author} {\bibinfo {author} {\bibfnamefont {M.}~\bibnamefont
  {Arzamasovs}}\ and\ \bibinfo {author} {\bibfnamefont {B.}~\bibnamefont
  {Liu}},\ }\bibfield  {title} {\bibinfo {title} {Tight-binding tunneling
  amplitude of an optical lattice},\ }\href@noop {} {\bibfield  {journal}
  {\bibinfo  {journal} {European Journal of Physics}\ }\textbf {\bibinfo
  {volume} {38}},\ \bibinfo {pages} {065405} (\bibinfo {year}
  {2017})}\BibitemShut {NoStop}%
\bibitem [{\citenamefont {Auerbach}(2012)}]{auerbach2012interacting}%
  \BibitemOpen
  \bibfield  {author} {\bibinfo {author} {\bibfnamefont {A.}~\bibnamefont
  {Auerbach}},\ }\href@noop {} {\emph {\bibinfo {title} {Interacting electrons
  and quantum magnetism}}}\ (\bibinfo  {publisher} {Springer Science \&
  Business Media},\ \bibinfo {year} {2012})\BibitemShut {NoStop}%
\bibitem [{\citenamefont {Young}\ and\ \citenamefont
  {Levitov}(2011)}]{levitov2011}%
  \BibitemOpen
  \bibfield  {author} {\bibinfo {author} {\bibfnamefont {A.~F.}\ \bibnamefont
  {Young}}\ and\ \bibinfo {author} {\bibfnamefont {L.~S.}\ \bibnamefont
  {Levitov}},\ }\bibfield  {title} {\bibinfo {title} {Capacitance of graphene
  bilayer as a probe of layer-specific properties},\ }\href
  {https://doi.org/10.1103/PhysRevB.84.085441} {\bibfield  {journal} {\bibinfo
  {journal} {Phys. Rev. B}\ }\textbf {\bibinfo {volume} {84}},\ \bibinfo
  {pages} {085441} (\bibinfo {year} {2011})}\BibitemShut {NoStop}%
\end{thebibliography}%
\onecolumngrid
\clearpage
\twocolumngrid

 \appendix*
  \setcounter{equation}{0}

\begin{center}
{\bf\large End Matter}
\end{center}

\section{Connection between different representations and realization of symmetries}
The continuum Hamiltonian for the $M$ point of the square lattice can be obtained following Refs. \cite{bistritzer2011moire,balents_scipost,10.21468/SciPostPhys.15.3.081} 
\begin{eqnarray}\label{Eqn:Main:H_unfolded}
H = 
\begin{pmatrix}
h(-i\boldsymbol{\nabla}^{-\theta/2}_{\bf x}+{\bf q}_-)  & T({\bf x})\\
T({\bf x})^* & h(-i\boldsymbol{\nabla}_{\bf x}^{\theta/2}+{\bf q}_+) \\
\end{pmatrix},
\end{eqnarray}
where $h({\bf k})$ is the dispersion of electrons in a single layer around $M$. The moir\'e reciprocal lattice vectors are given by ${\bf g}_{1,2}=R_{\theta/2}{\bf G}_{1,2}-R_{-\theta/2}{\bf G}_{1,2}$ (where ${\bf G}_{1,2}$ are the reciprocal vectors of a single unrotated layer), while ${\bf q}_\pm=R_{\pm \theta/2}{\bf M}-{\bf M}$. Eq. \eqref{Eqn:Main:H_leading} is obtained in the leading order in the twist angle, where the momentum origin has been shifted by ${\bf q}_-$, we used the expansion $h(-i\boldsymbol{\nabla}_{\bf x}) \approx \mu\nabla_{\bf x}^2$ and ${\bf q}_M \equiv {\bf q}_+-{\bf q}_- = ({\bf g}_1+{\bf g}_2)/2$ is the moir\'e Brillouin zone corner  momentum.

Analysis of symmetries in Eq. \eqref{Eqn:Main:H_unfolded}  is complicated by the fact that point-group symmetries are realized nontrivially due to the momentum offset between two layers. As an example, consider the action of spinless time-reversal symmetry $\mathcal{T}$ on \eqref{Eqn:Main:H_unfolded}. Indeed, a complex conjugation $K$ does not bring the diagonal of Eqn. \ref{Eqn:Main:H_unfolded} into itself, reflecting that the time-reversal invariant points for two layers are different: ${\bf M}_{\theta/2}$ \& ${\bf M}_{-\theta/2}$. 
Instead, the proper action of $\mathcal{T}$ has to compensate for this difference resulting in $\mathcal{T}=\mathcal{U}K\mathcal{U}^{\dagger}=K(\mathcal{U}^{\dagger})^2$, where 
\begin{eqnarray}\label{Eqn:Main:U_folding}
\mathcal{U}=\begin{pmatrix}
e^{i{\bf q}_+\cdot{\bf x}} & 0 \\
0 & e^{i{\bf q}_-\cdot{\bf x}}
\end{pmatrix} ,
\end{eqnarray}
(note that $R_{\mp \theta/2} {\bf q}_\pm = -{\bf q}_\mp$ and $\mathcal{U}^{\dagger}=\mathcal{U}^*$). This procedure holds more generally for determining the action of any point-group symmetry $g$ of the twisted bilayer on Eqn. \eqref{Eqn:Main:H_unfolded}, i.e $\mathcal{U}g\mathcal{U}^{\dagger}$ \cite{Note1}.

It follows then that all point-group operations will be realized conventionally (as just $g$) after the following unitary transformation of \eqref{Eqn:Main:H_unfolded}: $H_{f} = \mathcal{U}^{\dagger}H\mathcal{U}$, which results in Eq. \eqref{Eqn:Main:H_folded}.

\section{Rectangular lattice case}
We now extend our continuum model to the $S$ point (Brillouin zone corner) on the rectangular lattice. As for the square case, there are four relevant tunneling processes. The mirror symmetries $({\bf G}_{1},{\bf G}_{2})\to({\bf G}_{1},-{\bf G}_{2});\;(-{\bf G}_{1},{\bf G}_{2})$ guarantee that the coefficients in the tunneling term are equal to one another. Therefore, the only difference with the square lattice case is the presence of anisotropy in the single-layer dispersion, already at the level of quadratic expansion in momentum around $S$: $h(-i\boldsymbol{\nabla}_{\bf x})  \approx \mu_x \partial_x^2+\mu_y \partial_y^2$. Performing the same steps as for the square lattice, the Hamiltonian after the unitary transformation, Eq. \eqref{Eqn:Main:U_folding}, absorbing the momentum offset between layers (analogous to Eq. \eqref{Eqn:Main:H_folded}) is:
\begin{equation}
H^{rec}_f \approx 
\mu_x \partial_x^2+\mu_y \partial_y^2 
+
(\mu_x -\mu_y) \theta \partial_x\partial_y \sigma_z
+ U({\bf x})\sigma_x,
\end{equation}
where $U({\bf x})$ is defined in the same way as for the square lattice and corrections being of order $O(\theta^2)$. The second term, breaking the sublattice symmetry is suppressed at low twist angles where it is of the order $(\mu_x-\mu_y) \theta k_x k_y \propto \theta^3 $ as compared to $\theta^5$ in the square case. It is still subleading with respect to other terms, such that $\mathcal{L}$-symmetry is well-defined. In that case, the system will be described by an effective moir\'e rectangular lattice with next-nearest neighbor hopping; application of $V$ would induce nearest-neighbor one. The band dispersion consistent with the symmetries is then: $\epsilon_{\bf k}=2\big(t_x(V)\cos(k_x)+t_y(V)\cos(k_y)\big) +2t'\Big(\cos\big(k_x+k_y\big)+\cos\big(k_x-k_y\big)\Big)$, where $t_{x,y}(V) \propto V$. Unlike the square case, there is no variable separation, so the values of $t,t'$ can not be found from the solutions of 1D Mathieu equation and need to be computed numerically; qualitatively, the only difference with the square lattice case is the built-in anisotropy $t_x(V)\neq t_y(V)$.

\section{Orbital degeneracy at the $M$ point}
\label{sec:qbt_moire}
\begin{figure*}[t]
\centering
  \includegraphics[width=.31\linewidth]{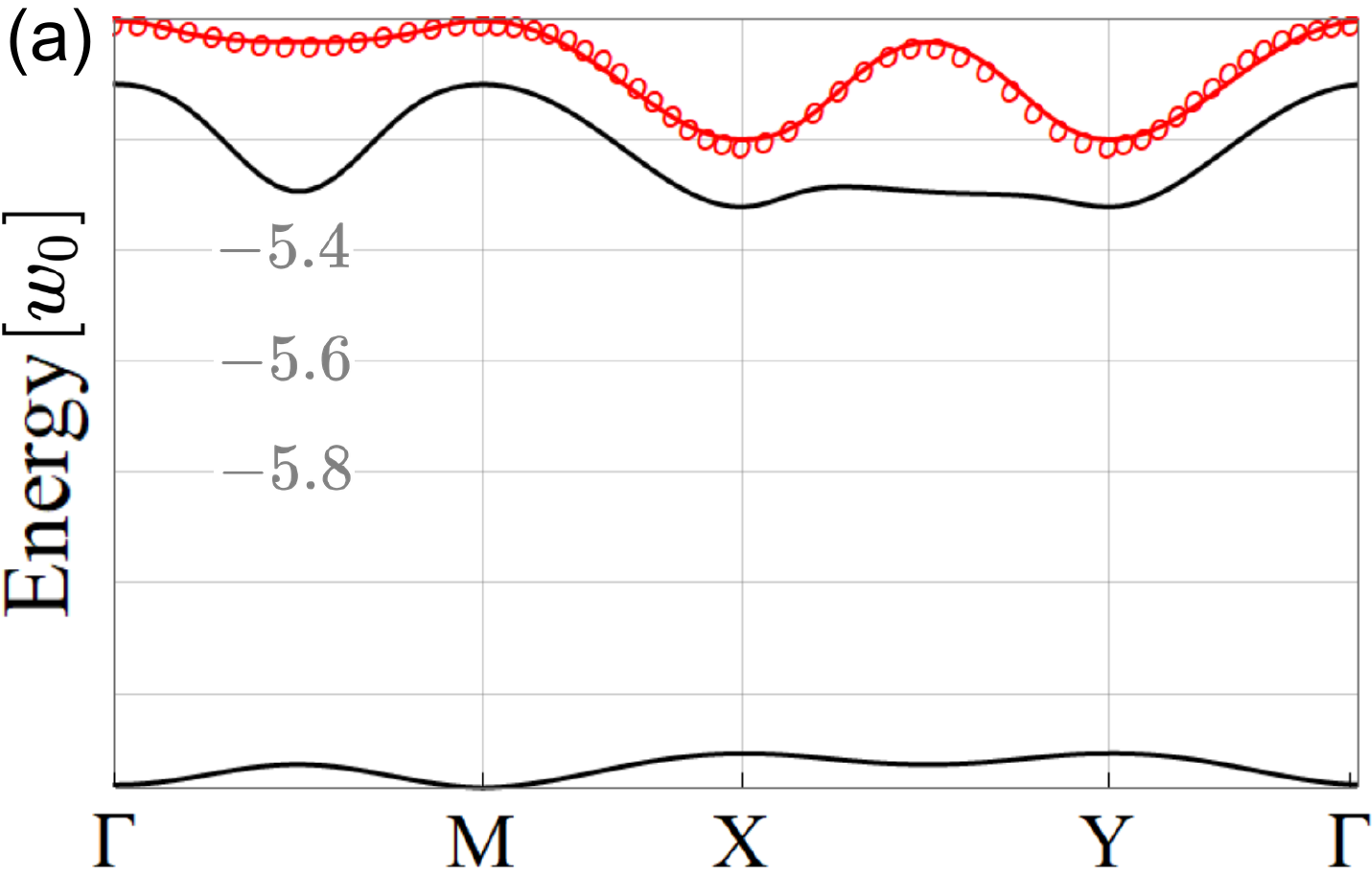}
  \includegraphics[width=.31\linewidth]{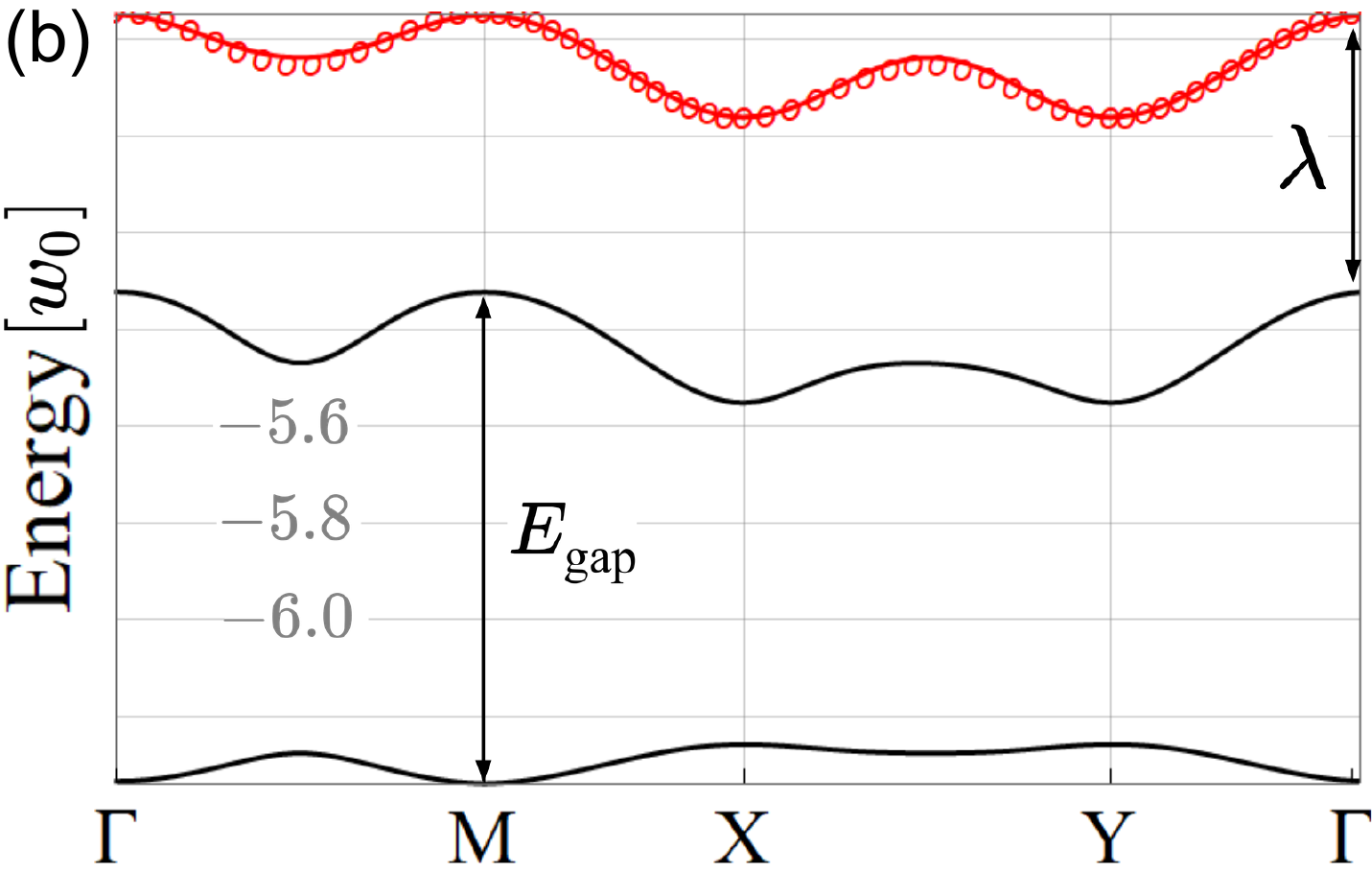}
  \includegraphics[width=.31\linewidth]{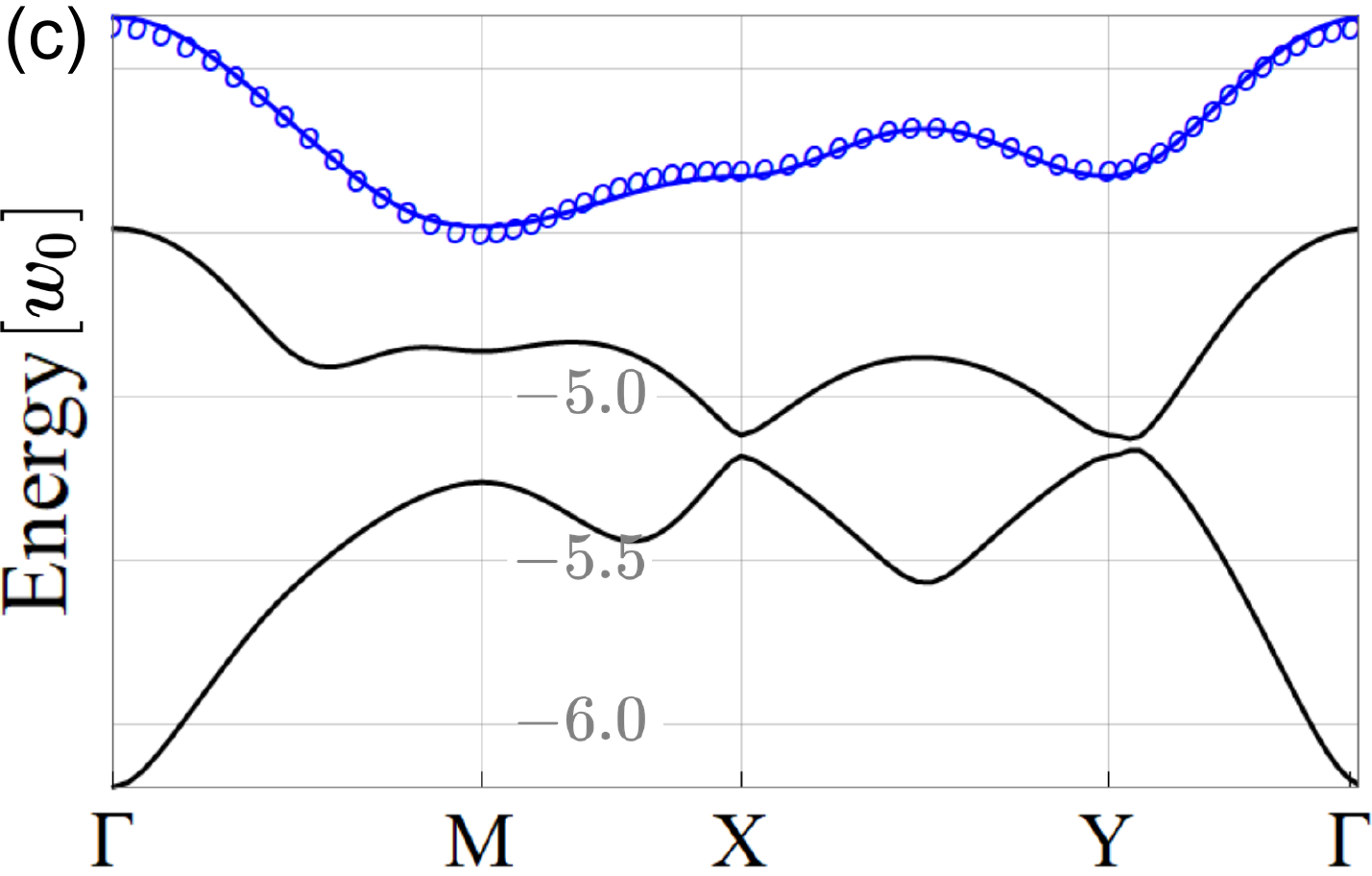}
  \caption{(a) Band structure (in units of interlayer tunneling amplitude $w_0$) of the model Eq. \eqref{Eqn:HGamma} with $\theta/\theta^*=.94/\sqrt{3}$ with $w_1=0$ as a function of spin-orbit coupling $\lambda$: 
    (a) $\lambda/w_0=.067$   (b,c) $\lambda/w_0=.33$. Circles represent the best fits to the top band of the extended tight-binding model \eqref{Eqn:Main:epsilonK2}. The resulting parameters are [given also in comparison to analytical expressions (\ref{Eqn:Main:tp},\ref{Eqn:Main:t}) for the single-orbital model in the main text] (a) $t=0$ \& $t'=0.041=.9t_{\text{analytic}}$ \& $t''=-0.013$;\;
    (b)$t=0$ \& $t'=0.041=.9t_{\text{analytic}}$ \& $t''=0$
 (c) $V/w_0=1.72$ with $t=0.118=.5t_{\text{analytic}}$ \& $t'=0.023=.5t_{\text{analytic}}$ \& $t''=-0.013$. 
 (Note that ${\bf k}\cdot{\bf p}$ invariants are taken from Ref \cite{10.21468/SciPostPhys.15.3.081}: $a=-3440.93$, $b=-a/2.5$, $\mu=-2830$, $l_a=\pi/.8$, \& $\theta=1.79^o$.)
 }
 \label{fig:2orb}
\end{figure*}

In this section we discuss the $M$ -point moir\'e materials with an orbital degeneracy. For a square lattice with $D_{4h}$ point group such a degeneracy arises without fine-tuning for electronic states belonging to doubly degenerate $E_u$ or $E_g$ representations. The $k\cdot p$ continuum Hamiltonian for a single layer takes the form (following conventions used for describing Fe-based superconductor materials \cite{cvetkovic2013,paul_fese}):
\begin{eqnarray}
h_{orb}(-i\boldsymbol{\nabla}) = -\mu\nabla^2-a\partial_x\partial_y\tau_z -b(\partial_x^2-\partial_y^2)\tau_x + \lambda\tau_y s_z,  
\end{eqnarray}
where $\tau_{\mu}$, $s_{\alpha}$ are the Pauli matrices in the orbital and spin space. The Hamiltonian of the twisted bilayer can be constructed following the usual approach \cite{paul_fese,bistritzer2011moire}; however, one needs to take into account the possibility of interorbital tunneling between the layers. The Hamiltonian after the unitary transformation, Eq. \eqref{Eqn:Main:U_folding}, takes the form:
\begin{equation}\label{Eqn:HGamma}
\begin{gathered}
H_{orb} = h_{orb}(-i\boldsymbol{\nabla})   
+ 2w_0\big(\cos({\bf q}_1\cdot{\bf x}) + \cos({\bf q}_2\cdot{\bf x})\big)\sigma_x
\\
+ 2w_1\big(\cos({\bf q}_1\cdot{\bf x}) - \cos({\bf q}_2\cdot{\bf x})\big)\sigma_x\tau_x,
\end{gathered}
\end{equation}
where $\sigma_{\nu}$ are Pauli matrices in layer space. The last term becomes allowed because both $\big(\cos({\bf q}_1\cdot{\bf x}) - \cos({\bf q}_2\cdot{\bf x})\big)$ and $\tau_x$ transform as $B_{2g}$ representation. $H_{orb}$ has the $\mathcal{L}$ symmetry $[H_{orb},\mathcal{L}]=0$, such that the eigenstates decouple into sectors with $\sigma_x=\pm 1$:
\begin{eqnarray}
H_{\pm} = h_{orb}(-i\boldsymbol{\nabla})  \pm \Big(U_0({\bf x}) + U_1({\bf x})\tau_x\Big), 
\end{eqnarray}
where $U_0({\bf x})=2w_0\big(\cos({\bf q}_1\cdot{\bf x}) + \cos({\bf q}_2\cdot{\bf x})\big)$ and $U_1({\bf x})=2w_1\big(\cos({\bf q}_1\cdot{\bf x}) - \cos({\bf q}_2\cdot{\bf x})\big)$. Remarkably, $U_{0,1}({\bf x}+{\bf R}_1) = -U_{0,1}({\bf x})$, such that two decoupled sectors are identical, related by a shift by one moir\'e lattice constant.
Therefore, the emergent sublattice symmetry also holds for the 2-orbital case. Interestingly, $H_+$ directly corresponds to Eq. (5) in Ref. \cite{10.21468/SciPostPhys.15.3.081}, which describes the moir\'e bands near the $\Gamma$ point of twisted bilayer FeSe.

As $+$ and $-$ sector produce identical bands, we focus on $H_+$. In the limit $\lambda\gg w_0,w_0, a q_M^2,b q_M^2$, the eigenstates of $H_+$ are approximately eigenstates of $\tau_y$, i.e $\psi({\bf x})=\phi({\bf x})(1,\pm i)^T$. Projecting $H_{\pm}$ onto these states cancels all terms whose orbital component is $\tau_x$ or $\tau_z$, such that contributions of $a,b$ and $U_1$ are suppressed by $1/\lambda$. The projected $H_{\pm}$ reduces to Eq. \eqref{Eqn:Main:HarmonicOscillators} of the main text shifted by $\pm \lambda$. Thus, in this limit the orbital degeneracy does not affect the physics. 

We further diagonalize Eq. \eqref{Eqn:HGamma} numerically, with the results shown in Fig. \ref{fig:2orb} (note that the original Brillouin zone is used). We fit the resulting band with an extended tight-binding model:
\begin{eqnarray}\label{Eqn:Main:epsilonK2}
&&\epsilon_{\bf k}=2t\big(\cos(k_x)+\cos(k_y)\big) \notag\\
&+&2t'\Big(\cos\big(k_x+k_y\big)+\cos\big(k_x-k_y\big)\Big) \\
&+&2t''\Big(\cos(2k_x)+\cos(2k_y)\Big).
\end{eqnarray}
Already for $\lambda/w_0=.067$ (Fig. \ref{fig:2orb} (a), corresponding to $\lambda/W\approx 0.3$ with $W\approx 0.2 w_0$) the $t''/t'$ ratio is about $0.3$ suggesting relatively well-localized Wannier functions. On increasing $\lambda/w_0$ to $0.33$ (Fig. \ref{fig:2orb} (b), $\lambda/W\approx 1.7$) $t''$ becomes effectively zero, and the value of $t'$ corresponds well to the analytical expression for the single-orbital model \eqref{Eqn:Main:tp}. Finally, adding interlayer potential difference $V$ (Fig. \ref{fig:2orb} (c)) allows to create a separated narrow band with dominant nearest-neighbor hopping $t/t'>5$ without closing the gap to the other bands. We note that in this case the agreement with analytical expression for the single orbital case is worse, but still within the right order of magnitude. Thus, the 2-orbital model with $\lambda\gtrsim W$ allows to reproduce the tunable Hubbard model physics of the single-orbital case already for moderate values of $\lambda$, $\lambda \approx W$.

The results of Fig. \ref{fig:2orb} neglected $w_1$. As was demonstrated by one of us in \cite{10.21468/SciPostPhys.15.3.081}, increasing $w_1/w_0$ ratio for a finite $\lambda$ leads to a topological transition for the top band of $H_+$. For values of $w_1/w_0$ beyond a critical one, the top band acquires a spin Chern number of $\pm 1$. Furthermore, for a fixed $w_1/w_0$, increasing $\lambda$ drives the system into $|C_s|=0$ phase, which can be understood as follows. The source of the Berry curvature for the non-trivial topology is the quadratic band touching at $\Gamma_M$ point \cite{10.21468/SciPostPhys.15.3.081}, which is gapped for $\lambda\neq 0$. Increasing $\lambda$ disperses the Berry curvature over a larger area in momentum space, eventually pushing it out from the first Brillouin zone. In that case, sustaining enough Berry curvature in the first Brillouin zone to produce $|C_s| = 1$ is impossible, such that a topological transition to $|C_s|=0$ phase occurs, leading to a Dirac touching at the $M_m$ emergent at the transition \cite{10.21468/SciPostPhys.15.3.081}.

In contrast to Ref. \onlinecite{10.21468/SciPostPhys.15.3.081}, $\mathcal{L}$-symmetry implies the presence of a decoupled $H_-$ sector with the same Chern number. Therefore, the total spin Chern number of the system would be $\pm 2$, such that $\mathbf{Z}_2$ invariant is formally zero. However, the presence of $\mathcal{L}$-symmetry forbids scattering between the two sectors (or sublattices). $\mathcal{L}$-symmetry can be also preserved at the edge; indeed, adding a layer-symmetric edge potential $U_{edge}({\bf x}) \sigma_0$ (that can be, e.g., defined electrostatically in an experiment) does not violate the $\mathcal{L}$-symmetry. Therefore, the edge modes of the two sectors would be protected by the $\mathcal{L}$-symmetry from hybridizing, realizing a quantum spin Hall insulator with 4 edge modes.

In summary, 2-orbital $M$-point square lattice moir\'e materials allow to realize tunable Hubbard model as well as a $\mathcal{L}$-symmetry protected topological spin Hall phase. Atomic spin-orbit coupling $\lambda$ and intra-orbital tunneling $w_0$ favor Hubbard physics, while the interorbital tunneling $w_1$ favors topological Chern bands.

\newpage
\clearpage
\onecolumngrid
\appendix

\renewcommand{\thefigure}{S\arabic{figure}}
\addtocounter{equation}{-12}
\addtocounter{figure}{-3}
\renewcommand{\theequation}{S\arabic{equation}}
\renewcommand{\thetable}{S\arabic{table}}

\begin{center}
\Huge{Supplemental Material}
\end{center}
\tableofcontents

This supplementary material is organized as follows. In Sec. \ref{Sec:symmetry} we present a detailed derivation of Eq. 1 in the main text and discuss the realization of symmetries. In particular, in \ref{Sec:symmetry:gen}, general properties of continuum moir\'e models are discussed, in \ref{Sec:symmetry:lat} we discuss the connections to the underlying lattice and in \ref{Sec:symmetry:deriv} we present a derivation of continuum hamiltonian starting from a microscopic model. The effects of subleading terms, suppressed at low $\theta$, are additionally  discussed in \ref{Sec:symmetry:double}. 

In Sec. \ref{sec:calc}, we present details of the calculations for Eqs. 8-12 in the main text for the low twist angle limit.

\section{A minimal picture of moir\'e tunneling}
\label{Sec:symmetry}

In the following subsections, we present a derivation of the moir\'e tunneling from the perspective of effective field theory and symmetry. The purpose of this presentation is to provide the reader with a top-down (starting from long-wavelength field theories) understanding of moir\'e, as opposed to bottom-up (starting from microscopics). The value in this minimal picture is that it highlights the essential ingredients leading to the emergence of periodicity in a system where microscopically no translation symmetry may exist, e.g. at an incommensurate twist angle. In other words, we wish to answer the questions: What ingredients in the low energy description must be present in order for moir\'e periodicity to emerge at arbitrary non-commensurate $\theta$? What does this low-energy description reveal to us about the sublattice structure governed by $\mathcal{L}$-symmetry? And at small $\theta$, how do subleading processes affect the microscopic translation symmetry at commensurate $\theta$?

\subsection{Boosts and the action of symmetries in different representations}
\label{Sec:symmetry:gen}

To start, consider a field theory with Hamiltonian $h(\hat{\bf p})=\hat{\bf p}^2$, which we argue to be the accurate low-energy description of a hypothetical system (a ``layer"). For simplicity, we will not consider spin nor valley degrees of freedom, such that $h$ labels all relevant low-energy modes. When we say that our layer has a symmetry, we mean that there is an operation $\hat{O}$ that leaves $h$ invariant, i.e $\hat{O}^{-1}h\hat{O}=h$ or equivalently $[\hat{O},h]=0$. One such operation is time-reversal $\mathcal{T}$, which by construction is an operation which reverses the motion of the system. This motion is described by $\hat{\bf p}=-i\boldsymbol{\nabla}$, which is the generator of translation. Since $\mathcal{T}$ leaves position invariant, this tells us that in order for $\mathcal{T}^{-1}\hat{\bf p}\mathcal{T}=-\hat{\bf p}$ one has to have $\mathcal{T}=K$ where $K$ is complex conjugation.


Consider now the unitary transformation $|\psi\rangle=\exp(i{\bf q}\cdot{\bf x})|\tilde{\psi}\rangle$, with constant vector ${\bf q}$, which has the effect of bringing us into a representation of our system whose wavefunction evolves via $i\partial_t|\tilde{\psi}\rangle=h(\hat{\bf p}+{\bf q})|\tilde{\psi}\rangle$. Such a Hamiltonian appears to describe a system whose center of momentum has been shifted to $-{\bf q}$. We may therefore feel obliged to relabel the total momentum as the operator $\hat{\boldsymbol{\Pi}} = \hat{\bf p}+{\bf q}$, and likewise expect $\mathcal{T}^{-1}\hat{\boldsymbol{\Pi}}\mathcal{T}=-\hat{\boldsymbol{\Pi}}$. However, if we take $\mathcal{T}=K$ as before, this does not work, because ${\bf q}$ is a real-valued vector which does not change sign under complex conjugation. This has the important consequence that $K^{-1}h(\hat{\bf p}+{\bf q})K= h(-\hat{\bf p}+{\bf q})\neq h(\hat{\bf p}+{\bf q})$, which appears to imply that $\mathcal{T}$ is not a symmetry. And yet this cannot be because we established that $\mathcal{T}$ was a physical symmetry from the outset, and the mathematical act of applying a unitary transformation shouldn't change this fact.

This apparent contradiction is resolved if we recognize that $\mathcal{T}\neq K$ in this new representation. Since the operation $\exp(i{\bf q}\cdot{\bf x})$ boosted our system by $-{\bf q}$, the proper action of time-reversal is to first boost into a representation in which it acts simply, apply $K$, then boost back \footnote{{We stress that the ``boost" operation described above, which we wrote $e^{i{\bf q}\cdot\hat{\bf x}}$, is formally the instantaneous ($t=0$) Galileian boost when referring to a system of like-mass particles. (See {\it Ballentine} \cite{Ballentine} for a complete treatise on the Galilei group.) It is not a symmetry of the Hamiltonian \cite{Soper_notes,Ballentine}, but an operation which connects physically identical representations of motion. It should not be confused with the notion of Galilean symmetry, which is an invariance of the Schr\"{o}dinger equation valid for free particles \cite{Ballentine}.}}. One can check that $\mathcal{T}=\exp(-i{\bf q}\cdot{\bf x})K\exp(i{\bf q}\cdot{\bf x})=K\exp(2i{\bf q}\cdot{\bf x})$ does just the trick. A gauge transformation, such as the phase twist described, has the effect of bringing us into a physically equivalent but (possibly) mathematical distinct description of our system. Regardless of our choice of gauge, if we demand that $\mathcal{T}$ be a symmetry of $h$, then there exists a representation of $\mathcal{T}$ such that $h=\mathcal{T}^{-1}h\mathcal{T}$ regardless of the shape of $h$.

Consider now the following (bilayer) Hamiltonian,  
\begin{eqnarray}\label{Eqn:Intro:H=h+T}
H=\begin{pmatrix}
h(\hat{\bf p}) & T({\bf x}) \\
T({\bf x})^*   & h(\hat{\bf p}+{\bf q}_0)
\end{pmatrix}
\end{eqnarray}
which describes a local tunneling between two identical layers. For the model in the main text, ${\bf q}_0 \equiv {\bf q}_M$. Note that as written, the tunneling appears to bring a particle from one layer into a boosted frame of an otherwise identical layer. Let us demand that $\mathcal{T}$ be a symmetry of $H$. In order for $H$ to be invariant under $\mathcal{T}$, 
\begin{eqnarray}\label{Eqn:Intro:timeRev[T]=T}
T({\bf x})^*e^{i(2{\bf q}_0)\cdot{\bf x}} = T({\bf x}) , 
\end{eqnarray}
where the phase comes from the non-trivial action of $\mathcal{T}$ in the boosted layer. Notice that Eqn \ref{Eqn:Intro:timeRev[T]=T} tells us the tunneling cannot be constant and preserve $\mathcal{T}$. The solution to Eqn \ref{Eqn:Intro:timeRev[T]=T} is a tunneling which is periodic under spatial shift reciprocal to ${\bf g}_1=2{\bf q}_0$. Which is to say that time-reversal symmetric tunneling between different boosted frames generates a periodic lattice in at least one direction. If additional symmetries are present (e.g $C_N$ rotations), then a lattice is generated which respects those symmetries (such as square for $C_4$). The action of any point-group operation $g$ on $H$ follows similarly to $\mathcal{T}$, i.e $\mathcal{U}g\mathcal{U}^*$ with $\mathcal{U}$ defined as (the difference with main text Eq. (5) is due to momentum origin shift in Eq. (1) compared to Eq. S1 above )
\begin{eqnarray}\label{Eqn:U}
\mathcal{U} = \begin{pmatrix}
1 & 0 \\
0 & e^{-i{\bf q}_0\cdot{\bf x}}
\end{pmatrix} . 
\end{eqnarray}

As demonstrated in the main text, there exists a unitary transformation which moves us into a representation without the relative boost between the diagonal terms of Eqn. \ref{Eqn:Intro:H=h+T}, but where the information of the relative boost has been pushed onto the off diagonal, 
\begin{eqnarray}
H_f=\mathcal{U}^*H\mathcal{U}=\begin{pmatrix}
h(\hat{\bf p}) & T({\bf x})e^{-i{\bf q}_0\cdot{\bf x}} \\
T({\bf x})^*e^{i{\bf q}_0\cdot{\bf x}} & h(\hat{\bf p}) 
\end{pmatrix} . 
\end{eqnarray}
This is equivalent to a folding of the original zone, which changes the translational symmetry of the eigenstates, such that they become Bloch states of an enlarged unit cell. Note that the translational symmetry is still present in this representation, but realized nontrivially (see main text).

\subsection{Relation of the continuum model to the underlying lattice model}
\label{Sec:symmetry:lat}

Here we provide definitions relating continuum operators of Sec. \ref{Sec:symmetry:gen} to operators defined on microscopic lattices of two layers. We will then show how continuum model is derived in Sec. \ref{Sec:symmetry:deriv}.

Let us start by defining a set of fixed ``lab" coordinates, ${\bf x}\in\{n\,{\bf a}_1+m\,{\bf a}_2|(n,m)\in\mathbb{Z}^2\}$, which describes a set of discrete periodic sites. Now consider a single crystal plane with lattice coordinates ${\bf r}_{\theta}\in\{n\,R_{\theta}{\bf a}_1+m\,R_{\theta}{\bf a}_2|(n,m)\in\mathbb{Z}^2\}$, or equivalently ${\bf r}_{\theta}=R_{\theta}{\bf x}$, which implicitly defines the reciprocal lattice ${\bf G}_{\theta}\in\{n\,R_{\theta}{\bf G}_1+m\,R_{\theta}{\bf G}_2|(n,m)\in\mathbb{Z}^2\}$ via the relationship $e^{i{\bf G}_{\theta}\cdot{\bf r}_{\theta}}=1$. We label the annihilation/creation operators for the tightly bound orbitals of this lattice $d({\bf r}_{\theta})$/$d^{\dagger}({\bf r}_{\theta})$. For our purposes, it is sufficient to consider one orbital per atomic cell, which corresponds to one band. Expanding $\hat{d}({\bf r}_{\theta})$ into crystal momenta 
\begin{eqnarray}
\hat{d}({\bf r}_{\theta}) = \sum_{\bf k}e^{i({\bf M}_{\theta}+{\bf k})\cdot{\bf r}_{\theta}} \hat{d}_{{\bf M}_{\theta}+{\bf k}} 
\equiv e^{i{\bf M}_{\theta}\cdot{\bf r}_{\theta}}\hat\psi({\bf r}_{\theta}) ,
\label{sup:eq:psi}
\end{eqnarray}
where ${\bf M}_{\theta}\equiv R_{\theta}{\bf M}$ is the crystal momenta corresponding to the rotated BZ corner ($M$-point). In doing this expansion, we implicitly chose a representation of the wavefunction $e^{i{\bf M}_{\theta}\cdot{\bf r}_{\theta}}=e^{i{\bf M}\cdot{\bf x}}$ which is an eigenstate of periodic shifts by ${\bf r}_{\theta}\rightarrow{\bf r}_{\theta}+R_{\theta}{\bf a}_{j}$. This is the natural representation of the crystal, in the sense that the periodicity of the wavefunction aligns with that of the physical crystal axis. We are not constrained to this representation, and can choose another, e.g. 
\begin{eqnarray}
\hat{d}({\bf r}_{\theta}) = \sum_{\bf k}e^{i({\bf M}+{\bf k})\cdot{\bf r}_{\theta}} \Big(e^{i({\bf M}_{\theta}-{\bf M})\cdot{\bf r}_{\theta}}\hat{d}_{{\bf M}_{\theta}+{\bf k}}\Big) 
\equiv e^{i{\bf M}\cdot{\bf r}_{\theta}}\hat\Psi({\bf x}) .  
\end{eqnarray}
If we define $\hat{d}_{{\bf M}+{\bf k}}({\bf x})\equiv e^{i({\bf M}_{\theta}-{\bf M})\cdot{\bf r}_{\theta}}\hat{d}_{{\bf M}_{\theta}+{\bf k}}$, we would see this is equivalent to expanding $\hat{d}({\bf r}_{\theta})$ into a BZ aligned with the the lab axis (${\bf x}$), and thus disaligned with the physical crystal axis (${\bf r}_{\theta}$). In this representation, the wavefunctions appear to be eigenstates of shifts by ${\bf r}_{\theta}\rightarrow{\bf r}_{\theta}+{\bf a}_j$; however, because ${\bf M}_{\theta}$ isn't the corner of this chosen zone, the annihilation operator necessarily picks up a spatial dependence in order to guarantee the proper transformation of physical symmetries. (See Ref \cite{theory_Vafek&Kang2018_TBGwannier} for an example of non-trivial point group operations in twisted bilayer graphene.)

In order to demonstrate this effect, it will be useful to define an operator (like momentum), and study how it transforms under symmetry in either representation. For this purpose, we construct a Bloch-periodic momentum operator, 
\begin{eqnarray}\label{Eqn:RotationBoosts:p=d*(-i*nabla)d}
\hat{\bf p}_{\text{B}} 
&=& \sum\limits_{{\bf G}_{\theta}}\int\limits_{{\bf r}_{\theta}} \hat{d}^{\dagger}({\bf r}_{\theta})\Big(-i\boldsymbol{\nabla}_{{\bf r}_\theta}+{\bf G}_{\theta}\Big)\hat{d}({\bf r}_{\theta}) \notag\\
&=& \sum\limits_{{\bf G}_{\theta}}\int\limits_{{\bf r}_{\theta}} \hat{\psi}^{\dagger}({\bf r}_{\theta})\Big(-i\boldsymbol{\nabla}_{{\bf r}_\theta}+{\bf M}_{\theta}+{\bf G}_{\theta}\Big)\hat{\psi}({\bf r}_{\theta}) , 
\end{eqnarray}
which is just the momentum operator projected onto Bloch states. (We employ the notation $\int_{\bf x}\equiv\int{d^2{\bf x}}$.) Notice that $\hat{\bf p}_{\text{B}}$ is defined modulo a reciprocal lattice vector aligned with the crystal axis (${\bf G}_{\theta}$), such that it is invariant under inserting the identity $e^{i{\bf G}_{\theta}\cdot{\bf r}_{\theta}}=1$. The action of $\mathcal{T}$ in this representation is the complex conjugate, which brings $\hat{\bf p}_{\text{B}}$ into 
\begin{eqnarray}
\mathcal{T}\big[\hat{\bf p}_{\text{B}}\big]
=\sum\limits_{{\bf G}_{\theta}}\int\limits_{{\bf r}_{\theta}} \hat{\psi}^{\dagger}({\bf r}_{\theta})\Big(+i\boldsymbol{\nabla}_{{\bf r}_\theta}+{\bf M}_{\theta}+{\bf G}_{\theta}\Big)\hat{\psi}({\bf r}_{\theta}) , 
\end{eqnarray}
but because there exists a ${\bf G}_{\theta}$ such that ${\bf M}_{\theta}={\bf G}_{\theta}-{\bf M}_{\theta}$, we can write  
\begin{eqnarray}
=\sum\limits_{{\bf G}_{\theta}}\int\limits_{{\bf r}_{\theta}} \hat{\psi}^{\dagger}({\bf r}_{\theta})\Big(+i\boldsymbol{\nabla}_{{\bf r}_\theta}-{\bf M}_{\theta}-{\bf G}_{\theta}\Big)\hat{\psi}({\bf r}_{\theta}) 
= -\hat{\bf p}_{\text{B}} . 
\end{eqnarray}
This is the expected action of momentum under $\mathcal{T}$. In the other ($\Psi$) representation, this is not the case. Noting the identity ${\bf x}\cdot\boldsymbol{\nabla}_{\bf x}={\bf r}_{\theta}\cdot\boldsymbol{\nabla}_{{\bf r}_{\theta}}$, we can write Eqn \ref{Eqn:RotationBoosts:p=d*(-i*nabla)d} as 
\begin{eqnarray}\label{Eqn:RotationBoosts:p=Psi*(-i*nabla+M)Psi}
\hat{\bf p}_{\text{B}}
=\sum\limits_{{\bf G}_{\theta}}\int\limits_{\bf x} \hat{\Psi}^{\dagger}({\bf x})\Big(-iR_{\theta}\boldsymbol{\nabla}_{\bf x}+{\bf M}+{\bf G}_{\theta}\Big)\hat{\Psi}({\bf x}) , 
\end{eqnarray}
which is centered on ${\bf M}$ and not ${\bf M}_{\theta}$. Since ${\bf M}$ is not connected to $-{\bf M}$ by some ${\bf G}_{\theta}$, we cannot employ that strategy in guaranteeing the physical condition $\mathcal{T}\big[\hat{\bf p}_{\text{B}}\big]=-\hat{\bf p}_{\text{B}}$. The solution is that, in addition to complex conjugation, $\hat\Psi({\bf x})$ picks up a phase under $\mathcal{T}$:
\begin{eqnarray}
\mathcal{T}\big[\hat\psi({\bf r}_{\theta})\big] &=& K\hat\psi({\bf r}_{\theta}) \\
\mathcal{T}\big[\hat\Psi({\bf x})\big] &=& Ke^{i2{\bf M}\cdot{\bf r}_{\theta}}\hat\Psi({\bf x}) = Ke^{i2({\bf M}-{\bf M}_{\theta})\cdot{\bf r}_{\theta}}\hat\Psi({\bf x}) . 
\label{eq:T}
\end{eqnarray}
The quantity $({\bf M}-{\bf M}_{\theta})\cdot{\bf r}_{\theta}={\bf M}\cdot(R_{\theta}{\bf x}-{\bf x})$ carries information about the displacement ${\bf u}_{\theta}\equiv R_{\theta}{\bf x}-{\bf x}$ of our rigidly-rotated layer from the lab frame \cite{balents_scipost}.

Because rotating a crystal rotates the crystal BZ, this change of representation into a rotated frame of reference is equivalent to a boost for modes labelled by ${\bf k}\neq 0$, such as for ${\bf k}={\bf M}$. Just as with $\mathcal{T}$, the choice of boosted frame has the effect of transforming pure point-group operations into space-group operations via the introduction of the position-dependent phase factors. The effect of the boost is trivial in the case of a single layer, as we are always free to choose the natural representation of the crystal. But in the case of twisted bilayers, the relative rotation between the crystals is physical, the effect of which is preserved regardless of our choice of gauge.

To understand this explicitly, consider the case of twisted bilayers with relative displacement ${\bf u}={\bf u}_{\theta/2}-{\bf u}_{-\theta/2}=R_{\theta/2}{\bf x}-R_{-\theta/2}{\bf x}$. We then write the Hamiltonian 
\begin{eqnarray}
\hat{H} = \int_{\bf x} \hat\Psi^{\dagger}({\bf x})
\begin{pmatrix}\label{Eqn:RotationBoosts:H}
h(-iR_{-\theta/2}\boldsymbol\nabla_{\bf x}+{\bf M}) & T^*({\bf x}) \\
T({\bf x}) & h(-iR_{\theta/2}\boldsymbol\nabla_{\bf x}+{\bf M})
\end{pmatrix}\hat\Psi({\bf x}) , 
\end{eqnarray}
where $\hat\Psi=(\hat\Psi_1,\hat\Psi_2)^T$ defines our layer-spinor space. Notice that by applying the gauge $\hat\Psi_l\rightarrow e^{-i{\bf M}\cdot{\bf x}}\hat\Psi_l$, we can recenter the layer's momenta about ${\bf M}=(\pi/a,\pi/a)$, which is the corner of the BZ aligned with our common ``unrotated" lab coordinates, producing 
\begin{eqnarray}
\hat{H} = \int_{\bf x} \hat\Psi^{\dagger}({\bf x})
\begin{pmatrix}\label{Eqn:RotationBoosts:H2}
h(-iR_{-\theta/2}\boldsymbol\nabla_{\bf x}+{\bf M}-{\bf M}_{-\theta/2}) & T^*({\bf x}) \\
T({\bf x}) & h(-iR_{\theta/2}\boldsymbol\nabla_{\bf x}+{\bf M}-{\bf M}_{\theta/2})
\end{pmatrix}\hat\Psi({\bf x}) 
= \int_{\bf x} \hat\Psi^{\dagger}\tilde{H}\hat\Psi
. 
\end{eqnarray}

For the low-energy description, we further constrain the superposition $\hat\Psi_{l}$ to be over those modes in the vicinity of ${\bf M}_{(2l-3)\theta/2}$ (where $l=1,2$ are the layer indices), which for our purposes are the relevant modes (i.e near the Fermi level). As is shown below in Sec. \ref{Sec:symmetry:deriv}, at low twist angles and long-wavelengths the relevant inter-layer tunnelings guarantee $T({\bf x})$ to be a local function of the common coordinate ${\bf x}$.

Moving into the $\hat\psi({\bf x})=\big(\hat\psi_1(R_{-\theta/2}{\bf x}),\hat\psi_2(R_{\theta/2}{\bf x})\big)^T$ representation (Eqn. \ref{sup:eq:psi}) removes the boost
\begin{eqnarray}
\hat{H} = \int_{\bf x} \hat\psi^{\dagger}({\bf x})
\begin{pmatrix}\label{Eqn:RotationBoosts:H2}
h(-iR_{-\theta/2}\boldsymbol\nabla_{\bf x}) & U({\bf x}) \\
U({\bf x}) & h(-iR_{\theta/2}\boldsymbol\nabla_{\bf x})
\end{pmatrix}\hat\psi({\bf x}) 
=\int_{\bf x} \hat\psi^{\dagger}H_f\hat\psi
, 
\end{eqnarray}
but pushes $e^{i({\bf M}_{\theta/2}-{\bf M}_{-\theta/2})\cdot{\bf x}}$ onto the tunneling $U({\bf x})=e^{i({\bf M}_{\theta/2}-{\bf M}_{-\theta/2})\cdot{\bf x}}T^*({\bf x})$. The wavevector ${\bf q}_M=-{\bf M}_{\theta/2}+{\bf M}_{-\theta/2}$ is {\it not} a moir\'e reciprocal lattice vector, and has the effect of doubling the moir\'e cell. Thus we can recognize $\hat\Psi$ and $\hat\psi$ as corresponding to the ``unfolded" and ``folded" representations of the moir\'e translation symmetry. The folding transformation is the unitary 
\begin{eqnarray}
\mathcal{U}_{\theta}=\begin{pmatrix}
e^{-i({\bf M}_{\theta/2}-{\bf M})\cdot{\bf x}} & 0 \\
0 & e^{-i({\bf M}_{-\theta/2}-{\bf M})\cdot{\bf x}}
\end{pmatrix} , 
\end{eqnarray}
which acts like $\mathcal{U}_{\theta}\psi=\Psi$. Because the space and translation groups are decoupled in folded representation, it can be convenient to treat $H_f$ as the ``seed" Hamiltonian, from which its non-trivial representations can be generated. (Noting: $\tilde{H}=\mathcal{U}^*H_f\mathcal{U}$ and $T^*({\bf x})=e^{-i({\bf M}_{\theta/2}-{\bf M}_{-\theta/2})\cdot{\bf x}}U({\bf x})$.) Starting from $H_f$ and successively applying $\mathcal{U}_{\theta}$ or $\mathcal{U}_{\theta}^*$, we can generate an infinite number of physically equivalent representations. The conjugate representation $H = \mathcal{U}H_f\mathcal{U}^* = H^*$ can be generated by the reverse application of $\mathcal{U}_{\theta}$, and can be understood as a freedom in the direction of unfolding. Written explicitly, 
\begin{eqnarray}
\hat{H} = \int_{\bf x} \Big(\mathcal{U}_{\theta}^*\hat\psi\Big)^{\dagger}
\begin{pmatrix}\label{Eqn:RotationBoosts:H2}
h(-iR_{-\theta/2}\boldsymbol\nabla_{\bf x}-{\bf M}+{\bf M}_{-\theta/2}) & T({\bf x}) \\
T^*({\bf x}) & h(-iR_{\theta/2}\boldsymbol\nabla_{\bf x}-{\bf M}+{\bf M}_{\theta/2})
\end{pmatrix}\mathcal{U}_{\theta}^*\hat\psi 
=\int_{\bf x} \Big(\mathcal{U}_{\theta}^*\hat\psi\Big)^{\dagger}H\mathcal{U}_{\theta}^*\hat\psi 
. 
\end{eqnarray}
This latter $H$ representation is the one used in the main text [Eqn. (1,2) of the main text]. The choice is arbitrary, differing only in the internal form of $\hat{H}$ and in the representation of the symmetry operations required to leave the Hamiltonian invariant.

\subsection{Deriving the local field theories from microscopics}

\label{Sec:symmetry:deriv}

Let us begin by writing the microscopic tight-binding Hamiltonian for the inter-layer tunneling (see Ref \cite{theory_Vafek&Kang2023_T(u)} for a more general formulation), 
\begin{eqnarray}
H_{\text{micro}} = \sum_{{\bf x}{\bf x}'}\hat{d}_1^{\dagger}(R_{-\theta/2}{\bf x})t(R_{-\theta/2}{\bf x}-R_{\theta/2}{\bf x}')\hat{d}_2(R_{\theta/2}{\bf x}') + \text{h.c} ,  
\end{eqnarray}
which we have written in the coordinate basis ${\bf x}$ where both crystal layers appear rotated. The Fourier transform follows 
\begin{eqnarray}
H_{\text{micro}} = \sum_{{\bf x}{\bf x}'}\Big(\sum_{\bf k}e^{-i{\bf k}\cdot(R_{-\theta/2}{\bf x})}\hat{d}_1^{\dagger}({\bf k})\Big)\Big(\int_{\bf q}\tilde{t}_{\bf q}e^{i{\bf q}\cdot(R_{-\theta/2}{\bf x}-R_{\theta/2}{\bf x}')}\Big)\Big(\sum_{\bf k'}e^{i{\bf k}'\cdot(R_{\theta/2}{\bf x}')}\hat{d}_2({\bf k'})\Big) +\text{h.c},    
\end{eqnarray}
where we employ the shorthand $\int_{\bf q}\equiv\int_{{\bf q}\in\mathbb{R}^2}\frac{d^2{\bf q}}{(2\pi)^2}$. Since the Brillouin zones of the layers are rotated, the operators in momentum space satisfy $\hat{d}_l({\bf k}+{\bf G}_{(2l-3)\frac{\theta}{2}})=\hat{d}_l({\bf k})$, $l=1,2$. Continuing we have 
\begin{eqnarray}
H_{\text{micro}} = \int_{\bf q}\sum_{{\bf k}{\bf k}'}\tilde{t}_{\bf q}\hat{d}_1^{\dagger}({\bf k})\hat{d}_2({\bf k}')
    \Big(\sum_{\bf x}e^{i{\bf x}\cdot(R_{\theta/2}{\bf q}-R_{\theta/2}{\bf k})}\Big)
    \Big(\sum_{\bf x'}e^{-i{\bf x}'\cdot(R_{-\theta/2}{\bf q}-R_{-\theta/2}{\bf k}')}\Big) 
    +\text{h.c}
\end{eqnarray}
the last two terms in parenthesis are delta functions mod ${\bf G}$, 
\begin{eqnarray}
H_{\text{micro}} = \int_{\bf q}\sum_{{\bf k}{\bf k}'}\tilde{t}_{\bf q}\hat{d}_1^{\dagger}({\bf k})\hat{d}_2({\bf k}')
    \sum_{{\bf G}{\bf G}'}\delta^2(R_{\theta/2}{\bf q}-R_{\theta/2}{\bf k}+{\bf G})\delta^2(R_{-\theta/2}{\bf q}-R_{-\theta/2}{\bf k}'+{\bf G}') 
    +\text{h.c} ,  
\end{eqnarray}
where ${\bf G}$ labels the unrotated reciprocal lattice, i.e $e^{i{\bf G}\cdot{\bf x}}=1$. For brevity, we have absorbed (and will continue to absorb) any (additional) normalization factors into the definition of $\tilde{t}_{\bf q}$. The product of delta functions sets the constraint 
\begin{eqnarray}
{\bf q} = {\bf k}-{\bf G}_{\frac{-\theta}{2}} = {\bf k}'-{\bf G}_{\frac{\theta}{2}}' . 
\end{eqnarray}
Evaluating the delta functions allows us to eliminate ${\bf q}$ and ${\bf k}'$,  
\begin{eqnarray}
H_{\text{micro}} = \sum_{{\bf k}}\sum_{{\bf G}}\tilde{t}_{{\bf k}-{\bf G}_{\frac{-\theta}{2}}}\hat{d}_1^{\dagger}({\bf k})\sum_{{\bf G}'}'\hat{d}_2({\bf k}+{\bf G}_{\frac{\theta}{2}}'-{\bf G}_{-\frac{\theta}{2}}) 
+\text{h.c} ,
\end{eqnarray}
which says that the moir\'e tunneling process connects any pair of momenta $({\bf k},{\bf k}+{\bf G}_{\frac{\theta}{2}}'-{\bf G}_{-\frac{\theta}{2}})$ from the BZ's of our rotated crystals. $\sum_{{\bf G}'}'$ imposes the constraint that the momenta ${\bf k}$ and ${\bf k'}={\bf k}+{\bf G}_{\frac{\theta}{2}}'-{\bf G}_{-\frac{\theta}{2}}$ are restricted to be within first Brillouin zone only. One can choose a common microscopic Brillouin zone (since the volume of the Brillouin zone does not change with twist) for both layers centered around the microscopic $M$ point, which allows to write the constraint as:
\begin{equation}
{\bf k}; {\bf k}+{\bf G}_{\frac{\theta}{2}}'-{\bf G}_{-\frac{\theta}{2}} \in BZ.
\label{sup:eq:bzconstraint}
\end{equation}
This in particular constrains the values of ${\bf G}$ and ${\bf G'}$; for $k\ll 2\pi/l_a$ one gets an order of magnitude estimate
\begin{equation}
    |{\bf G}_{\frac{\theta}{2}}'-{\bf G}_{-\frac{\theta}{2}}|\lesssim l_a^{-1}.
    \label{sup:eq:estconstr}
\end{equation}

We are only interested in the physics near the Fermi level, which here means the band extrema at ${\bf M}$. In other words, we only need to keep processes involving ${\bf k}$ (mod ${\bf G}$) such that the single-layer dispersion at ${\bf k}$ can be expanded in Taylor series with lowest term being a quadratic $h_l({\bf k})\simeq\mu ({\bf k}-R_{(2l-3)\theta/2}{\bf M})^2$, $l=1,2$. Therefore, the relevant moir\'e tunneling processes are those involving a hopping to ${\bf k}\simeq {\bf M}_{-\theta/2}$ in layer 1, which we can get by shifting ${\bf k}={\bf M}_{-\frac{\theta}{2}}+\delta{\bf k}$, 
\begin{eqnarray}
H_{\text{micro}} \simeq \sum_{\delta{\bf k}}\sum_{{\bf G}}\tilde{t}_{\delta{\bf k}+{\bf M}_{-\frac{\theta}{2}}-{\bf G}_{\frac{-\theta}{2}}}\hat{d}_1^{\dagger}(\delta{\bf k}+{\bf M}_{-\frac{\theta}{2}})\sum_{{\bf G}'}'\hat{d}_2(\delta{\bf k}+{\bf M}_{-\frac{\theta}{2}}+{\bf G}_{\frac{\theta}{2}}'-{\bf G}_{-\frac{\theta}{2}}) +\text{h.c} , \notag\\ 
\end{eqnarray}
and keeping only small $\delta{\bf k}$. For an infinitely large system, we can replace $\sum_{\bf k}\rightarrow\int_{\bf k}$. The function $\tilde{t}_{\delta{\bf k}+{\bf M}_{-\frac{\theta}{2}}-{\bf G}_{\frac{-\theta}{2}}}$ is analytic in $\delta{\bf k}$. We can therefore approximate $\tilde{t}_{\delta{\bf k}+{\bf M}_{-\frac{\theta}{2}}-{\bf G}_{\frac{-\theta}{2}}}\simeq \tilde{t}_{{\bf M}_{-\frac{\theta}{2}}-{\bf G}_{\frac{-\theta}{2}}}$ to leading order in powers of small $|\delta{\bf k}|\ll|{\bf G}_1|$, 
\begin{eqnarray}\label{Eqn:Suppl:H_micro_intermediateStep}
H_{\text{micro}} \simeq  \sum_{{\bf G}}\tilde{t}_{{\bf M}_{-\frac{\theta}{2}}-{\bf G}_{\frac{-\theta}{2}}}
\int_{\delta{\bf k}}
\hat{d}_1^{\dagger}(\delta{\bf k}+{\bf M}_{-\frac{\theta}{2}})\sum_{{\bf G}'}'\hat{d}_2(\delta{\bf k}+{\bf M}_{-\frac{\theta}{2}}+{\bf G}_{\frac{\theta}{2}}'-{\bf G}_{-\frac{\theta}{2}}) +\text{h.c} .  
\end{eqnarray}
The latter step results in our tunneling potential being local in real space. In the previous section we learned 
\begin{eqnarray}
\hat{d}_1(\delta{\bf k}+{\bf M}_{-\frac{\theta}{2}}) &=& \int_{\bf x}e^{i{\bf k}\cdot{\bf x}}\hat\psi_1(R_{-\theta/2}{\bf x}) \\
\hat{d}_2(\delta{\bf k}+{\bf M}_{-\frac{\theta}{2}}+{\bf G}_{\frac{\theta}{2}}'-{\bf G}_{-\frac{\theta}{2}}) &=& \int_{\bf x'}e^{i{\bf k}\cdot{\bf x}'}e^{i({\bf M}_{-\frac{\theta}{2}}-{\bf M}_{\frac{\theta}{2}})\cdot{\bf x}'}e^{i({\bf G}_{\frac{\theta}{2}}'-{\bf G}_{-\frac{\theta}{2}})\cdot{\bf x}'}\hat\psi_2(R_{\theta/2}{\bf x}') . 
\end{eqnarray}
Plugging these into Eqn \ref{Eqn:Suppl:H_micro_intermediateStep} reveals 
\begin{eqnarray}
H_{\text{micro}} \simeq \int_{\bf x}\hat{\psi}_1^{\dagger}(R_{-\theta/2}{\bf x})\Bigg(\sum_{{\bf G}} \sum_{{\bf G}'}'\tilde{t}_{{\bf M}_{-\frac{\theta}{2}}-{\bf G}_{\frac{-\theta}{2}}}e^{i({\bf M}_{-\frac{\theta}{2}}-{\bf M}_{\frac{\theta}{2}})\cdot{\bf x}}e^{i({\bf G}_{\frac{\theta}{2}}'-{\bf G}_{-\frac{\theta}{2}})\cdot{\bf x}}\Bigg)\hat{\psi}_2(R_{\theta/2}{\bf x}) 
\equiv H_{\text{continuum}} ,   
\end{eqnarray}
which is the continuum limit in the folded ($\psi$) representation. The unfolded ($\Psi$) representation is identical but without the factor $e^{i({\bf M}_{-\frac{\theta}{2}}-{\bf M}_{\frac{\theta}{2}})\cdot{\bf x}}$, 
\begin{eqnarray}
H_{\text{continuum}} = \int_{\bf x}\hat{\Psi}_1^{\dagger}({\bf x})\Bigg(\sum_{{\bf G}}\sum_{{\bf G}'}'\tilde{t}_{{\bf M}_{-\frac{\theta}{2}}-{\bf G}_{\frac{-\theta}{2}}}e^{i({\bf G}_{\frac{\theta}{2}}'-{\bf G}_{-\frac{\theta}{2}})\cdot{\bf x}}\Bigg)\hat{\Psi}_2({\bf x}) + \text{h.c} .   
\end{eqnarray}
The term in parenthesis is a sum over local superlattice potentials, i.e 
\begin{eqnarray}
\sum_{{\bf G}{\bf G}'}\tilde{t}_{{\bf M}_{-\frac{\theta}{2}}-{\bf G}_{\frac{-\theta}{2}}}e^{i({\bf G}_{\frac{\theta}{2}}'-{\bf G}_{-\frac{\theta}{2}})\cdot{\bf x}} 
= T({\bf x}) + \sum_{\bf G}\tilde{T}_{\bf G}({\bf x}) , 
\end{eqnarray}
where 
\begin{eqnarray}
T({\bf x}) = \sum_{{\bf G}}\tilde{t}_{{\bf M}_{-\frac{\theta}{2}}-{\bf G}_{\frac{-\theta}{2}}}e^{i({\bf G}_{\frac{\theta}{2}}-{\bf G}_{-\frac{\theta}{2}})\cdot{\bf x}}
\end{eqnarray}
\begin{eqnarray}
\tilde{T}_{\bf G}({\bf x}) = \sum_{{\bf G}'\neq{\bf G}}'\tilde{t}_{{\bf M}_{-\frac{\theta}{2}}-{\bf G}_{\frac{-\theta}{2}}}e^{i({\bf G}_{\frac{\theta}{2}}'-{\bf G}_{-\frac{\theta}{2}})\cdot{\bf x}}  
\label{sup:eq:tg}
\end{eqnarray}
are respectively the intra (${\bf G}={\bf G}'$) and inter-zone (${\bf G}\neq{\bf G}'$) tunneling potentials. 

We note that the restriction on momenta ${\bf G}'$ leads to $\tilde{T}_{\bf G}({\bf x})\ll T({\bf x})$ at low twist angles. Indeed, let us decompose ${\bf G}' = {\bf G}+ {\bf G}''$ in \eqref{sup:eq:estconstr}. For small $\theta$, $|{\bf G}_{\frac{\theta}{2}}-{\bf G}_{-\frac{\theta}{2}}|\sim |{\bf G}| \theta$. However $G_{\frac{\theta}{2}}'' \gtrsim 1/l_a$, and therefore $G \sim 1/(l_a \theta)$ is required to satisfy \eqref{sup:eq:estconstr} and \eqref{sup:eq:bzconstraint}. However, since the $\tilde{t}_{{\bf M}_{-\frac{\theta}{2}}-{\bf G}_{\frac{-\theta}{2}}}$ is expected to decay exponentially for $G\gtrsim l_a^{-1}$ \cite{bistritzer2011moire}, all allowed terms in $\tilde{T}_{\bf G}({\bf x})$ are suppressed exponentially with respect to those in $T({\bf x})$. 

Note that the primitive moir\'e wavevectors are ${\bf g}_k=R_{\frac{\theta}{2}}{\bf G}_{k}-R_{-\frac{\theta}{2}}{\bf G}_{k}$ for $k=1,2$. Therefore the vector ${\bf g}={\bf G}_{\frac{\theta}{2}}-{\bf G}_{-\frac{\theta}{2}}$, where ${\bf g}\in\{n{\bf g}_1+m{\bf g}_2|(n,m)\in\mathbb{Z}^2\}$, labels the emergent moir\'e reciprocal lattice. Thus we can recognize $T({\bf x})$ as being the moir\'e-periodic tunneling derived from symmetry in the main text. For $\tilde{T}_{\bf G}({\bf x})$ coming from the ${\bf G}\neq{\bf G}'$ terms, we can rewrite ${\bf G}_{\frac{\theta}{2}}'-{\bf G}_{-\frac{\theta}{2}}={\bf G}_{\frac{\theta}{2}}-{\bf G}_{-\frac{\theta}{2}}+{\bf G}_{\frac{\theta}{2}}'-{\bf G}_{\frac{\theta}{2}} = {\bf g}+{\bf G}_{\frac{\theta}{2}}'-{\bf G}_{\frac{\theta}{2}}$. Thus 
\begin{eqnarray}
e^{i({\bf G}_{\frac{\theta}{2}}'-{\bf G}_{-\frac{\theta}{2}})\cdot{\bf x}} 
= e^{i{\bf g}\cdot{\bf x}}\times e^{i({\bf G}_{\frac{\theta}{2}}'-{\bf G}_{\frac{\theta}{2}})\cdot{\bf x}}
\end{eqnarray}
is a product over a moir\'e-periodic wave $e^{i{\bf g}\cdot{\bf x}}$ and a wave $e^{i({\bf G}_{\frac{\theta}{2}}'-{\bf G}_{\frac{\theta}{2}})\cdot{\bf x}}$ which varies rapidly at the {\it atomic} scale. Because these two wavevectors aren't guaranteed to be commensurate for arbitrary (or any) $\theta$, the corresponding periodicity of the $\tilde{T}_{\bf G}({\bf x})$ terms {\it isn't} guaranteed to be the moir\'e one. (More on this in the following subsection.)

\subsection{Moir\'e unit cell doubling from subleading interlayer tunnelings}

\label{Sec:symmetry:double}

We now demonstrate that the tunneling terms \eqref{sup:eq:tg}, suppressed at low $\theta$ as discussed above lead to the doubling of the moir\'e unit cell. In the main text, we neglected these terms completely, which implies that their weak effects at low twist angles is not the origin of decoupled sublattices.

We first demonstrate that \eqref{sup:eq:tg} is not consistent with the conventional moir\'e periodicity. If there exists an angle $\theta=\theta_c$ such that the atomic wavevector $R_{\theta/2}{\bf G}_1$ is a commensurate fraction of moir\'e wavevectors, i.e 
\begin{eqnarray}\label{Eqn:commensurate_angle_unfolded}
R_{\theta/2}{\bf G}_1 = n{\bf g}_1 + m{\bf g}_2
\end{eqnarray}
for integers $n,m$; then that angle is a commensurate angle, at which translation is a good microscopic symmetry. Starting from the definition of ${\bf G}_1=2\pi l_{a}^{-1}(1,0)$ \& ${\bf G}_2=2\pi l_{a}^{-1}(0,1)$, Eqn \ref{Eqn:commensurate_angle_unfolded} tells us the following must be true:
\begin{eqnarray}
\begin{pmatrix}
\cos(\theta/2) \\ \sin(\theta/2)
\end{pmatrix}
=2\sin(\theta/2)\begin{pmatrix}
m \\ n 
\end{pmatrix} . 
\end{eqnarray}
However clearly, the bottom components implies the relationship $1=2n$, which cannot be satisfied for integer $n$. Therefore, no $\theta$ exists at which the emergent moir\'e translations are a microscopic symmetry.

This does not mean there is no commensurate $\theta_c$, only that at $\theta_c$ the microscopic translation symmetry isn't the moir\'e one. Instead it is the {\it double} moir\'e cell which is microscopically commensurate at $\theta_c$, which is to say that the {\it folded} moir\'e zone is commensurate with the atomic one, i.e 
\begin{eqnarray}\label{Eqn:commensurate_angle_folded}
R_{\theta/2}{\bf G}_1 = n{\bf g}_{1,f} + m{\bf g}_{2,f} , 
\end{eqnarray}
where the folded reciprocal vectors at ${\bf g}_{1,f}={\bf q}_M$ and ${\bf g}_{2,f}=C_4{\bf q}_M$. Eqn \ref{Eqn:commensurate_angle_folded} says 
\begin{eqnarray}
\begin{pmatrix}
\cos(\theta/2) \\ \sin(\theta/2)
\end{pmatrix}
=\sin(\theta/2)\begin{pmatrix}
n+m \\ n-m 
\end{pmatrix} ,  
\end{eqnarray}
which demands $n=m+1$ and 
\begin{eqnarray}
\frac{\cos(\theta/2)}{\sin(\theta/2)} = 2m+1 . 
\end{eqnarray}
The commensurate angles are those which satisfy $\theta_c=2\arccos(2m+1)$ for integer $m$. We numerically checked this formula for $\theta_c$ against the one in Ref \cite{Toshi}, and found them to be identical in the domain $m>0$.

\section{Tight-binding in the asymptotic limit of small $\theta$}
\label{sec:calc}

Here we present the details of analysis of Eqn. (7) of the main text leading to Eqns. (8-12). In particular, Sec. \ref{Sec:HarmonicOscillators} performs a harmonic-oscillator limit analysis appropriate to calculate Coulomb interactions projected to flat bands. To calculate the band dispersions, this is not sufficient; in Sec. \ref{Sec:Mathieu} we present the approach via exact solutions and in Sec. \ref{sec:wannier} and \ref{Sec:WKB} we show the details of asymptotic analytic calculations using WKB method.

\subsection{Notation}
The following notations are used in the proceeding subsections: The superlattice period is denoted $l_{\mathcal{S}}=l_{a}/\theta$. We write the ground state wavefunction of the harmonic oscillator $\psi_0({\bf x})$, which is only used in Sec \ref{Sec:HarmonicOscillators} for calculating the projected Coulomb potential $U(r)$. The Mathieu wavefunctions are defined $\psi_{\pm}({\bf x})$ with respect to the full solution to the Mathieu equations in Sec \ref{Sec:Mathieu}, where $\pm$ distinguishes the layer bonded/antibonded moir\'e sublattices. The Wannier functions corresponding to the Mathieu solutions are written $\omega_{\bf R}({\bf x},\pm)$.

In Sec \ref{Sec:WKB}, we extend a lengthy calculation worked out in the supplement of Ref \cite{glazman2011}. In order to stay consistent with their derivation, we introduced their notation in Sec \ref{Sec:WKB}. This includes writing $\omega_{{\bf 0}}({\bf x},-)=\psi(\tilde{x}-l_{\mathcal{S}})\psi(\tilde{y}-l_{\mathcal{S}})$, where $\psi$ is the Wannier function for the ground state of the 1D Mathieu equation.

Any additional definitions are defined with respect to the above in their specific subsections.

\subsection{Harmonic oscillator limit}\label{Sec:HarmonicOscillators}

To start, note that the definition of ``layer-bonded" verse ``layer-antibonded" is an arbitrary distinction. If we are given a layer-bonded state $|\text{top}\rangle+|\text{bottom}\rangle$, we can convert it to a layer-antibonded one by a redefinition of $|\text{bottom}\rangle'=-|\text{bottom}\rangle$.

The Hamiltonian for a layer-bonded sector follows from Eqn. (7) of the main text, 
\begin{eqnarray}\label{Eqn:Suppl:HarmonicOscillatorDecouples1D}
H_{+}({\bf x},\boldsymbol{\nabla}) = -\mu\Big(\partial_{\tilde{x}}^2+\partial_{\tilde{y}}^2\Big)+2w_0\Big(\cos\big(q_M\tilde{x}\big)+\cos\big(q_M\tilde{y}\big)\Big) 
= H_{\tilde{x}} + H_{\tilde{y}} . 
\label{sup:eq:osc}
\end{eqnarray}
Written in terms of coordinates axes which connect next-nearest neighbor sites, $\tilde{x}=(x+y)/\sqrt{2}$ \& $\tilde{y}=(-x+y)/\sqrt{2}$, $H_{\text{lb}}$ decouples into orthogonal 1D Hamiltonians. The expansion of the potential about its minima 
\begin{eqnarray}
H_{\tilde{x}} = -\mu\partial_{\tilde{x}}^2 -w_0q_{M}^2\tilde{x}^2 + \cdots
\end{eqnarray}
can be truncated at quadratic order by taking $l_o/l_{\mathcal{S}}\rightarrow 0$, where we can interpret $l_o=\sqrt{2}(\mu w_0^{-1})^{\frac{1}{4}}q_M^{-{\frac{1}{2}}}$ as measuring the charge spread of the ground state, i.e $\psi_0(\tilde{x})\propto e^{-(\tilde{x}/l_o)^2}$. The 2D ground state follows from separation of variables 
\begin{eqnarray}
\psi_0(\tilde{x},\tilde{y}) = \frac{\sqrt{2/\pi}}{l_o} \exp\Bigg(-\frac{\tilde{x}^2+\tilde{y}^2}{l_o^2}\Bigg) . 
\end{eqnarray}
By comparison with the canonical form $H_{x}=-\hbar^2/(2m)\partial_x^2+m\omega^2x^2/2$, we have $\hbar^2/(2m)\equiv\mu$ and $m\omega^2/2\equiv w_0q_M^2$. The gap in this limit is the oscillator gap $E_{\text{gap}}=\hbar\omega=4\sqrt{w_0 E_C}$, where $E_C\equiv \mu q_M^2/4.$ (The $E_C$ notation is relevant for comparisons with Ref \cite{glazman2011} in Sec \ref{Sec:WKB}.)

Consider density-density interactions between sites connected by some arbitrary distance vector ${\bf d}=(d_x,d_y)$, and wavefunction normalization $N\equiv l_o^{-1}\sqrt{2/\pi}$:
\begin{eqnarray}
&&(4\pi\epsilon/e^2) U({\bf d}) = \frac{1}{2}\int_{{\bf r}{\bf r'}} \Big(\psi_{o}({\bf r}-{\bf d})\Big)^2 \frac{1}{|{\bf r}-{\bf r}'|} \Big(\psi_{o}({\bf r'})\Big)^2 
= N^4\frac{1}{2}\int_{{\bf r}{\bf r'}} e^{-2l_o^{-2}({\bf r}-{\bf d})^2} e^{-2l_o^{-2}{\bf r'}^2} \frac{1}{|{\bf r}-{\bf r}'|} . 
\end{eqnarray}
Which decouples into center-of-mass (${\bf R}$) and relative coordinate parts (${\boldsymbol\rho}$), via the map ${\bf r}={\bf R}+\frac{1}{2}\boldsymbol{\rho}$ \& ${\bf r}'={\bf R}-\frac{1}{2}\boldsymbol{\rho}$, 
\begin{eqnarray}
&&= N^4\frac{1}{2}\int_{{\bf R}}\int_0^{\infty}{d\rho}\int_{-\pi}^{\pi}{d\phi}\; e^{-2l_o^{-2}({\bf R}+\frac{1}{2}{\boldsymbol\rho}-{\bf d})^2} e^{-2l_o^{-2}({\bf R}-\frac{1}{2}{\boldsymbol\rho})^2} \notag\\
&&= \frac{N^4}{2}e^{-2l_o^{-2} d^2}\Bigg(\int_{0}^{\infty}{dR}\;Re^{-4l_o^{-2} R^2} \int_{-\pi}^{\pi}{d\theta}\; e^{4l_o^{-2}{\bf d}\cdot{\bf R}}\Bigg)\Bigg(\int_0^{\infty}{d\rho}\; e^{-l_o^{-2}\rho^2} \int_{-\pi}^{\pi}{d\phi}\; e^{2l_o^{-2}{\bf d}\cdot\boldsymbol{\rho}}\Bigg) \notag\\
&&= \frac{N^4}{2}e^{-2l_o^{-2} d^2}\Bigg(\frac{\pi}{4l_o^{-2}}e^{l_o^{-2} d^2}\Bigg)\Bigg(\frac{\pi^{3/2}}{\sqrt{l_o^{-2}}}e^{\frac{l_o^{-2} d^2}{2}}I_0\Big(\frac{l_o^{-2} d^2}{2}\Big)\Bigg) \notag\\
&&=\frac{N^4\pi^{5/2}}{8l_o^{-3}}e^{-\frac{l_o^{-2} d^2}{2}}I_0\Big(\frac{l_o^{-2} d^2}{2}\Big) , 
\end{eqnarray}
where in going to the second-to-last line, we used the fact that 
\begin{eqnarray}
\int_0^{\infty}{d\rho}\; e^{-\alpha\rho^2} \int_{-\pi}^{\pi}{d\phi}\; e^{2(\beta_x\cos{\phi}+\beta_y\sin{\phi})\rho} = \frac{\pi^{\frac{3}{2}}}{\sqrt{\alpha}} e^{\frac{\beta^2}{2\alpha}} I_{0}\Big(\frac{\beta^2}{2\alpha}\Big) . 
\end{eqnarray}
Or equivalently 
\begin{eqnarray}
I_{0}\Big(\frac{\beta^2}{2\alpha}\Big) = \frac{\sqrt{\alpha}}{\pi^{\frac{3}{2}}} e^{-\frac{\beta^2}{2\alpha}} \int_0^{\infty}{d\rho}\; e^{-\alpha\rho^2} \int_{-\pi}^{\pi}{d\phi}\; e^{2(\beta_x\cos{\phi}+\beta_y\sin{\phi})\rho} , 
\end{eqnarray}
is the zeroth modified Bessel function ($I_{0}(x)=J_{0}(ix)$), which satisfies the Bessel differential equation 
\begin{eqnarray}
x^2\frac{d^2I_0}{dx^2}+x\frac{dI_0}{dx}-x^2I_0=0 . 
\end{eqnarray}

\subsection{Mathieu equation and wavefunctions}\label{Sec:Mathieu}

Here we demonstrate that Eqn. (7) of the main text can be exactly solved using 1D Mathieu functions. Without loss of generality, we consider here the $+$ sector (the degenerate solutions in "-" sector are obtained by ${\bf x} \to {\bf x}+{\bf R}_1$), for which the wavefunction satisfies:
\begin{eqnarray}
-\mu\Big(\partial_{\tilde{x}}^2+\partial_{\tilde{y}}^2\Big)+2w_0\Big(\cos\big(q_M\tilde{x}\big)+\cos\big(q_M\tilde{y}\big)\Big)\psi_{+}(\tilde{x},\tilde{y}) = E\psi_{+}(\tilde{x},\tilde{y}) ,
\notag
\end{eqnarray}
where $q_M = \sqrt{2} \pi/l_a$ and $\tilde{x},\tilde{y}$ are oriented along ${\bf g}_1+{\bf g}_2$ and ${\bf g}_1-{\bf g}_2$, respectively. One notices that a solution can be found with variable separation ansatz $\psi_{+}(\tilde{x},\tilde{y}) = X(\tilde{x})Y(\tilde{y})$, such that result satisfies:
\begin{eqnarray}
E=E_X+E_Y,
\\
-\mu\partial_{\tilde{x}}^2X(\tilde{x})+2w_0\cos\big(q_M\tilde{x}\big)X(\tilde{x})= E_X X(\tilde{x}) ,
\\
-\mu\partial_{\tilde{y}}^2Y(\tilde{y})+2w_0\cos\big(q_M\tilde{y}\big)Y(\tilde{y})= E_Y Y(\tilde{y}).
\end{eqnarray}

The equations for $X(\tilde{x})$ and $Y(\tilde{y})$ can be reduced to the following dimensionless form:
\begin{equation}
    \xi''(z) + [a - 2 q \cos(2 z)]\xi(z) = 0,
\end{equation}
where $q = w_0/(\mu q_M^2/4) = \left(\frac{\theta^*}{\theta}\right)^2$ and $a = E_{X,Y}/(\mu q_M^2/4)$. The equation above is the Mathieu equation. The eigenenergies correspond to the lowest band are found from characteristic values $a_r(q)$, where $r\in(-1,1)$ is the Mathieu characteristic exponent corresponding to momentum in the Brillouin zone. Corresponding real symmetric eigenfunctions (which correspond to superpositions between $k$ in $-k$) are $ce_r(z,q)$ - the periodic Mathieu functions, normalized to $\int_0^{2\pi} dz ce_r(z,q)^2 = \pi$ .

The bandwidth $E^{max}-E^{min}$ is equal to $E_X^{max}-E_X^{min}+E_Y^{max}-E_Y^{min} = 2(\mu q_M^2/4)[a_1\left(1,q\right)-a_1\left(0,q\right)] = 2w_0 (\theta/\theta^*)^2[a_1\left(1,q\right)-a_1\left(0,q\right)]$ leading to (8) of the main text using $W=8t'$ in $V=0$ case.

For the wavefunction overlaps, Eqn. (9) of the main text, $\langle\psi_{+}^{\bf k=0}|\psi_{-}^{\bf k=0}\rangle$ one gets: $\langle\psi_{+}^{\bf k=0}|\psi_{-}^{\bf k=0}\rangle = \langle X^{\bf k=0}|X^{\bf k=0}(\tilde{x}+{\bf R_1}^{\tilde{x}})\rangle\langle Y^{\bf k=0}|Y^{\bf k=0}(\tilde{y}+{\bf R_1}^{\tilde{y}})\rangle$, where ${\bf R_1}^{\tilde{x}/\tilde{y}}$ is the component of ${\bf R_1}$ along $\tilde{x}/\tilde{y}$. The latter is expressed via the Mathieu functions as $\left(\int_0^\pi dx \, ce_0(x,q)\, ce_0(x+\pi/2,q)\right)^2/(\pi^2/4)$, where we used the $\pi$-periodicity of the $r=0$ Mathieu functions.

\subsection{Wannier orbitals and tight-binding formalism}
\label{sec:wannier}

We begin by demonstrating how the relevant Wannier orbitals are related to the basis states $|{\bf x},l\rangle$ of the continuum Hamiltonian Eqn. (6,7) of the main text, where $l=1,2$ is the layer index. Note that this means we are choosing $|{\bf x},l\rangle$ to be in the ``folded" representation, such that $\hat{\Psi}_l({\bf x})^{\dagger}|0\rangle=|{\bf x},l\rangle$. Since we know that Eqn. (6) is diagonal in the layer-bonding/antibonding basis, it will be more convenient for us to start by first rotating into this basis, defining 
\begin{eqnarray}\label{Eqn:Suppl:|x,lambda> def}
|{\bf x},\pm\rangle = \frac{|{\bf x},1\rangle\pm|{\bf x},2\rangle}{\sqrt{2}} , 
\end{eqnarray}
which we will write as $|{\bf x},\lambda\rangle$ where $\lambda=\pm$ is the index. The Fourier transform of $|{\bf x},\lambda\rangle$ are plane waves labelled by momenta ${\bf p}$, which we integrate over in the limit of an infinitely large system, 
\begin{eqnarray}\label{Eqn:Suppl:|x,lambda>}
\langle{\bf x},\lambda| = \int_{\bf p}e^{i{\bf x}\cdot{\bf p}}\langle{\bf p},\lambda| . 
\end{eqnarray}
Note that $H_f$ (Eqn. (6) of the main text) is a folded representation of $H$ (Eqn. (1) of the main text), which is periodic under an enlarged cell defined by basis vectors ${\bf R}_{f,1}=({\bf R}_1+{\bf R}_2)/2$ and $C_4{\bf R}_{f,2}$. Because of this, we choose to relabel our continuum momenta ${\bf p}={\bf k}_f+{\bf g}_f$, where ${\bf g}_f$ spans the discrete set of folded reciprocal lattice vectors, and ${\bf k}_f$ is a continuous variable which lies within the first folded zone. With this choice Eqn \ref{Eqn:Suppl:|x,lambda>} becomes 
\begin{eqnarray}\label{Eqn:Suppl:|x,lambda> 2}
 = \int_{{\bf k}_f}\sum_{{\bf g}_f}e^{i{\bf x}\cdot({\bf k}_f+{\bf g}_f)}\langle{\bf k}_f+{\bf g}_f,\lambda| . 
\end{eqnarray}
Translational invariance means that ${\bf k}_f$ is a conserved quantity, which has the consequence that the matrix structure of $H_f$ is block diagonal in the $\langle{\bf k}_f+{\bf g}_f,\lambda|$ basis, where each independent block is labelled by ${\bf k}$. Additionally, without additional perturbations ($V$ or $(B_x,B_y)$), $H_f$ is diagonal in the $\lambda$ basis, such that the Bloch states are layer-bonded/antibonded eigenstates. Therefore each value of $({\bf k}_f,\lambda)$ has a unitary transformation $u_{{\bf g}_f,\xi}({\bf k}_f,\lambda)$ which diagonalizes the blocks (the $u$ here should not to be confused with the displacement defined below equation \eqref{eq:T} in Appendix \ref{Sec:symmetry}), i.e $\langle{\bf k}_f+{\bf g}_f,\lambda|=\sum_{\xi}u_{{\bf g}_f,\xi}({\bf k}_f,\lambda)\langle{\bf k}_f,\xi,\lambda|$, where index $\xi$ labels the discrete set of energy bands at each value of $({\bf k}_f,\lambda)$, and ${\bf g}_f$ indexes the infinitely-many continuum plane waves with flavour $\lambda$. Rotating into the band basis allows us to reorganize Eqn \ref{Eqn:Suppl:|x,lambda> 2} into the form 
\begin{eqnarray}\label{Eqn:Suppl:|x,lambda> 2}
 = \sum_{\xi}\int_{{\bf k}_f}e^{i{\bf x}\cdot{\bf k}_f}\Big(\sum_{{\bf g}_f}e^{i{\bf x}\cdot{\bf g}_f}u_{{\bf g}_f,\xi}({\bf k}_f,\lambda)\Big)\langle{\bf k}_f,\xi,\lambda| 
 \equiv \sum_{\xi}\int_{{\bf k}_f}e^{i{\bf x}\cdot{\bf k}_f}u_{\xi,{\bf k}_f}({\bf x},\lambda)\langle{\bf k}_f,\xi,\lambda| . 
\end{eqnarray}
One may recognize $u_{\xi,{\bf k}_f}({\bf x},\lambda)$ as being the periodic part of the Bloch function, satisfying $u_{\xi,{\bf k}_f}({\bf x}+{\bf R}_f,\lambda)=u_{\xi,{\bf k}_f}({\bf x},\lambda)$ for any ${\bf R}_f\in\{n{\bf R}_{f,1}+m{\bf R}_{f,2}|(n,m)\in\mathbb{Z}^2\}$. The full Bloch wavefunction is therefore $\Psi_{\xi,{\bf k}_f}({\bf x},\lambda)=e^{i{\bf x}\cdot{\bf k}_f}u_{\xi,{\bf k}_f}({\bf x},\lambda)$.

For our purposes, the relevant bands are those near the Fermi level, of which there are two in the folded representation. We therefore fix $\xi=1$, which we choose to label the moire band in either $\lambda$ sector. Since the index is irrelevant from here out, we will drop it, with the understanding that $\lambda$ now distinguishes the two degenerate low-energy bands.

Because we have only a single band per $\lambda$, their Fourier transformation are localized Wannier orbitals, i.e $|{\bf k}_f,\lambda\rangle=\sum_{{\bf R}_f}e^{i{\bf R}_f\cdot{\bf k}_f}|{\bf R}_f,\lambda\rangle$. Thus Eqn \ref{Eqn:Suppl:|x,lambda> 2} becomes 
\begin{eqnarray}\label{Eqn:Suppl:|x,lambda> 3}
\langle{\bf x},\lambda| = \sum_{{\bf R}_f}\Big(\int_{{\bf k}_f}e^{-i{\bf R}_f\cdot{\bf k}_f}\Psi_{{\bf k}_f}({\bf x},\lambda)\Big)\langle{\bf R}_f,\lambda| 
 = \sum_{{\bf R}_f}\omega_{{\bf R}_f}({\bf x},\lambda)\langle{\bf R}_f,\lambda| . 
\end{eqnarray}
or equivalently 
\begin{eqnarray}
\omega_{{\bf R}_f}({\bf x},\lambda) = \sum_{{\bf k}_f}e^{-i{\bf R}_f\cdot{\bf k}_f}\Psi_{{\bf k}_f}({\bf x},\lambda) .
\end{eqnarray}
These Wannier functions $\omega_{{\bf R}_f}({\bf x},\lambda)$ are the Fourier transform of the ground states of the 2D Mathieu equations, the densities of which are centered at ${\bf R}_f$ (or ${\bf R}_f+{\bf R}_1$) for $\lambda=+1$ ($-1$). Thus we write $\omega_{{\bf R}_f}({\bf x})\equiv\omega_{{\bf R}_f}({\bf x},+)$ and $\omega_{{\bf R}_f+{\bf R}_1}({\bf x})=\omega_{{\bf R}_f}({\bf x}-{\bf R}_1)\equiv\omega_{{\bf R}_f}({\bf x},-)$, where again ${\bf R}_1=({\bf R}_{f,1}-{\bf R}_{f,2})/2$ is the moir\'e lattice vector.

What are the Wannier functions in the unfolded representation? The Wannier functions in the folded/unfolded representations cannot be the same. This is because for fixed ${\bf R}_f$, the pair of Wannier functions for different $\lambda$ correspond to different sublattices, and thus may have a finite overlap. However, in the unfolded representation, the pair of localized sublattices is replaced with a single set of orthogonal Wannier functions, which necessarily have vanishing overlaps. To construct the unfolded set of Wannier functions, we can take advantage of the transformation $\mathcal{U}$, Eqn. \ref{Eqn:U}) to shift one layer band relative from the other by momentum ${\bf q}_M$. For our purposes here, we will choose to use a modified version of Eqn. (5), which fixes the first layer at ${\bf k}=0$ while boosting the second. (The choice amounts to an arbitrary shift in the definition of the moir\'e BZ boundary.)
\begin{eqnarray}
\begin{pmatrix}
|{\bf x},1\rangle' \\ |{\bf x},2\rangle'
\end{pmatrix} = 
\begin{pmatrix}
1 & 0 \\
0 & e^{i{\bf x}\cdot{\bf q}_M}
\end{pmatrix}
\begin{pmatrix}
|{\bf x},1\rangle \\ |{\bf x},2\rangle
\end{pmatrix} . 
\end{eqnarray}
Which according to Eqn \ref{Eqn:Suppl:|x,lambda> def} means, 
\begin{eqnarray}
\langle{\bf x},\lambda|' = \frac{1}{\sqrt{2}}\Big(\langle{\bf x},1|+\lambda e^{-i{\bf x}\cdot{\bf q}_M}\langle{\bf x},2|\Big) . 
\end{eqnarray}
Inverting Eqn \ref{Eqn:Suppl:|x,lambda> def} for $\langle{\bf x},l|$ and plugging back into the above produces 
\begin{eqnarray}
=\frac{1}{2}\sum_{{\bf R}_f}\Big(\omega_{{\bf R}_f}({\bf x},+)(1+\lambda e^{-i{\bf x}\cdot{\bf q}_M})\langle{\bf R}_f,+|+\omega_{{\bf R}_f}({\bf x},-)(1-\lambda e^{-i{\bf x}\cdot{\bf q}_M})\langle{\bf R}_f,-|\Big) . 
\end{eqnarray}
Note that we are free to relabel $\langle{\bf R}_f|\equiv\langle{\bf R}_f,+|$ \& $\langle{\bf R}_f+{\bf R}_1|\equiv\langle{\bf R}_f,-|$, 
\begin{eqnarray}
=\frac{1}{2}\sum_{{\bf R}_f}\Big(\omega_{{\bf R}_f}({\bf x})(1+\lambda e^{-i{\bf x}\cdot{\bf q}_M})\langle{\bf R}_f|+\omega_{{\bf R}_f+{\bf R}_1}({\bf x})(1-\lambda e^{-i{\bf x}\cdot{\bf q}_M})\langle{\bf R}_f+{\bf R}_1|\Big) , 
\end{eqnarray}
and using the fact that $e^{i{\bf q}_M\cdot{\bf R}_{f,1}}=+1$ \& $e^{i{\bf q}_M\cdot{\bf R}_1}=-1$, 
\begin{eqnarray}
=\frac{1}{2}\sum_{{\bf R}}\omega_{{\bf R}}({\bf x})(1+\lambda e^{-i({\bf x}-{\bf R})\cdot{\bf q}_M})\langle{\bf R}| , 
\end{eqnarray}
taking note that the sum now runs over the moir\'e lattice sites (i.e ${\bf R}$ w/o the subscript $f$). It then follows that the unfolded Wannier functions are 
\begin{eqnarray}
\tilde\omega_{{\bf R}}({\bf x},\lambda)=\frac{1+\lambda e^{-i({\bf x}-{\bf R})\cdot{\bf q}_M}}{2}\omega_{{\bf R}}({\bf x}) . 
\end{eqnarray}
Unlike the folded set of Wannier functions $\omega_{{\bf R}}({\bf x})$, the unfolded Wannier functions are two-component vectors $\boldsymbol{\tilde\omega}_{{\bf R}}({\bf x})=(\tilde\omega_{{\bf R}}({\bf x},+),\tilde\omega_{{\bf R}}({\bf x},-))^T$ in the bonded/antibonded space. Thus we can see that the Wannier function peaked at ${\bf x}={\bf 0}$, i.e $\boldsymbol{\tilde\omega}_{{\bf 0}}({\bf x})$, is entirely bonding at its center ${\bf x}={\bf 0}$, and becoming antibonding at ${\bf x}={\bf R}_1$. Further, one can check that the unfolded set is orthogonal between different ${\bf R}$ \& ${\bf R}'$, either because: (i) ${\bf R}$ \& ${\bf R}'$ lie on the same folded sublattice, where $\omega_{{\bf R}}({\bf x})$ are guaranteed orthogonal by construction; or (ii) ${\bf R}$ \& ${\bf R}'$ are on different sublattices, but where $\boldsymbol{\tilde\omega}_{{\bf R}}({\bf x})^{\dagger}\boldsymbol{\tilde\omega}_{{\bf R}'}({\bf x})$ vanishes because of the phase. Despite this subtle distinction between the folded/unfolded sets of Wannier functions, their densities are identically: $\boldsymbol{\tilde\omega}_{{\bf R}}({\bf x})^{\dagger}\boldsymbol{\tilde\omega}_{{\bf R}}({\bf x}) = |\omega_{{\bf R}}({\bf x})|^2$.

\subsubsection{Hoppings: $t$,$t'$, \& $t_{x/y}$}

We start by writing the Hamiltonian in the folded representation, 
\begin{eqnarray}
H = \sum\limits_{\lambda\lambda'}\int\limits_{\bf x} |{\bf x},\lambda\rangle 
\begin{pmatrix}
h(-i\boldsymbol{\nabla}_{\bf x})+U({\bf x}) & \Delta h \\
\Delta h & h(-i\boldsymbol{\nabla}_{\bf x})-U({\bf x})
\end{pmatrix}_{\lambda\lambda'}
\langle{\bf x},\lambda'| , 
\end{eqnarray} 
which is diagonal in the bonding-antibonding basis modulo perturbations $\Delta h\equiv (-V-2\mu{\bf A}\cdot i\boldsymbol{\nabla})$. The hoppings within the tight-binding formalism are the projection of the Hamiltonian onto the localized Wannier functions, 
\begin{eqnarray}
\langle{\bf R}_f,\lambda|H|{\bf R}_f',\lambda'\rangle = \int\limits_{\bf x} \langle{\bf R}|{\bf x},\lambda\rangle 
\begin{pmatrix}
h(-i\boldsymbol{\nabla}_{\bf x})+U({\bf x}) & \Delta h \\
\Delta h & h(-i\boldsymbol{\nabla}_{\bf x})-U({\bf x})
\end{pmatrix}_{\lambda\lambda'}
\langle{\bf x},\lambda'|{\bf R}'\rangle ,  
\end{eqnarray}
where $\langle{\bf x},\lambda'|{\bf R}_f',\lambda'\rangle=\omega_{{\bf R}_f'}({\bf x},\lambda')$ follows from Eqn \Ref{Eqn:Suppl:|x,lambda> 3}. (Remember that the difference between $\langle{\bf R}|$ and $\langle{\bf R}_f,\lambda|$ is just a relabelling, i.e $\langle{\bf R}_f,\lambda|\equiv\langle{\bf R}_f+\frac{1}{2}(1-\lambda){\bf R}_1|$.) Thus we have 
\begin{eqnarray}
t'=\int\limits_{\bf x} \omega_{{\bf R}_{f,1}}({\bf x},+) 
\Big(h(-i\boldsymbol{\nabla}_{\bf x})+U({\bf x})\Big)
\omega_{{\bf 0}}({\bf x},+)
\end{eqnarray}
and 
\begin{eqnarray}
t_x=\int\limits_{\bf x} \omega_{{\bf 0}}({\bf x},-) 
\Delta h\,
\omega_{{\bf 0}}({\bf x},+) . 
\end{eqnarray}
The real part of $t_x=t_x^r+it_x^i$, 
\begin{eqnarray}
t_x^r=-V\int\limits_{\bf x} \omega_{{\bf 0}}({\bf x},-) 
\omega_{{\bf 0}}({\bf x},+)  
\end{eqnarray}
comes from the application of $V$, which does not break $C_4$ such that $t_x^r=t_y^r$. Its imaginary part is generated by in-plane field, 
\begin{eqnarray}
t_x^i=-2\mu\int\limits_{\bf x} \omega_{{\bf 0}}({\bf x},-) 
\big({\bf A}\cdot\boldsymbol{\nabla}\big)
\omega_{{\bf 0}}({\bf x},+) .
\end{eqnarray}
As demonstrated by Eqn \ref{Eqn:Suppl:HarmonicOscillatorDecouples1D}, the 2D Mathieu equations decouple into 1D problems such that we can write $\omega_{{\bf 0}}({\bf x},+)=\psi(\tilde{x})\psi(\tilde{y})$ and $\omega_{{\bf 0}}({\bf x},-)=\psi(\tilde{x}-l_{\mathcal{S}})\psi(\tilde{y}-l_{\mathcal{S}})$, where $\psi$ is the Wannier function for the ground state of the 1D Mathieu equation. Plugging this back into the formula 
\begin{eqnarray}
&&t_x^i=-2\mu\int\limits_{\bf x} \psi(\tilde{x}-l_{\mathcal{S}})\psi(\tilde{y}-l_{\mathcal{S}}) 
\big(A_{\tilde{x}}\partial_{\tilde{x}}+A_{\tilde{y}}\partial_{\tilde{y}}\big)
\psi(\tilde{x})\psi(\tilde{y}) \\
&=&-2\mu A_{\tilde{x}}\Big(\int\limits_{\tilde{x}}\psi(\tilde{x}-l_{\mathcal{S}})\partial_{\tilde{x}}\psi(\tilde{x})\Big)\Big(\int\limits_{\tilde{y}}\psi(\tilde{y}-l_{\mathcal{S}})\psi(\tilde{y})\Big)
-2\mu A_{\tilde{y}}\Big(\int\limits_{\tilde{x}}\psi(\tilde{x}-l_{\mathcal{S}})\psi(\tilde{x})\Big)\Big(\int\limits_{\tilde{y}}\psi(\tilde{y}-l_{\mathcal{S}})\partial_{\tilde{y}}\psi(\tilde{y})\Big)
\end{eqnarray}
The direct overlaps of the 1D Wannier functions are dimensionless, but the overlaps of the gradient carry the dimensions of the gradient. We therefore non-dimensionalize by defining $\varphi\equiv q_{M}\tilde{x}$, where $\partial_{\tilde{x}}=q_{M}\partial_{\varphi}$, giving 
\begin{eqnarray}
&=&-2\mu A_{\tilde{x}}q_M\Big(\int\limits_{\varphi}\psi(\varphi-\pi)\partial_{\varphi}\psi(\varphi)\Big)\Big(\int\limits_{\varphi'}\psi(\varphi'-\pi)\psi(\varphi')\Big)
-2\mu A_{\tilde{y}}q_M\Big(\int\limits_{\varphi}\psi(\varphi-\pi)\psi(\varphi)\Big)\Big(\int\limits_{\varphi'}\psi(\varphi'-\pi)\partial_{\varphi'}\psi(\varphi')\Big) \\
&=&-2\mu (A_{\tilde{x}}+A_{\tilde{y}})q_M\Big(\int\limits_{\varphi}\psi(\varphi-\pi)\partial_{\varphi}\psi(\varphi)\Big)\Big(\int\limits_{\varphi'}\psi(\varphi'-\pi)\psi(\varphi')\Big) \\
&=&-2\mu (A_{\tilde{x}}+A_{\tilde{y}})q_M\mathcal{O}_0\mathcal{O}_1 , 
\end{eqnarray}
where $\mathcal{O}_0\equiv\int\limits_{\varphi}\psi(\varphi-\pi)\psi(\varphi)$ and $\mathcal{O}_1\equiv\int\limits_{\varphi}\psi(\varphi-\pi)\partial_{\varphi}\psi(\varphi)$ are worked out in the next section. Now similarly, 
\begin{eqnarray}
t_y^i=-2\mu\int\limits_{\bf x} \psi(\tilde{x}-l_{\mathcal{S}})\psi(\tilde{y}+l_{\mathcal{S}}) 
\big(A_{\tilde{x}}\partial_{\tilde{x}}+A_{\tilde{y}}\partial_{\tilde{y}}\big)
\psi(\tilde{x})\psi(\tilde{y}) . 
\end{eqnarray}
Using the fact that $\psi(\varphi)=\psi(-\varphi)$, we have $\int\limits_{\varphi}\psi(\varphi+\pi)\partial_{\varphi}\psi(\varphi)
= -\int\limits_{\varphi}\psi(\varphi-\pi)\partial_{\varphi}\psi(\varphi)$, which we can use to get $t_y^i$. Together, 
\begin{eqnarray}
t_x^i&=&-2\mu (A_{\tilde{x}}+A_{\tilde{y}})q_M\mathcal{O}_0\mathcal{O}_1 \\
t_y^i&=&-2\mu (A_{\tilde{x}}-A_{\tilde{y}})q_M\mathcal{O}_0\mathcal{O}_1 .  
\end{eqnarray}
Noting that in our original coordinates: $A_{\tilde{x}}+A_{\tilde{y}} = \sqrt{2}A_x$ and $A_{\tilde{x}}-A_{\tilde{y}} = \sqrt{2}A_y$.

\subsection{WKB derivation of the overlaps}\label{Sec:WKB}

Because of the simplicity afforded by the folded representation, we were able to find analytical forms for our hoppings in the limit of flat isolated bands. Naively, one might try to calculate $t$ \& $t'$ analytically using a linear combination of atomic orbitals (LCAO), where the atomic orbitals are taken to be the Gaussian wavefunctions of the 2D harmonic oscillator, which is the correct wavefunction in the atomic (i.e $\theta\rightarrow 0$) limit of Eqn. (7) of teh main text. While quantities which depend only on densities -- such as $E_{\text{gap}}$ -- can be derived from the harmonic oscillator, tunnelings which depend on the tails of the Wannier functions cannot \cite{arzamasovs2017tight}. This is because the wavefunction inside the potential barrier (i.e the classically forbidden region), where the overlap between neighboring sites is highest, is non-Gaussian. As was demonstrated for 1D \cite{arzamasovs2017tight}, the asymptotic solutions to the Mathieu equations \cite{arzamasovs2017tight}, as well as semi-classical solutions which account for the classical turning points \cite{arzamasovs2017tight,glazman2011}, capture the correct form of the hopping.

Because Eqn. \ref{Eqn:Suppl:HarmonicOscillatorDecouples1D} decouples into two 1D Mathieu equations along orthogonal lines connecting next-to-nearest neighbor moir\'e sites, we were able to directly exploit the 1D solution \cite{arzamasovs2017tight} when writing $t'$ (Eqn. (8) of the main text). However, to obtain closed-form analytical results for $t$ (Eqn. (9,12) of the main text), we calculate the semi-classical approximation to the wavefunction in the overlapping region. We follow the derivation in Appendix B in Ref \cite{glazman2011} (where the semi-classical result for $t'$ is obtained).

\pv{In presence of finite $V$, eigenstates of Eq. \eqref{Eqn:Suppl:HarmonicOscillatorDecouples1D}  split at ${\bf k}=0$ by $\Delta E({\bf k}=0) = 2 V\langle\psi_{+}^{\bf k=0}|\psi_{-}^{\bf k=0}\rangle$. For $V
\ll E_{gap}$ (see below), mixing with other bands can be neglected. Since Eq. (5) of the main text folds the ${\bf \Gamma}_M$ and ${\bf M}_M$ points onto one another, this splitting is equal to $8t$. The wavefunction overlap can be obtained from Mathieu functions 
\cite{Note1,arzamasovs2017tight,glazman2011}.}

To keep consistent with the notation in \cite{glazman2011}, we'll write $E_C= \mu q_{M}^2/4$, $E_J=2w_0$, $\sqrt{\frac{E_J}{E_C}} = \sqrt{2} \frac{\theta^*}{\theta}$ and $\varphi\equiv q_{M}\tilde{x}$. Using this notation, the semi-classical momentum is 
\begin{eqnarray}
p(\varphi)=\frac{1}{\sqrt{4E_C}}\sqrt{V(\varphi)-E} 
=\frac{1}{\sqrt{4E_C}}\sqrt{-E+E_J(1+\cos(\varphi))} 
\end{eqnarray}
in the region $|\varphi|<\pi$. (Using $V(\varphi)=-E_j(1+\cos\varphi)$ in line with Ref \cite{glazman2011}.) In the classically allowed region, $|\varphi|<a$ the ground state is the ground state of the Harmonic oscillator with energy $E=E_{\text{gap}}/2-2E_J=\sqrt{2E_JE_C}-2E_J$; and where $a=\arccos(1-\sqrt{2E_C/E_J})$, which is approximately $a\simeq 0$ in the asymptotic (small twist angle) limit $\sqrt{E_C/E_J}\rightarrow 0$. The semi-classical approximation in the region $a<\varphi<\pi$ has the form 
\begin{eqnarray}
\psi(\varphi) \simeq \frac{C_0}{2\sqrt{p(\varphi)}}\exp\Bigg(-\int\limits_{a}^{\varphi}d\varphi'p(\varphi')\Bigg) . 
\end{eqnarray}
\pv{In what follows, we will need to approximate the wavefunction in vicinity of $\varphi=\pi/2$.} Note that $p(\varphi)$ can be rewritten as the sum of the following integrals, 
\begin{eqnarray}
\int\limits_{a}^{\varphi}d\varphi'p(\varphi')
=\int\limits_{a}^{\pi}d\varphi'p(\varphi')
-\int\limits_{\pi/2}^{\pi}d\varphi'p(\varphi')
-\int\limits_{\varphi}^{\pi/2}d\varphi'p(\varphi') . 
\end{eqnarray}
The solution to the first integral is in Ref \cite{glazman2011} -- see B14. In the limit $\sqrt{E_C/E_J}\rightarrow 0$, it becomes 
\begin{eqnarray}
\int\limits_{a}^{\pi}d\varphi'p(\varphi') \simeq 
\sqrt{\frac{2E_J}{E_C}}\Bigg(1-\frac{1}{2}\sqrt{\frac{E_C}{2E_J}}\log\Bigg[4\Big(\frac{2E_J}{E_C}\Big)^{1/4}\Bigg]-\frac{1}{4}\sqrt{\frac{E_C}{2E_J}}\Bigg) . 
\end{eqnarray}
The second integral has the exact form 
\begin{eqnarray}
\int\limits_{\pi/2}^{\pi}d\varphi'p(\varphi') 
=\sqrt{\frac{E_J}{4E_C}}\Bigg(2+\sqrt{2}\sqrt{\frac{E_C}{2E_J}}\log\Big(\tan\big(\frac{\pi}{8}\big)\Big)\Bigg) .  
\end{eqnarray}
This leaves the third integral, which carries the $\varphi$ dependence: 
\begin{eqnarray}
\int\limits_{\varphi}^{\pi/2}d\varphi'p(\varphi')
&=&\sqrt{\frac{E_J}{4E_C}}\int\limits_{\varphi}^{\pi/2}d\varphi'\sqrt{1-\sqrt{\frac{2E_C}{E_J}}-\cos\varphi'} \notag\\
&=&\sqrt{\frac{E_J}{4E_C}}\int\limits_{\varphi}^{\pi/2}d\varphi'\Big(\sqrt{1-\cos\varphi'}-\frac{1}{2}\frac{\sqrt{2E_C/E_J}}{\sqrt{1-\cos\varphi'}}\Big) + \mathcal{O}\Big(\sqrt{\frac{E_C}{E_J}}\Big) . 
\end{eqnarray}
The higher order corrections are irrelevant in the asymptotic limit. Dropping these and the remaining integral gives  
\begin{eqnarray}
=\sqrt{\frac{E_J}{4E_C}}\Bigg(\frac{2}{3}\Big(1-(1+\frac{\pi}{2}-\varphi)^{3/2}\Big)-\sqrt{\frac{2E_C}{E_J}}\Big(1-\sqrt{1+\frac{\pi}{2}-\varphi}\Big)\Bigg) . 
\end{eqnarray}
We keep the relevant contributions and now have 
\begin{eqnarray}\label{Eqn:Suppl:psi(varphi)}
\psi(\varphi \approx \pi/2) \simeq \frac{C_0e^{\frac{1}{4}}\sqrt{\tan(\pi/8)}}{\sqrt{p(\varphi)}}\Big(\frac{E_C}{2E_J}\Big)^{-\frac{1}{8}}e^{-\sqrt{\frac{2E_J}{E_C}}(1-\frac{1}{\sqrt{2}})}e^{-\frac{1}{3\sqrt{2}}\sqrt{\frac{2E_J}{E_C}}\big(-1+(1+\frac{\pi}{2}-\varphi)^{3/2}\big)}e^{-\frac{1}{\sqrt{2}}\big(1-\sqrt{1+\frac{\pi}{2}-\varphi}\big)} .
\end{eqnarray}

\subsubsection{Overlap $\mathcal{O}_0$}
The nearest-neighbor tunneling $t$ depends on the overlap, 
\begin{eqnarray}
\int d\varphi\; \psi(\varphi)\psi(\pi-\varphi) 
&&=\Bigg(\frac{C_0e^{\frac{1}{4}}\sqrt{\tan(\pi/8)}}{e^{\sqrt{\frac{2E_J}{E_C}}(1-\frac{1}{\sqrt{2}})}}\Bigg)^2\Big(\frac{E_C}{2E_J}\Big)^{-\frac{1}{4}} \pv{e^{-2\sqrt{\frac{2E_J}{E_C}}(1-\frac{1}{\sqrt{2}})}} \notag\\
&&\times\int d\varphi\; \frac{e^{-\frac{1}{3\sqrt{2}}\sqrt{\frac{2E_J}{E_C}}\big(-2+(1+\frac{\pi}{2}-\varphi)^{3/2}+(1-\frac{\pi}{2}+\varphi)^{3/2}\big)}e^{\frac{1}{\sqrt{2}}\big(-2+\sqrt{1+\frac{\pi}{2}-\varphi}+\sqrt{1-\frac{\pi}{2}+\varphi}\big)}}{\sqrt{p(\varphi)p(\pi-\varphi)}}
\end{eqnarray}
The peak of the integrand is at $\varphi=\pi/2$, so we change variables $z\equiv (\frac{\pi}{2}-\varphi)$ and expand the argument of the first exponential about $z=0$, 
\begin{eqnarray}
=\Bigg(\frac{C_0e^{\frac{1}{4}}\sqrt{\tan(\pi/8)}}{e^{\sqrt{\frac{2E_J}{E_C}}(1-\frac{1}{\sqrt{2}})}}\Bigg)^2\Big(\frac{E_C}{2E_J}\Big)^{-\frac{1}{4}} \pv{e^{-2\sqrt{\frac{2E_J}{E_C}}(1-\frac{1}{\sqrt{2}})}} \int dz\; \frac{e^{-\frac{1}{3}\sqrt{\frac{E_J}{E_C}}\big(\frac{3}{4}z^2+\mathcal{O}(z^4)\big)}e^{\frac{1}{\sqrt{2}}\big(-2+\sqrt{1+z}+\sqrt{1-z}\big)}}{\sqrt{p(\frac{\pi}{2}-z)p(\frac{\pi}{2}+z)}} . 
\end{eqnarray}
Rescaling $z'=(E_J/E_C)^{1/4}z$ brings out a factor of $(E_C/E_J)^{1/4}$ in front of the integral, the integrand of which can now be seen to be Gaussian, peaked at $z=0$. This Gaussian dominates in the asymptotic limit, such that we can treat it as a Gaussian integral in the range $-\infty<z<\infty$, with all other subleading contributions to the integrand evaluating at $z=0$, i.e 
\begin{eqnarray}
&=&\Bigg(\frac{C_0e^{\frac{1}{4}}\sqrt{\tan(\pi/8)}}{\sqrt{p(\frac{\pi}{2})}}\Bigg)^2\Big(\frac{E_C}{2E_J}\Big)^{-\frac{1}{4}}e^{-2\sqrt{\frac{2E_J}{E_C}}(1-\frac{1}{\sqrt{2}})}\Big(\frac{E_C}{E_J}\Big)^{1/4}\int\limits_{-\infty}^{\infty} dz\; e^{-\frac{1}{4}z^2} \notag\\
&=& \Bigg(\frac{C_0e^{\frac{1}{4}}\sqrt{\tan(\pi/8)}}{\sqrt{p(\frac{\pi}{2})}}\Bigg)^2\Big(\frac{E_C}{2E_J}\Big)^{-\frac{1}{4}}e^{-2\sqrt{\frac{2E_J}{E_C}}(1-\frac{1}{\sqrt{2}})}\Big(\frac{E_C}{E_J}\Big)^{1/4}\sqrt{4\pi} . 
\end{eqnarray}
The constant $C_0=\sqrt{(\pi e)^{-1/2}\sqrt{E_J/(2E_C)}}$ is worked out in \cite{glazman2011}. Plugging this in, 
\begin{eqnarray}
\mathcal{O}_0\equiv \int d\varphi\; \psi(\varphi)\psi(\pi-\varphi) = \frac{\tan(\pi/8)}{2^{-7/4}}e^{-(2-\sqrt{2})\sqrt{\frac{2E_J}{E_C}}} . 
\end{eqnarray}

\subsubsection{Overlap of the gradient $\mathcal{O}_1$}

For clarity, let us rewrite Eqn \ref{Eqn:Suppl:psi(varphi)} as $\psi(\varphi)=\frac{N}{\sqrt{p(\varphi)}}e^{-\frac{1}{3\sqrt{2}}\sqrt{\frac{2E_J}{E_C}}\big(-1+(1+\frac{\pi}{2}-\varphi)^{3/2}\big)}e^{-\frac{1}{\sqrt{2}}\big(1-\sqrt{1+\frac{\pi}{2}-\varphi}\big)}$, where all $\varphi$-independent terms have been coalesced into the prefactor $N$. As written, the $\varphi$-dependence of the wavefunction is a product of three parts, the most relevant of which is the first exponential carrying the argument scaled by $\sqrt{E_J/E_C}$. The overlap of the gradient 
\begin{eqnarray}
-\mathcal{O}_{1}\equiv\int d\varphi\; \psi(\pi-\varphi)(-\partial_{\varphi})\psi(\varphi)
\end{eqnarray}
is dominated by the derivative of this exponential, so long as it does not vanish (which we will see it does not). Thus 
\begin{eqnarray}
= \int d\varphi\; \psi(\pi-\varphi)\psi(\varphi)\partial_{\varphi}\Bigg(\frac{1}{3\sqrt{2}}\sqrt{\frac{2E_J}{E_C}}\big(-1+(1+\frac{\pi}{2}-\varphi)^{3/2}\big)\Bigg) ,  
\end{eqnarray}
where we have dropped the sub-leading corrections. Continuing with the derivative, 
\begin{eqnarray}
= \frac{1}{2}\sqrt{\frac{E_J}{E_C}}\int d\varphi\; \psi(\pi-\varphi)\psi(\varphi)\sqrt{1+\frac{\pi}{2}-\varphi} .  
\end{eqnarray}
As demonstrated previously, the overlap $\psi(\varphi)\psi(\pi-\varphi)$ is dominated by the product of their dominate exponentials, which forms a strong Gaussian peak centered on $\varphi=\pi/2$. We then have  
\begin{eqnarray}
= \frac{1}{2}\sqrt{\frac{E_J}{E_C}}\int d\varphi\; \psi(\pi-\varphi)\psi(\varphi) ,    
\end{eqnarray}
or equivalently 
\begin{eqnarray}
\mathcal{O}_1 = -\frac{1}{2}\sqrt{\frac{E_J}{E_C}}\mathcal{O}_0 . 
\end{eqnarray}

\subsubsection{Direct and superexchange interactions}

In this section, we use the WKB wavefunction to compute direct and superexchange interactions in the Mott insulating state. We will also neglect here the separation between the layers, assuming it to be much smaller than the moir\'e lattice scale. For nearest-neighbor sites, the Coulomb interaction leads to a direct exchange between spins: $-J_D {\bf S}_{\bf R} {\bf S}_{{\bf R}+{\bf R_1}}$ , where $J_D$ is given by \cite{auerbach2012interacting}:
\begin{equation}
    \begin{gathered}
        J_D = \int d {\bf x} d{\bf x}' \frac{e_0^2}{\epsilon |{\bf x} - {\bf x}'|} \omega_{{\bf R}}({\bf x})
        \omega^*_{{\bf R}+{\bf R}_1}({\bf x})
        \omega_{{\bf R}+{\bf R}_1}({\bf x}')
        \omega^*_{{\bf R}}({\bf x}')
        =
        \\
       = \frac{2\pi e_0^2 \theta}{\sqrt{2} \epsilon l_a}
        \int d \varphi_1 d \varphi_2 d \varphi_1' d \varphi_2'
        \frac{1}{|{\vec \varphi} - {\vec \varphi}'|} 
        \psi(\varphi_1)\psi(\pi-\varphi_1)
        \psi(\varphi_2)\psi(\pi-\varphi_2)
        \psi(\varphi_1')\psi(\pi-\varphi_1')
        \psi(\varphi_2')\psi(\pi-\varphi_2')=
        \\
        =
        \frac{2\pi e_0^2 \theta}{\sqrt{2} \epsilon l_a}
        \int \frac{d {\bf q}}{(2\pi)^2} \frac{2\pi}{|{\bf q}|} I(q_1)I(q_2)I(-q_1)I(-q_2),
    \end{gathered}
\end{equation}
where 
\begin{equation}
    \begin{gathered}
I(q) = \int d\varphi  \psi(\varphi)\psi(\pi-\varphi) e^{i q (\varphi -\pi/2)}
=
\mathcal{O}_0 \cdot  e^{-q^2 \sqrt{E_J/E_C}}.
    \end{gathered}
\end{equation}
We thus arrive at:
\begin{equation}
    \begin{gathered}
        J_D = \mathcal{O}_0^4 \frac{2\pi e_0^2 \theta}{\sqrt{2} \epsilon l_a}
        \int  \frac{d {\bf q}}{(2\pi)^2}  \frac{2\pi}{|{\bf q}|} e^{-2 {\bf q}^2 \sqrt{E_C/E_J}} 
        =\mathcal{O}_0^4 \frac{2\pi e_0^2 \theta}{\sqrt{2} \epsilon l_a}
        \sqrt{\frac{\pi}{2 \sqrt{2}}\frac{\theta^*}{\theta}} \approx
        \\
        \approx
        17.6 
        \sqrt{\theta^*\theta} 
        e^{-8(2-\sqrt{2})\frac{\theta^*}{\theta} }
        \frac{e_0^2}{\epsilon l_a}.
    \end{gathered}
    \label{sup:eq:jd}
\end{equation}

The nearest-neighbor superexchange is equal to \cite{auerbach2012interacting} $\frac{4t^2}{U_0}$, i.e.
\begin{equation}
  J_{SE} = 4 V^2 \mathcal{O}_0^4 /\left(\frac{e_0^2}{4\pi\epsilon l_a}\frac{\pi^{3/2}}{2^{3/2}}\sqrt{\theta\theta^*}\right)
  \approx
 96.2  e^{-8(2-\sqrt{2})\frac{\theta^*}{\theta} } \frac{V^2}{\sqrt{\theta^*\theta} \frac{e_0^2}{\epsilon l_a}}
\end{equation}
such that 
\begin{equation}
  J_{D}/J_{SE}
    =
    \left( \frac{e_0^2}{\epsilon l_a V} \right)^2 \theta^* \theta \cdot (\pi/4)^2 /2^{7/4} \approx 0.18 \theta^* \theta \left( \frac{e_0^2}{\epsilon l_a V} \right)^2
\end{equation}

With the same approach, we can calculate the next-nearest-neighbor exchange interactions. For the direct exchange, we have:

\begin{equation}
    \begin{gathered}
        J_D' = \frac{2\pi e_0^2 \theta}{\sqrt{2} \epsilon l_a}
        \int d \varphi_1 d \varphi_2 d \varphi_1' d \varphi_2'
        \frac{1}{|{\vec \varphi} - {\vec \varphi}'|} 
        \psi(\varphi_1)\psi(2\pi-\varphi_1)
        \psi(\varphi_2)\psi(\varphi_2)
        \psi(\varphi_1')\psi(2\pi-\varphi_1')
        \psi(\varphi_2')\psi(\varphi_2')=
        \\
        =
        \frac{2\pi e_0^2 \theta}{\sqrt{2} \epsilon l_a}
        \int \frac{d {\bf q}}{(2\pi)^2} \frac{2\pi}{|{\bf q}|} I_1(q_1)I_2(q_2)I_1(-q_1)I_2(-q_2),
    \end{gathered}
\end{equation}
where 
\begin{equation}
    \begin{gathered}
I_1(q) = \int d\varphi  \psi(\varphi)\psi(2\pi-\varphi) e^{i q (\varphi -\pi)} 
\approx
\frac{2}{\sqrt{2 \pi}} \left(\frac{8 E_J}{E_C}\right)^{1/4} e^{-2\sqrt{2 E_J/E_C}}
\int_0^\infty d\varphi  \cos(q \varphi) e^{-\sqrt{\frac{E_J}{2E_C}}\frac{\varphi^3}{24} }=
\\
=
\frac{2}{\sqrt{2 \pi}} \left(\frac{8 E_J}{E_C}\right)^{1/4} 
\frac{e^{-2\sqrt{2 E_J/E_C}}}{(E_J/2 E_C)^{1/6}}
f\left(q\left(\frac{2 E_C}{E_J}\right)^{1/6}\right);
\\
f(a) = \int_0^\infty dz \cos(a z) e^{-\frac{z^3}{24}} = \int_-\infty^\infty dz e^{i a z} e^{-\frac{|z|^3}{24}} ,
\\
I_2(q) = \int d\varphi  \psi^2(\varphi) e^{i q \varphi} =  e^{-\sqrt{\frac{E_C}{2 E_J}} q^2}.
    \end{gathered}
\end{equation}
To evaluate $I_2$, we used the harmonic oscillator approximation to the wavefunction near $\varphi=0$\cite{glazman2011}: $\psi(\varphi)\approx \frac{1}{(2\sqrt{2} \pi)^{1/4}} \left(\frac{E_J}{E_C}\right)^{1/8} e^{-\frac{\varphi^2}{2} \sqrt{\frac{E_J}{8 E_C}} }$.

Now we can evaluate $J_D'$:
\begin{equation}
    \begin{gathered}
        J_D' = 
        \frac{e_0^2 \theta}{\sqrt{2} \epsilon l_a}
        \frac{2}{\pi} \left(\frac{8 E_J}{E_C}\right)^{1/2} 
\frac{e^{-4\sqrt{2 E_J/E_C}}}{(E_J/2 E_C)^{1/3}}
        \int \frac{d q_1 d q_2} {\sqrt{q_1^2+q_2^2}}
 e^{-2\sqrt{\frac{E_C}{2 E_J}} q_2^2}
        f^2\left[q_1\left(\frac{2 E_C}{E_J}\right)^{1/6}\right].
    \end{gathered}
\end{equation}
In the limit $E_J/E_C\gg 1$ one notices that the integrand is cut off at $q_2\sim(E_J/E_C)^{1/4}$ and $q_1\sim (E_J/E_C)^{1/6}\ll (E_J/E_C)^{1/4}$. Thus we can approximate the integral (with logarithmic accuracy) as follows:
\begin{equation}
\begin{gathered}
            \int \frac{d q_1 d q_2} {\sqrt{q_1^2+q_2^2}}
 e^{-2\sqrt{\frac{E_C}{2 E_J}} q_2^2}
        f^2\left[q_1\left(\frac{2 E_C}{E_J}\right)^{1/6}\right]
        \approx
\int_{|q_2|\gtrsim (E_J/E_C)^{1/6} } \frac{d q_2} {q_2}
 e^{-2\sqrt{\frac{E_C}{2 E_J}} q_2^2}
 \int d q_1
        f^2\left[q_1\left(\frac{2 E_C}{E_J}\right)^{1/6}\right]=
        \\
        =2 \left( \log \left[ \frac{E_J}{E_C}\right]^{1/12} +O(1)  \right) 
        \left(\frac{E_J}{2 E_C}\right)^{1/6} 4\pi (2/3)^{2/3} \Gamma[1/3],
        \end{gathered}
\end{equation}
where we used:
\begin{equation}
\begin{gathered}
     \int d x f^2(x) = \int dx \int dz e^{i x z} e^{-\frac{|z|^3}{24}}
     \int dz' e^{i x z'} e^{-\frac{|z'|^3}{24}}=
     \\
     =
     2\pi \int dz dz' \delta(z-z')  e^{-\frac{|z|^3}{24}} e^{-\frac{|z'|^3}{24}} = 2\pi  \int dz  e^{-\frac{|z|^3}{12}} = 4\pi (2/3)^{2/3} \Gamma[1/3].
     \end{gathered}
\end{equation}
Collecting all the terms above, we can bring the answer to the following form:
\begin{equation}
    \begin{gathered}
        J_D' \approx 
        \frac{2 \Gamma[1/3] (2/3)^{5/3} e_0^2 \theta}{\sqrt{2} \epsilon l_a}
     \left(\frac{8 E_J}{E_C}\right)^{1/2} 
\frac{e^{-4\sqrt{2 E_J/E_C}}}{(E_J/2 E_C)^{1/6}}
\log \left[ \frac{E_J}{E_C}\right]
\approx
\\
\approx
7.7
       \frac{ e_0^2 \theta^{1/3} (\theta^*)^{2/3} }{\epsilon l_a}
e^{-8 \frac{\theta^*}{\theta}}
\log \left[ \frac{\theta^*}{\theta}\right].
    \end{gathered}
\end{equation}

The expression for next-nearest neighbor superexchange is:
\begin{equation}
  J_{SE}' = 4 t'^2 /\left(\frac{e_0^2}{4\pi\epsilon l_a}\frac{\pi^{3/2}}{2^{3/2}}\sqrt{\theta\theta^*}\right)
  \approx
  4096 \sqrt{2/\pi^3}
  \frac{\theta}{\theta^*}
\frac{w_0^2}{\sqrt{\theta^*\theta} \frac{e_0^2}{\epsilon l_a}} e^{-8\frac{\theta^*}{\theta}} 
\approx
10^3  \frac{\theta}{\theta^*} \frac{w_0^2}{\sqrt{\theta^*\theta} \frac{e_0^2}{\epsilon l_a}} e^{-8\frac{\theta^*}{\theta}}.
\end{equation}
Already for $e^2/l_a\sim 1$ eV, $w_0\sim 10$ meV and $\theta\sim \theta^*$ one finds $J_D'/J_{SE}'\sim 10^{2} (\theta^*)^{13/6} \theta^{-1/6}/\epsilon^2$. As $\theta,\theta^*\sim 10^{-1\div -2}$ this suggests that next-nearest-neighbor direct exchange can always be neglected.

This is not the case for direct nearest-neighbor exchange \eqref{sup:eq:jd}, as the exponential factor is weaker in that case. For the same parameters, $J_D/J_{SE}'\sim 10^{2} (\theta^*)^2  e^{8(\sqrt{2}-1)\frac{\theta^*}{\theta} } /\epsilon^2$. The exponential factor quickly becomes large, such that at $\theta = 0.7 \theta^*$ one has $e^{8(\sqrt{2}-1)\frac{\theta^*}{\theta} }\approx 100$. Therefore, at sufficiently low twist angles, direct nearest-neighbor exchange would be the dominant spin-spin interaction. Interestingly, since direct exchange and superexchange have opposite signs, one can use the displacement field $V$ to tune effective nearest-neighbor interactions to zero and make them undergo a sign change.

\pv{\section{Interlayer capacitance from flat bands}
The intrinsic (quantum) interlayer capacitance is defined as $C_q=-e \frac{d(n_{top} - n_{bot})}{d V}$ \cite{levitov2011}. We assume a fixed chemical potential of the bilayer (e.g., due to connection to an external lead). We first find $n_{top} - n_{bot} = \langle \sigma_z\rangle = \langle \frac{d \hat{H}}{d V}\rangle$. 
\\
Let us first find the contribution of the top narrow bands to $n_{top} - n_{bot}$. 
The Hamiltonian projected to flat band has been found to take the form $c^\dagger_{\bf k} \epsilon_{\bf k}(V) c_{\bf k}$, where $\epsilon_{\bf k}(V)=2 t(V) (\cos (k_x) + \cos(k_y))+4t'\cos(k_x)\cos(k_y);\;t(V) \approx \alpha(\theta) V$, and $\alpha(\theta)\approx_{_{\theta\ll \theta^*}}\frac{\tan^2\left(\frac{\pi}{8}\right)}{2^{-7/2}} e^{-4\left(2-\sqrt{2}\right) \frac{\theta^*}{\theta}} $. As a result, the contribution of the flat bands reads $\langle \frac{d \hat{H}}{d V}\rangle_{FB} = \frac{d t}{dV} \sum_{\bf k} (\cos(k_x)+\cos(k_y)) n(\epsilon_{\bf k}-\mu)$, where $n(\epsilon_{\bf k}-\mu)$ is the Fermi distribution. The capacitance at $T\to 0$  is then equal to:
\begin{equation}
\begin{gathered}
    \frac{C_q^{FB}}{S} =
    \sum_{\bf k} e^2\alpha^2(\theta) 2 (\cos(k_x)+\cos(k_y))^2 \delta(\epsilon_{\bf k}-\mu)
    =
    e^2\alpha^2(\theta) \frac{\theta^2}{l_a^2} I_C(\mu,t(V),t'),
    \\
    I_C(\mu,t(V),t') = 2\int \frac{d {\bf k}}{(2\pi)^2} (\cos(k_x)+\cos(k_y))^2 \delta(\epsilon_{\bf k}-\mu).
\end{gathered}
\end{equation}
where $(k_x,k_y) \in (-\pi,\pi)$. A useful comparison can be made with the density of states:
\begin{equation}
\begin{gathered}
    \nu = 2 \int \frac{d {\bf k}}{(2\pi)^2} \delta(\epsilon_{\bf k}-\mu).
\end{gathered}
\end{equation}
\\
\begin{figure}
    \centering
    \begin{minipage}[h]{0.47\linewidth}
\begin{center}
\includegraphics[width=1\linewidth]{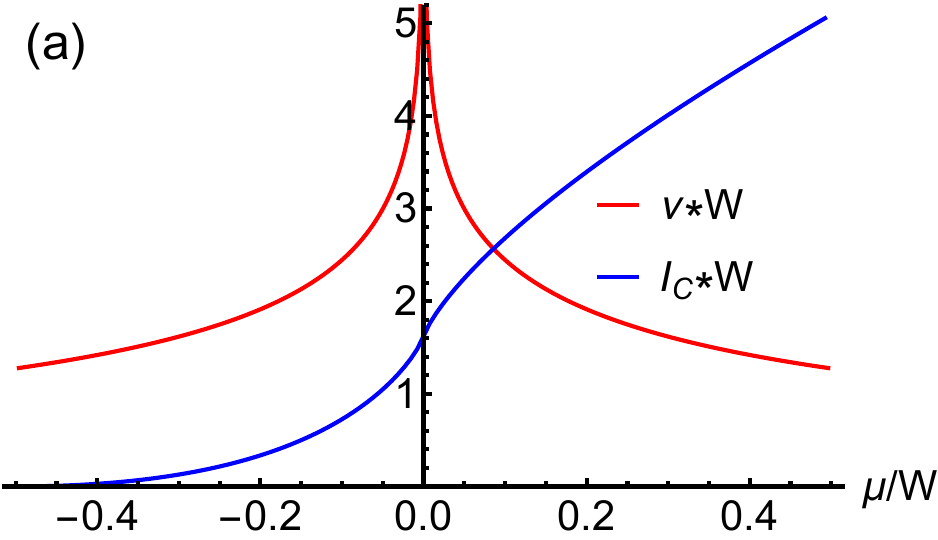} 
\end{center} 
\end{minipage}
\hfill
\vspace{0.2 cm}
\begin{minipage}[h]{0.47\linewidth}
\begin{center}
\includegraphics[width=1\linewidth]{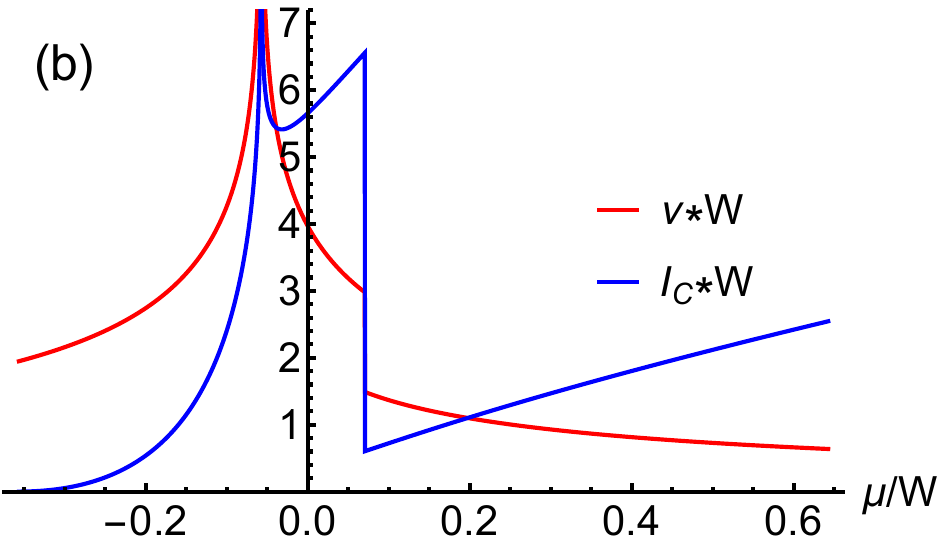} 
\end{center}
\end{minipage}
\vfill
\vspace{0.2 cm}
\begin{minipage}[h]{0.47\linewidth}
\begin{center}
\includegraphics[width=1\linewidth]{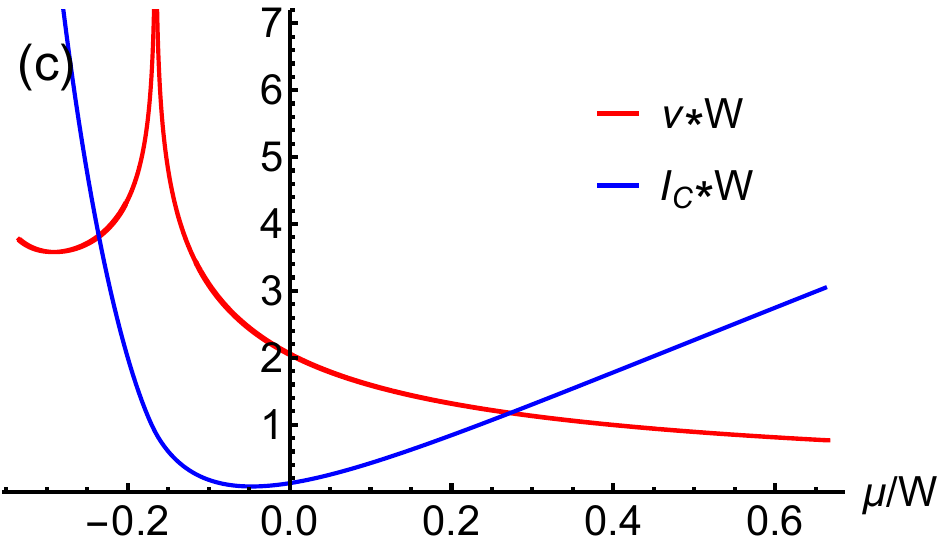} 
\end{center}
\end{minipage}
\hfill
\begin{minipage}[h]{0.47\linewidth}
\begin{center}
\includegraphics[width=1\linewidth]{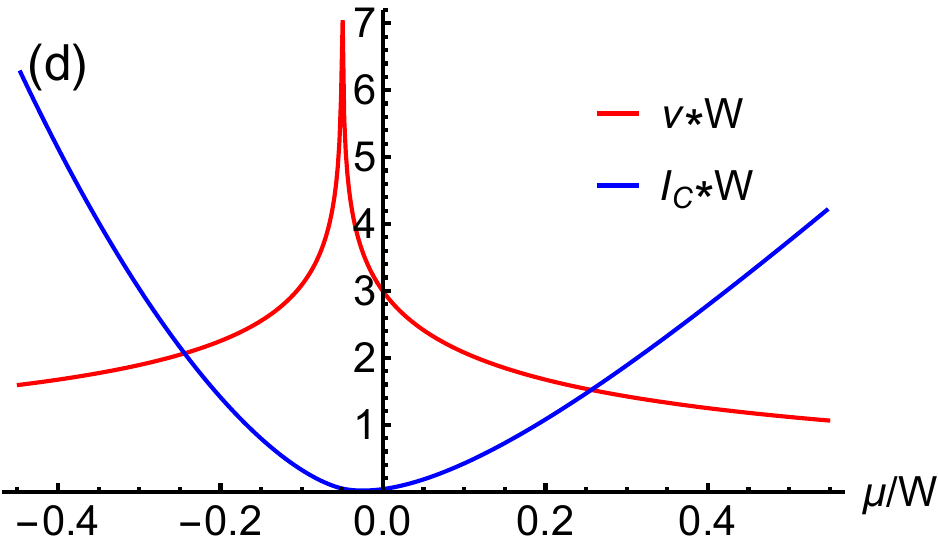} 
\end{center}
\end{minipage}
    \caption{(a) t/t'=0 (b) t/t' = 0.8 t/t'=3 t/t'=10}
    \label{fig:dosints}
\end{figure}
In Fig. \ref{fig:dosints} we present $W*\nu(\mu,t,t')$ and $W* I_\nu(\mu,t,t')$ normalized to the bandwidth $W$ for several values of $t/t'$. Overall, both quantities are typically of order $1$, but the details differ. Most notably, the Van Hove singularity in the density of states appears only for intermediate $t/t'$ (panel (b)). This can be understood as follows. At $t=0$ the saddle points of the energy bands are at ${\bf k} = (\pm \pi/2, \pm; \mp \pi/2)$ where $\cos(k_x)+\cos(k_y)=0$. Similarly for $|t|>2|t'|$ they are at $\vec{k} = (\pi,0); (0,\pi)$. However, for $0<|t|<2|t'|$, saddle points of the dispersion are at ${\bf k} = (\pm \arccos(t/2t'), \pm; \mp \arccos(t/2t'))$, where $\cos(k_x)+\cos(k_y)$ doesn't generically vanish. 
\\
An upper estimate on the contributions of the other bands can be obtained assuming an empty top band (so that $V$ smaller than the band gap does not change the occupation numbers for a fixed chemical potential $\mu$) as $\delta C = - \frac{d E^2}{d V^2} \approx 4 \sum_{{\bf k},n} \frac{|\langle \psi_{{\bf k},FB}|\sigma_z|\psi_{{\bf k},n'}\rangle|^2}{E_{FB}({\bf k}) - E_{n}({\bf k}) }$. For low-energy states the tight-binding picture holds; in the limit $\theta\ll \theta^*$ we can obtain an estimate neglecting the bandwidth. Then, the eigenstates can be taken as just the Wannier functions of each band and one expects $|\langle \psi_{{\bf k},FB}|\sigma_z|\psi_{{\bf k},n'}\rangle|$ to be suppressed with the same exponential factor as for the top flat band due to localization of wavefunction of two sectors on different sublattices. On the other hand, the denominator is $\gtrsim E_{gap}$, so that overall $\delta C/ C^{FB}_q \sim W/E_{gap}\ll 1$ is expected at low twist angles.
\\
At higher binding energies $|E|\gtrsim w_0$ the states are no longer well localized and are better represented by plane waves. The overlap $|\langle \psi_{{\bf k},FB}|\sigma_z|\psi_{{\bf k},n'}\rangle|^2$ can be then estimated from the overlap of a plane wave and ground-state oscillator wavefunction around the minimum of the potential term in Eq. \eqref{sup:eq:osc}. The result $\frac{2}{\pi}\frac{\theta}{\theta^*} e^{-{\bf k}^2\frac{\sqrt{\mu} l_a}{\sqrt{2w_0} \pi \theta}} \lesssim\frac{2}{\pi}\frac{\theta}{\theta^*} e^{-\frac{\theta^*}{\theta}} $, where we assumed $\mu k_{x,y}^2  = w_0$ in the final estimate. Thus the $\delta C/ C_q^{FB} \sim \left(\frac{\theta}{\theta^*} e^{-\frac{\theta^*}{\theta}}\right) /(\alpha^2(\theta)/W) \propto \left(\frac{\theta}{\theta^*} \right)^{(3/2)}  e^{(-1+4(3-2\sqrt{2}))\frac{\theta^*}{\theta}}$ where we took $W\propto t'$ and used expressions (8,9) from the main text. As $-1+4(3-2\sqrt{2}) \approx -0.3$, together with the prefactor $\left(\frac{\theta}{\theta^*} \right)^{(3/2)}$ this suggest that the contribution of these states is also subleading at low twist angles.}

\end{document}